\documentclass[aps,twocolumn,prd,nofootinbib,showpacs]{revtex4-1}
\usepackage{amsmath}
\usepackage{amssymb}
\usepackage[aps]{mrv}
\usepackage{graphicx}
\usepackage{hyperref}
\usepackage{mathrsfs}
\usepackage{bm}

\makeatletter
\gdef\@fpheader{}
\g@addto@macro\bfseries{\boldmath}
 \makeatother
\newlength{\figw}
\newlength{\figmaxw}
\newlength{\figh}
\newcommand{\heaviside}[1]{{\Theta}\!\left( #1 \right)}
\newcommand{\heavisideb}[1]{{\Theta}\!\left[ #1 \right]}
\newcommand{\sigmam}{\sigma_{-}}
\newcommand{\sigmap}{\sigma_{+}}

\newcommand{\usr}{\mathrm{sr}}
\newcommand{\utr}{\mathrm{tr}}
\newcommand{\ustart}{\mathrm{start}}
\newcommand{\uuv}{\mathrm{uv}}

\newcommand{\uFL}{\mathrm{FL}}
\newcommand{\umat}{\mathrm{mat}}
\newcommand{\ubbn}{\eisss{\mathrm{BBN}}}
\newcommand{\utot}{\mathrm{tot}}

\newcommand{\uKL}{\mathrm{KL}}
\newcommand{\uGHS}{\eisss{\mathrm{GHS}}}

\newcommand{\expo}{\ee}
\newcommand{\Nx}{N_{\times}}

\newcommand{\vol}{\mathscr{V}}
\newcommand{\dist}{\mathscr{D}}
\newcommand{\orb}{\mathscr{C}}

\newcommand{\lfi}{\mathrm{LFI}}
\newcommand{\si}{\mathrm{SI}}
\newcommand{\hi}{\mathrm{HI}}
\newcommand{\sfi}{\mathrm{SFI}}
\newcommand{\gdwi}{\mathrm{GDWI}}
\newcommand{\dwi}{\mathrm{DWI}}
\newcommand{\cwi}{\mathrm{CWI}}
\newcommand{\ccsi}{\mathrm{CCSI}}

\newcommand{\phip}{\Phi}
\newcommand{\phipini}{\phip_\uini}
\newcommand{\phipmax}{\phip_{\max}}
\newcommand{\phipx}{\phip_{\times}}
\newcommand{\phiuv}{\phi_{\uuv}}

\newcommand{\phiVmax}{\phi_{\iVmax}}
\newcommand{\Gammasr}{\Gamma_{\usr}}
\newcommand{\Gammaini}{\Gamma_{\uini}}
\newcommand{\Gammax}{\Gamma_{\times}}
\newcommand{\Vx}{V_{\times}}
\newcommand{\xx}{x_{\times}}

\newcommand{\DGamma}{\Gamma_{\negthickspace,N}}

\newcommand{\fieldinf}{\textsc{FieldInf}}
\newcommand{\Hend}{H_\uend}
\newcommand{\sixty}{\mathcal{N}_\uinf}

\newcommand{\bmk}{\boldmathsymbol{k}}
\newcommand{\bmx}{\boldmathsymbol{x}}
\newcommand{\bmy}{\boldmathsymbol{y}}
\newcommand{\bmu}{\boldmathsymbol{u}}
\newcommand{\bmF}{\boldmathsymbol{F}}

\newcommand{\calK}{\mathcal{K}}
\newcommand{\calC}{\mathcal{C}}
\newcommand{\calD}{\mathcal{D}}
\newcommand{\calM}{\mathcal{M}}
\newcommand{\calN}{\mathcal{N}}
\newcommand{\OmegaK}{\Omega_{\calK}}
\newcommand{\OmegaKo}{\Omega_{\calK_0}}
\newcommand{\calLmax}{\calL_{\max}}
\newcommand{\Cmax}{\calC_{\max}}
\newcommand{\Cmin}{\calC_{\min}}
\newcommand{\Cflat}{\calC_{\flat}}
\newcommand{\Minf}{\calM_{\uinf}}
\newcommand{\MFL}{\calM_{\uFL}}

\newcommand{\DKL}{D_{\uKL}}

\newcommand{\varpibbn}{\varpi_{\ubbn}}
\newcommand{\varpiini}{\varpi_{\uini}}
\newcommand{\varpio}{\varpi_{0}}
\newcommand{\Ninf}{N_\uinf}
\newcommand{\Ninfbar}{\bar{N}_\uinf}
\newcommand{\Nrad}{N_\urad}
\newcommand{\Nthird}{N_{1/3}}
\newcommand{\Nmat}{N_\umat}
\newcommand{\Ntot}{N_\utot}
\newcommand{\Nflat}{N_\flat}
\newcommand{\barN}{\bar{N}}
\newcommand{\barn}{\bar{n}}

\newcommand{\Binf}{B_\uinf}

\newcommand{\wbar}{\overline{w}}
\newcommand{\wbarrad}{\wbar_{\urad}}
\newcommand{\wbarmat}{\wbar_{\umat}}
\newcommand{\No}{N_0}

\newcommand{\rhobbn}{\rho_{\ubbn}}
\newcommand{\zbbn}{z_\ubbn}
\newcommand{\zeq}{z_\ueq}

\newcommand{\Ham}{\mathscr{H}}
\newcommand{\Surf}{\mathscr{S}}
\newcommand{\adim}{\mathscr{A}}

\newcommand{\RN}[1]{\textup{\uppercase\expandafter{\romannumeral#1}}}
\hypersetup{
    unicode=false,          % non-Latin characters in Acrobat’s bookmarks
    pdftoolbar=true,        % show Acrobat’s toolbar?
    pdfmenubar=true,        % show Acrobat’s menu?
    pdffitwindow=false,     % window fit to page when opened
    pdfstartview={FitH},    % fits the width of the page to the window
    pdfkeywords={keyword1} {key2} {key3}, % list of keywords
    pdfnewwindow=true,      % links in new window
    colorlinks=true,       % false: boxed links; true: colored links
    linkcolor=red,          % color of internal links
    citecolor=cyan,        % color of links to bibliography
    filecolor=magenta,      % color of file links
    urlcolor=blue,           % color of external links
    linktocpage=true
}
\begin{document}

\title{Inflation after Planck: Judgment Day}

\author{Debika Chowdhury} \email{debika@theory.tifr.res.in}
\affiliation{Department of Physics, Indian Institute of Technology
  Madras, Chennai 600036, India}
\affiliation{Department of Theoretical Physics,
   Tata Institute of Fundamental Research, Mumbai
   400005, India}
% \altaffiliation[Current address: ]{Department of Theoretical Physics,
%   Tata Institute of Fundamental Research, Homi Bhabha Road, Mumbai
%   400005, India}

\author{J\'er\^ome Martin} \email{jmartin@iap.fr}
\affiliation{Institut d'Astrophysique de Paris, UMR 7095-CNRS,
Universit\'e Pierre et Marie Curie, 98 bis boulevard Arago, 75014
Paris, France}

\author{Christophe Ringeval} \email{christophe.ringeval@uclouvain.be}
\affiliation{Cosmology, Universe and Relativity at Louvain,
  Institute of Mathematics and Physics, Louvain University, 2 Chemin
  du Cyclotron, 1348 Louvain-la-Neuve, Belgium}

\author{Vincent Vennin} \email{vincent.vennin@apc.univ-paris7.fr}
\affiliation{Laboratoire Astroparticule et Cosmologie, Universit\'e
  Denis Diderot Paris 7, 10 rue Alice Domon et L\'eonie Duquet, 75013
  Paris, France}

\date{\today}

\begin{abstract}
  Inflation is considered as the best theory of the early universe by
  a very large fraction of cosmologists. However, the validity of a
  scientific model is not decided by counting the number of its
  supporters and, therefore, this dominance cannot be taken as a proof
  of its correctness. Throughout its history, many criticisms have
  been put forward against inflation. The final publication of the
  Planck cosmic microwave background data represents a benchmark time
  to study their relevance and to decide whether inflation really
  deserves its supremacy. In this paper, we categorize the criticisms
  against inflation, go through all of them in the light of what is
  now observationally known about the early universe, and try to infer 
  and assess the scientific status of inflation. Although we find that
  important questions still remain open, we conclude that the
  inflationary paradigm is not in trouble but, on the contrary, has
  rather been strengthened by the Planck data.
\end{abstract}

\maketitle

%%%%%%%%%%%%%%%%%%%%%%%%%%%%%%%%%%%%%%%%%%%%%%%%%%%%%%%%%%%%%%%%%%%%%%%%%%%%
\section{Introduction}
\label{sec:intro}
%%%%%%%%%%%%%%%%%%%%%%%%%%%%%%%%%%%%%%%%%%%%%%%%%%%%%%%%%%%%%%%%%%%%%%%%%%%%

Inflation refers to a period of accelerated
expansion of the universe~\cite{Starobinsky:1979ty,
  Starobinsky:1980te, Guth:1980zm, Linde:1981mu, Albrecht:1982wi,
  Linde:1983gd, Mukhanov:1981xt, Mukhanov:1982nu, Starobinsky:1982ee,
  Guth:1982ec, Hawking:1982cz, Bardeen:1983qw}. It is a paradigm aimed
at overcoming the various difficulties associated with the
conventional hot big bang model of Friedmann and Lema\^{\i}tre, such
as the horizon problem and the flatness problem. Furthermore, the
inflationary scenario also provides a natural mechanism for generating
primordial perturbations that subsequently act as seeds for the
formation of large-scale structures. According to inflation, they are
the unavoidable quantum fluctuations of the inflaton and gravitational
fields, amplified by gravitational instability and stretched by the
cosmic expansion.

Although no sociological data are available, it seems fair to say that
inflation is viewed as the best paradigm for the early universe by a
vast majority of scientists working in the field of
cosmology. However, the validity of a scientific theory shall not be
decided by a democratic vote but by a careful study of its content and
predictions. Throughout its history, inflation has received various
criticisms on its different aspects. This is certainly sound since a
healthy scientific process for validating a theory implies a skeptical
and critical approach. Recently, the final Planck Cosmic Microwave
Background (CMB) data have been released, and it therefore seems
especially timely to take stock of these criticisms and to assess the
status of inflation in light of these new CMB measurements. This is
the main goal of this article.

Let us mention that other works have addressed this topic from various
perspectives; see, for instance, Refs.~\cite{Guth:2013sya,
  Linde:2014nna, Mishra:2018dtg, Marsh:2018fsu}. The present paper
aims at being exhaustive, and presents various new results that shed
new light on some of the commonly discussed issues. The article is
organized as follows. We have identified nine different broad classes
of criticisms that we discuss one by one. The first type concerns the
initial conditions needed to start inflation in a homogeneous and
isotropic situation. In particular, \Ref{Ijjas:2013vea} has argued
that the Planck data precisely single out models for which this issue
is most problematic. This question is treated in \Sec{sec:beyondI}. The
second type of criticisms, addressed in \Sec{sec:beyondFL}, concerns
the ability of inflation to make the universe isotropic and
homogeneous, the question being whether inflation requires some
extent of homogeneity to begin with. In \Sec{sec:tpl}, we briefly
mention the trans-Planckian problem of inflation, which is also an
initial condition problem but, this time, for the perturbations. In
\Sec{sec:qminflation}, we discuss how the inflationary mechanism for
structure formation is impacted by foundational issues of Quantum
Mechanics. In \Sec{sec:probainf}, we consider another type of
criticisms related to the likeliness of inflation. It is sometimes
argued that, if the extended phase-space is endowed with a ``proper''
measure, there is very little chance to start inflation. This requires
first to define what ``proper'' exactly means and, in this section, we
discuss this issue. In \Sec{sec:measure}, we consider a related
question, namely how the choice of a measure in the field phase-space
affects the initial condition problem of \Sec{sec:beyondI}. In
particular, we pay attention to how the existence of an attractor
depends on what we assume about the measure. In \Sec{sec:building}, we
summarize the model building question. Implementing inflation in
high-energy physics indeed presents important challenges. In
\Sec{sec:flatness}, we discuss another class of criticisms consisting
in challenging the basic motivations of the inflationary scenario,
{\ie}, the hot big bang model problems. We focus on the flatness problem
since this point has been recurrently pushed forward in the recent
years. Finally, in \Sec{sec:multiverse}, we briefly comment on a
criticism of different nature, namely the supposedly unavoidable
presence of a multiverse. In the last section, \Sec{sec:conclusion},
we present our conclusions and summarize the current status of
inflation. Throughout this article, we consider single-scalar-field
implementations of inflation only, both for simplicity and since this
is the framework in which the criticisms we will address are usually
formulated.

\section{Initial conditions beyond inflation}
\label{sec:beyondI}

As mentioned in \Sec{sec:intro}, a first class of criticisms against
inflation concerns the initial conditions that are needed to start a
phase of accelerated expansion. It has indeed been argued in various
works that, generically, it is difficult to start inflation and that,
consequently, inflation is unlikely. Notice that this initial
condition problem is a multifaceted question. One can indeed study it
in the most general setup but one can also specify the microphysics of
inflation as being that of \eg~a self-gravitating single field $\phi$,
with standard kinetic terms, in an isotropic, homogeneous and
spatially flat Friedmann-Lema\^itre-Robertson-Walker spacetime
(FLRW).  In this section, we mostly investigate this case since this
is the one considered in many articles on the subject. In fact, the
criticisms put forward in this context are well exemplified and
summarized in \Ref{Ijjas:2013vea} and, in this section, we therefore
consider this paper as representative of this type of arguments. The
impact of anisotropies, inhomogeneities and initial spatial curvature
on the initial conditions, which is a much more difficult problem, is
discussed in \Sec{sec:beyondFL}.

A lexical warning is also in order before we start. In this section,
by ``fine-tuned'' initial conditions, we mean a set of initial
conditions that occupies a tiny fraction of phase-space. Of course,
this implicitly assumes a measure on phase-space. The papers
discussing the initial conditions problem usually do not specify any
measure, hence they implicitly assume the ``naive'', or ``flat'', one,
namely proportional to the field and its velocity. This is why we also
assume the flat measure in this section, before examining in
\Sec{sec:measure} how the results derived below depend on the choice
of the measure.

In the next subsection, we introduce the tools needed to discuss
phase-space trajectories outside the slow-roll regime. Our approach
differs from the seminal work~\cite{Goldwirth:1989pr,Goldwirth:1991rj}
and, we argue, is more efficient. Then, endowed with these appropriate
tools, we apply our formalism to well-known examples and discuss in
detail the criticisms put forward in \Ref{Ijjas:2013vea}.

\subsection{Phase-space trajectories}
\label{sec:pspacetrajec}

By definition, (slow-roll) inflation requires the kinetic energy
stored in the inflaton field to remain (much) smaller than the
potential energy. In this section, we exhaustively explore
single-field dynamics on flat FLRW space times, and
determine the conditions under which a suitable inflationary phase
takes place.

\subsubsection{Equations of motion}
\label{sec:eom}

In an isotropic and homogeneous situation, the system is controlled by
three equations, namely the Friedmann-Lema\^itre and the Klein-Gordon
equations
\begin{align}
  \label{eq:hubble}
  H^2 &= \dfrac{1}{6\Mp^2}\left[ \Pi^2 + 2 V(\phi)\right], \\
  \label{eq:hubbledot}
  H^2 + \dot{H} & = -\dfrac{1}{6\Mp^2} \left[2 \Pi^2 - 2 V(\phi) \right], \\
  \label{eq:kg}
  \dot{\Pi} & + 3H\Pi+ \dfrac{\ud V(\phi)}{\ud \phi} = 0,
\end{align}
where $\Pi\equiv \dot{\phi}$, a dot denoting a derivative with respect
to cosmic time $t$ and $\Mp$ is the reduced Planck mass. The quantity $H$
is the Hubble parameter, $V(\phi)$ is the potential function. One of
these three equations is redundant as imposed by the stress tensor
conservation. As such, the Cauchy problem is solved by setting initial
conditions on $(\phi_\uini,\Pi_\uini)$, from which there is a unique
solution $\phi(t)$, $\Pi(t)$ and $H(t)$.

The system of equations \eqref{eq:hubble}-\eqref{eq:kg} can be
decoupled by changing from the cosmic time variable to the number
of e-folds $N\equiv \ln a$, where $a(t)$ is the FLRW scale factor. In
that situation, one can define a dimensionless field velocity,
measured in e-folds, by
\begin{equation}
\Gamma \equiv \dfrac{\ud \phip}{\ud N} = \dfrac{1}{\Mp}
\dfrac{\ud \phi}{\ud N} = \dfrac{\Pi}{\Mp H}\,,
\label{eq:Gamma}
\end{equation}
where we have defined
\begin{equation}
\phip \equiv \dfrac{\phi}{\Mp}\,.
\end{equation}
The field dynamics is equivalently described in the phase-space
$(\phip,\Gamma)$. From \Eqs{eq:hubble} and \eqref{eq:hubbledot} one
obtains
\begin{equation}
  H^2 = \dfrac{2}{\Mp^2} \dfrac{V(\phi)}{6 - \Gamma^2}\,, 
\qquad \epsilon_1 \equiv
  -\dfrac{\ud \ln H}{\ud N} = \dfrac{1}{2} \Gamma^2.
\label{eq:flefold}
\end{equation}
The Hubble parameter is thus completely determined from $V(\phi)$ and
$\Gamma$. The quantity $\epsilon_1$ is the first Hubble-flow
function~\cite{Schwarz:2001vv, Schwarz:2004tz}. Because inflation
requires $\ddot{a} > 0$, {\ie}, $\epsilon_1<1$, this translates into the
condition $\Gamma^2 < 2$. Let us notice that slow roll would further
require $\Gamma^2 \ll 1$. Moreover, assuming that $V(\phi) \ge 0$,
\Eq{eq:flefold} shows that the field velocity is always bounded by
\begin{equation}
-\sqrt{6} \le \Gamma \le \sqrt{6}\,.
\label{eq:Gammabounds}
\end{equation}
In the limit $\Gamma^2 \rightarrow 6$, a regime that we refer to as
``kination'', the kinetic energy of the field becomes infinite, such
that the condition~(\ref{eq:Gammabounds}) encompasses all the possible
kinetic regimes for a single scalar field. Expressing the Klein-Gordon
equation \eqref{eq:kg} in terms of e-folds, and using the first of
\Eqs{eq:flefold}, gives
\begin{equation}
\dfrac{2}{6 - \Gamma^2} \dfrac{\ud \Gamma}{\ud N} + \Gamma = - 
\dfrac{\ud \ln V}{\ud \phip}\,.
\label{eq:kgefold}
\end{equation}
This equation is much easier to deal with than the original coupled
system. Moreover, the field trajectory in the phase-space $(\phi,\Pi)$
can always be explicitly recovered from \Eqs{eq:Gamma} and
\eqref{eq:flefold}, which yield
\begin{equation}
\Pi = \sqrt{\frac{2 V(\phi)}{6 - \Gamma^2}} \, \Gamma.
\label{eq:PiofGamma}
\end{equation}

The functional form of \Eq{eq:kgefold} already gives an answer to the
question of whether a large kinetic energy may prevent
inflation from occurring. It is indeed similar to the differential equation
describing a relativistic particle of speed $\Gamma$, accelerated by a
force deriving from the potential $W=\ln V$, in the presence of a constant
friction term (the equivalent of the speed of light would be
$c^2=6$). For all initial velocities, and for a force term not varying
too fast, after a transient acceleration, the particle settles at the
friction-driven terminal velocity, namely
\begin{equation}
\Gamma \simeq \Gammasr \equiv -\dfrac{\ud \ln V}{\ud \phip}\,.
\label{eq:Gammasr}
\end{equation}
The above expression is actually the approximation used within the
slow-roll inflationary regime.

In the following we show that kination is indeed a repeller for
not-too-steep potentials. New non-perturbative solutions for the
transition from kination to inflation are derived, and we also recover
the ultra-slow-roll regime~\cite{Kinney:2005vj, Martin:2012pe,
  Motohashi:2014ppa} together with the usual slow roll.

\subsubsection{Sustained kination?}
\label{sec:asol}

All of the field dynamics are described by \Eq{eq:kgefold}, which we can
rewrite in a more convenient way:
\begin{equation}
\dfrac{1}{\sqrt{6}} \dfrac{\ud}{\ud N}\left[ \ln \left(\dfrac{\sqrt{6}
    + \Gamma}{\sqrt{6}-\Gamma} \right) \right] + \Gamma = - \dfrac{\ud
  \ln V}{\ud \phip} \equiv \Gammasr(\phip)\,.
\label{eq:DGamma}
\end{equation}
This expression shows that large deviations from slow roll, defined as
$\Gamma \simeq \Gammasr$, are present as soon as
$|\Gamma| \simeq \sqrt{6}$, {\ie}, in kination. The
first question we address is whether a field starting deeply in the
kination regime can sustainably remain in this state.

To this end, we solve \Eq{eq:DGamma} perturbatively. Assuming that
$\Gammaini \simeq +\sqrt{6}$, one can define
\begin{equation}
0<\gamma \equiv \sqrt{6} - \Gamma \ll 1.
\label{eq:gammadef}
\end{equation}
Remarking that $\ud/\ud N = \Gamma \ud / \ud \phip$, and plugging
\Eq{eq:gammadef} into \Eq{eq:DGamma} gives
\begin{equation}
\dfrac{1}{\gamma} \dfrac{\ud \gamma}{\ud \phip} - \sqrt{6} =
\dfrac{\ud \ln V}{\ud \phip}\,,
\end{equation}
at leading order in $\gamma$. The solution reads
\begin{equation}
\gamma(\phip) = \left(\sqrt{6} - \Gammaini \right) \dfrac{\ee^{\sqrt{6}
    \phip} V(\phip)}{\ee^{\sqrt{6} \phipini} V(\phipini)}\,.
\end{equation}
As a result, $\gamma$ is always positive and, because the system
evolves toward larger field values ($\Gammaini>0$), the exponential
term in the numerator implies that kination is \emph{generically} a
repeller and cannot be sustained. In order for $\gamma(\phip)$ to be a
decreasing function, one would need
\begin{equation}
\dfrac{V(\phip)}{V(\phipini)} < \ee^{-\sqrt{6}\left(\phip - \phipini\right)}\,,
\end{equation}
{\ie}, the potential function would have to decrease faster than
$\ee^{-\sqrt{6} \phi/\Mp}$. Such a potential is too steep to support
slow-roll inflation at all.

The symmetric situation obtained when kination comes from an initial
negative field velocity, $\Gammaini \simeq -\sqrt{6}$, gives the same
result: defining $\gamma \equiv \sqrt{6} - \left|\Gamma\right|$, one
finds at leading order
\begin{equation}
\gamma(\phip) = \left(\sqrt{6} - \left|\Gammaini\right| \right) 
\dfrac{\ee^{-\sqrt{6}
    \phip} V(\phip)}{\ee^{-\sqrt{6} \phipini} V(\phipini)}\,.
\end{equation}
Again, kination toward decreasing field values is sustained only if
the potential function increases faster than
$\ee^{\sqrt{6} \phi/\Mp}$.

We conclude that, in any potential region allowing for inflation to
settle, kination is a repeller and cannot be
sustained~\cite{Goldwirth:1991rj}. In the next section, we discuss the
relaxation from kination toward inflation.

\subsubsection{Relaxation toward inflation}
\label{sec:relax}

There is no exact analytical solution of \Eq{eq:DGamma}, but the two
terms in the left-hand side of this equation encode all the effects
coming from the kinetic term and the friction term while the
right-hand side is the ``force term'' that will be driving
slow roll. In the previous section, we have studied the regime
$\Gamma^2 \simeq 6$ for which the kinetic term dominates everything
else. Let us now study the transition regime in which the field leaves
kination and enters into inflation. If we assume that the potential is
flat enough, then, for most of the transitional phase, the ``force
term'' remains small with respect to the kinetic and friction terms,
{\ie}, we have
\begin{equation}
  |\Gamma| \gg |\Gammasr|.
\label{eq:forcesmall}
\end{equation}
Let us stress again that inflation occurs for $|\Gamma| < \sqrt{2}$
whereas slow-roll inflation occurs only for $\Gamma \simeq
\Gammasr$. Therefore, the inflationary regimes explored under this
approximation are necessarily non slow rolling and \Eq{eq:DGamma}
becomes
\begin{equation}
\dfrac{1}{\sqrt{6}} \Gamma \dfrac{\ud}{\ud \phip}
\left[ \ln \left(\dfrac{\sqrt{6}
    + \Gamma}{\sqrt{6}-\Gamma} \right) \right] + \Gamma \simeq 0.
\label{eq:DGammaApprox}
\end{equation}

This equation admits two exact solutions in phase-space,
\begin{equation}
\Gamma(\phip) = \sqrt{6}\, \dfrac{\Gammaini - \sqrt{6}
  \tanh \negthinspace \left[\dfrac{\sqrt{6}}{{2}}\left(\phip -
    \phipini\right)\right]}{\sqrt{6} - \Gammaini
  \tanh \negthinspace \left[\dfrac{\sqrt{6}}{{2}}\left(\phip -
    \phipini\right)\right]}\,,
\label{eq:GammaApprox}
\end{equation}
and $\Gamma = 0$. This is a new non-perturbative solution of the field
dynamics in phase-space that describes the transition from kination,
in which $\Gamma^2 \simeq 6$, to inflation when $\Gamma^2 < 2$.

The inflationary regime reached for $|\Gamma| \lesssim \sqrt{2}$ and
$|\Gamma| > |\Gammasr|$ partially encompasses various kinetically
driven inflationary regimes discussed in the
literature~\cite{Linde:2001ae, Handley:2014bqa}. The field excursion
and the number of e-folds can actually be derived when
\Eq{eq:forcesmall} is satisfied. If we define the field excursion by
\begin{equation}
\Delta \phip \equiv \phip - \phipini,
\end{equation}
given $(\phipini,\Gammaini)$, one can immediately derive $\Delta
\phipx$ such that $\Gamma$ relaxes from $\Gammaini$ to a given value
$\Gammax$ (still assuming that $|\Gammax| > |\Gammasr|$). Solving for
$\Gamma=\Gammax$ in \Eq{eq:GammaApprox} yields
\begin{equation}
\Delta \phipx = \dfrac{1}{\sqrt{6}}
\ln\left(\dfrac{1+\dfrac{\Gammaini}{\sqrt{6}}
      \dfrac{1-\dfrac{\Gammax}{\Gammaini}}{1 -
        \dfrac{\Gammax\Gammaini}{6}}}{
    1 - \dfrac{\Gammaini}{\sqrt{6}} \dfrac{1 -
      \dfrac{\Gammax}{\Gammaini}}{1 - \dfrac{\Gammax \Gammaini}{6}}} \right).
\label{eq:deltaphipx}
\end{equation}
The logarithmic dependence shows that, in terms of Planckian field
excursion, the relaxation from kination to inflation is relatively
``short''\footnote{If a sub-Planckian vacuum expectation value (vev),
  say $\mu$, fixes the typical scale of $\phi$, or the size of the
  inflating domain, then \Eq{eq:deltaphipx} may actually correspond to
  a large field excursion in terms of $\mu$. This is further discussed
  in \Sec{sec:sfi}.}. Because $\Gamma = \ud \phip/\ud N$, the number
of e-folds associated with a field excursion $\Delta\phip$ is obtained
by a direct integration of \Eq{eq:GammaApprox} and reads
\begin{equation}
\begin{aligned}
  \Delta N   \equiv N - \Nini & = 
  \frac{1}{6}\Bigg\lbrace \ln\left(2 \Gammaini^2\right)  -\ln\left[1
    + \cosh\left(\sqrt{6}\Delta\phip\right)\right] 
  \\ &
    -2\ln \left[\left|\Gammaini
      - \sqrt{6} \tanh\left(\dfrac{\sqrt{6}}{2} \Delta\phip\right)
      \right|\right]\Bigg\rbrace\, .
\end{aligned}
\label{eq:invtraj}
\end{equation}

Plugging the value of $\Delta \phipx$ given by \Eq{eq:deltaphipx} into
\Eq{eq:invtraj} gives
\begin{equation}
\Delta \Nx \equiv \Nx - \Nini = \dfrac{1}{6} \ln \left[\dfrac{1 -
    \dfrac{\Gammaini^2}{6}\left( \dfrac{1-\dfrac{\Gammax}{\Gammaini}}{1 -
      \dfrac{\Gammax\Gammaini}{6}}\right)^2 }{\left(1 -
    \dfrac{1-\dfrac{\Gammax}{\Gammaini}}{1 -
      \dfrac{\Gammax\Gammaini}{6}} \right)^2} \right].
\label{eq:deltaNx}
\end{equation}
The logarithmic dependence of $\Delta \Nx$ with respect to $\Gammaini$
is again showing that the field trajectory usually spends only a very
few number of e-folds in this regime. However, with some amount of
tuning, the number of e-folds spent in the transitional regime from
kination to inflation can become large. To boost the number of e-folds
spent in kination, $\Gammaini$ can be taken very close to
$\pm\sqrt{6}$. Similarly, the number of transitional inflationary
e-folds can be increased by pushing $\Gammax$ to very small
values. There is indeed a logarithmic divergence of the denominator in
\Eq{eq:deltaNx} for $\Gammax \to 0$, which corresponds to
$\Delta \Nx \to \infty$ and $\Delta\phipx \to \Delta\phipmax$ with
\begin{equation}
\Delta \phipmax = \dfrac{1}{\sqrt{6}} \ln \left(\dfrac{1+
    \dfrac{\Gammaini}{\sqrt{6}}}{1 - \dfrac{\Gammaini}{\sqrt{6}}}
  \right).
\label{eq:deltaphipmax}
\end{equation}
Notice that taking the limit $\Gammax \to 0$ while ensuring our
working hypothesis in Eq.~\eqref{eq:forcesmall} requires $|\Gammasr|
\to 0$, namely the potential should be extremely flat.

In fact, the transitional inflationary regime taken in the small
$|\Gamma|$ limit, while enforcing the condition
$|\Gamma|\gg|\Gammasr|$, is the so-called ultra-slow roll
regime~\cite{Kinney:2005vj, Martin:2012pe, Motohashi:2014ppa}. Again,
it is contained in \Eqs{eq:GammaApprox}, \eqref{eq:invtraj} and
\eqref{eq:deltaNx}. This can be explicitly shown by Taylor expanding
\Eq{eq:GammaApprox} at field values for which $|\Gamma| \ll \sqrt{6}$,
namely for $\phip \simeq \phipmax$. One gets $\Gamma = -3 (\phip -
\phipmax) + \orderb{(\phip - \phipmax)^2}$ and, thus, $\phip -
\phipmax \propto \exp[-3 (N- \Nmax)]$. Following
\Ref{Pattison:2018bct}, one can define the field acceleration
parameter (in cosmic time) $f \equiv - \ddot{\phi}/(3H \dot{\phi})$,
which, in terms of $\Gamma$ simplifies to
\begin{equation}
f = 1 -\dfrac{\Gammasr}{\Gamma} \left(1 - \dfrac{\Gamma^2}{6} \right).
\end{equation}
This parameter is close to unity in kination, but also in inflation
when $|\Gamma| \gg |\Gammasr|$. Therefore, if $|\Gammasr|$ is very
small, the transitional inflationary regime lands on
ultra-slow roll. The question of knowing if inflation can remain
locked within the ultra-slow-roll regime has recently been addressed
in \Ref{Pattison:2018bct}. For some potentials, and for a set of
particular initial conditions, this can indeed be the case, see below.

In the next section, we show that this transitional inflationary
regime, when not locked into ultra-slow-roll, actually evolves and
relaxes to slow roll.

\subsubsection{``Non-relativistic'' inflationary regimes}

To discuss the field evolution in the regime for which
$|\Gammasr|$ can no longer be neglected, one must go back to the exact
Eq.~\eqref{eq:DGamma}. However, this time, assuming that
$\Gamma^2 \ll 6$, we can take the ``non-relativistic'' limit, namely
Taylor expanding the kinetic term in $\Gamma/\sqrt{6}$, without
neglecting the friction term and the right-hand side
$\Gammasr(\phip)$. One gets
\begin{equation}
\dfrac{1}{3} \dfrac{\ud \Gamma}{\ud N} + \Gamma = \Gammasr.
\end{equation}
This equation can be exactly solved by remarking that, for
$\Gamma \ne 0$,
\begin{equation}
\Gammasr(N) = - \dfrac{1}{\Gamma(N)} \dfrac{\ud \ln V}{\ud N}\,.
\end{equation}
One gets a non-homogeneous first order differential equation with
constant coefficient
\begin{equation}
\dfrac{\ud \Gamma^2}{\ud N} + 6 \Gamma^2 = - 6 \dfrac{\ud
  \ln V}{\ud N}\,,
\label{eq:GammaNR}
\end{equation}
whose solution reads
\begin{equation}
\Gamma^2(N) = \Gammax^2 e^{-6(N-\Nx)} - 6 \int_{\Nx}^N e^{6(n-N)}
\dfrac{\ud \ln V}{\ud n} \ud n.
\label{eq:NRsol}
\end{equation}
Here, we have started the integration at an e-fold number $\Nx$ for
which $\Gamma(\Nx) = \Gammax$, assuming only $|\Gammax| \ll
\sqrt{6}$.
We have used the same notation as in the previous section, precisely
because this value can be chosen to match both regimes (see next
section). To explicitly show the attractive behavior of
slow roll, one can integrate by part \Eq{eq:NRsol} to pull the
potential derivative out of the integral. Defining
\begin{equation}
\Delta\barN \equiv N - \Nx,
\end{equation}
after some algebra, one gets
\begin{equation}
\begin{aligned}
  \Gamma^2(N) &= \left(\Gammax^2 + \left.\dfrac{\ud \ln V}{\ud
  N}\right|_{\Nx} \right) e^{-6 \Delta\barN} - \dfrac{\ud \ln V}{\ud
    N} \\ & + \int_0^{\Delta\barN} e^{6\left(\barn -
    \Delta\barN \right)} \dfrac{\ud^2 \ln V}{\ud \barn^2} \ud
  \barn.
\end{aligned}
\label{eq:NRsolSR}
\end{equation}
The first term in the right-hand side is a transient associated with
the initial conditions. It is damped by the exponential term. The
second term is precisely the slow-roll solution since
$-\ud\ln V/\ud N = \Gamma \Gammasr$, so
$\Gamma^2 \simeq -\ud\ln V/\ud N $ implies that
$\Gamma \simeq \Gammasr$ for $\Gamma \ne 0$. We recover the well-known
result that slow roll is the attractor provided the last integral
remains negligible. As discussed in \Ref{Pattison:2018bct}, there are
situations in which this is not the case. The integral
\begin{equation}
\calC(N) \equiv \int_0^{\Delta\barN} e^{6\left(\barn -
    \Delta\barN \right)} \dfrac{\ud^2 \ln V}{\ud \barn^2}\ud
  \barn
\end{equation}
is a convolution of a damped exponential kernel with the second
logarithmic derivative of the potential. Therefore, it is possible to
have $\calC$ larger than $\Gammasr$ by approaching a point of
vanishing gradient, but not vanishing curvature in $W=\ln V$.
The precise conditions for this to happen are derived in
\Ref{Pattison:2018bct}, where it is also stressed that, because
$\calC$ is a convolution, it necessarily retains some dependence on
the initial conditions. So even when this regime is stable, it is not
an attractor in the dynamical sense.

If $W=\ln V$ is sufficiently regular, one can also keep on integrating
by part \Eq{eq:NRsolSR} to infinite order. One gets for the
convolution integral the following expression:
\begin{equation}
\begin{aligned}
  \calC(N) & = \left[\sum_{k=2}^{+\infty}
  \left(-\dfrac{1}{6}\right)^{k-1} \left.\dfrac{\ud^k \ln V}{\ud
    N^k}\right|_{\Nx} \right] e^{-6 \Delta\barN} \\ & -
  \sum_{k=2}^{+\infty} \left(-\dfrac{1}{6}\right)^{k-1} \dfrac{\ud^k
    \ln V}{\ud N^k}.
\end{aligned}
\end{equation}
The first summation features the initial conditions, as
announced\footnote{There is an infinite number of terms and, although
  each is exponentially damped, one should be careful in their
  evaluation. For some very peculiar potentials, the sum may not
  converge or could be dominated by very high-order terms.}. The second
summation shows that, in principle, any higher-order derivative of the
potential can take over the slow-roll term $-\ud \ln V/\ud N$. Let us
stress, however, that in practice, this can happen only around peculiar
points in a potential for which the gradient vanishes while one, or
more, higher-order derivatives are large.

Finally, because \Eq{eq:NRsol} only assumes $\Gamma^2 \ll 6$, it can
also be applied to not so flat potentials. In that case, an
expansion in logarithmic derivatives may no longer be well defined
but, demanding only an integrable logarithmic potential, one can
integrate \Eq{eq:NRsol} by parts by pulling the exponential term out
of the integral. One gets another (equivalent) expression for the
solution which reads
\begin{equation}
\begin{aligned}
  \Gamma^2(N) & = \left(\Gammax^2 + 6 \ln \Vx\right) e^{-6
    \Delta\barN} -6 \ln V\\ & + 36 \int_0^{\Delta\barN} e^{6 \left(\barn -
    \Delta\barN \right)} \ln V \ud \barn.
\end{aligned}
\label{eq:gammalnV}
\end{equation}
This expression makes explicit that the field actually evolves in the
effective potential $W = \ln V$ as opposed to $V$. In particular
\Eq{eq:gammalnV} is relevant to describe the end of inflation. Indeed,
the slow-roll regime does not last forever, the field rolling along
the potential's gradient, it will ultimately reach larger slopes for
which $|\Gammasr|$ is no longer a small quantity and thereby build
again kinetic energy. This is the graceful exit of inflation which
occurs for $\Gamma^2 = 2$ and for which \Eq{eq:gammalnV} is still
valid. Past this point, one has to use a full numerical integration
to describe the field evolution around the potential minimum and this
is discussed in \Sec{sec:method}.

\subsubsection{Matching solutions}
\label{sec:matching}

From the previous discussion, the field trajectory can be separated
into two regimes. The initial regime is transitional, from kination to
inflation. The field trajectory in the phase-space
$(\phip,\Gamma)$ does not depend on the potential and is given by
\Eq{eq:GammaApprox}. The trajectory $\phip(N)$, with respect to the
e-fold number, can be explicitly obtained by inverting
\Eq{eq:invtraj}.

The second regime is described by \Eq{eq:NRsolSR}, which relaxes to
slow roll if one can neglect the convolution integral $\calC(N)$.  As
a result, the complete trajectory $\Gamma(\phip)$ can simply be
obtained by matching the two regimes at a crossing value $\Gammax$
that should verify
\begin{equation}
|\Gammasr| \ll |\Gammax| \ll \sqrt{6}\,.
\end{equation}
In view of the previous results, we conclude that if the initial field
velocity is not strongly fine-tuned to $\pm \sqrt{6}$, and if the
potential supports slow-roll inflation, the field trajectory
generically relaxes toward the slow-roll attractor. The only other
alternative would be a relaxation toward ultra-slow roll, but this
requires specific potential shapes and initial conditions. In all
cases, however, the initial kinetic energy stored in a homogeneous
field cannot, alone, prevent inflation to start.

When ultra-slow roll is not present, one can neglect $\calC(N)$ in
\Eq{eq:NRsolSR}, and the terms depending on $\Gammax$ are
exponentially damped. As a result, there is a cruder, but still good
approximation of the whole phase-space trajectory which consists in
choosing $\Gammax \equiv \pm |\Gammasr|$ depending on the sign of
$\Gamma$. If $\Gamma$ and $\Gammasr$ are of the same sign, this boils
down to extrapolating \Eq{eq:GammaApprox} directly onto the slow-roll
attractor $\Gamma=\Gammasr(\phip)$. If $\Gamma$ and $\Gammasr$ are of
opposite sign, it means that we extrapolate $\Gamma$ until it matches
$-\Gammasr$, and the transition from $-\Gammasr$ to $\Gammasr$ is
performed at constant $\phip$ value. This is again a very good
approximation provided $|\Gammasr| \ll 1$ because $|\Gamma| \ll 1$ and
thus $\phip(N)$ can only remain constant (this does not say anything
on the number of e-folds spent in that regime, it could be large).

In the next sections, these findings are confirmed by exact numerical
integration of the field trajectory in various potentials. We also
numerically explore the situation in which the initial kinetic energy
is very large in domains where the potential is steep enough \emph{not
  to} support inflation, which our previous approximations did not
allow us to study. This allows us to discuss how the UV completion of
the various inflationary models could affect the initial conditions
necessary to trigger inflation.

\subsubsection{Comparison with previous works}
\label{subsec:comparison}

In the previous sections, we have shown how the trajectories of the
system can be worked out in the entire phase-space (hence, possibly,
outside the slow-roll regime). In fact, the first systematic study of
this question was performed, long ago, in the
article~\cite{Goldwirth:1991rj}, a classic reference on the
subject. It is therefore interesting to compare the methods of
\Ref{Goldwirth:1991rj} to our approach. Reference~\cite{Goldwirth:1991rj} starts
with rewriting the Hubble and Klein-Gordon \Eqs{eq:hubble} and
\eqref{eq:kg} in the following manner:
\begin{align}
\label{eq:fried}
H^2 &=\frac{\Pi^2}{2\Mp^2 \epsilon_1}, \\
\label{eq:kg-eqn}
\frac{\dd}{\dd N}\left(\ln \frac{\Pi}{\Mp^2}\right)
+3\left[1+\calP\left(\phi,\Pi\right)\right]&=0,
\end{align}
where the quantity $\calP(\phi,\Pi)$ is defined by the
following expression
\begin{align}
\calP\left(\phi,\Pi\right)&\equiv \frac{V_\phi}{3H\Pi}= 
\frac13 \sqrt{\frac{\epsilon_{1V}}{\epsilon_1}}
\left(3-\epsilon_1\right),
\end{align}
$\epsilon_{1V}\equiv \Mp^2\left(V_{\phi}/V\right)^2/2$ being the first
potential slow-roll parameter. The functional dependence of
${\mathcal P}\left(\phi,\Pi\right)$ on $\Pi$ is fixed while its
dependence on $\phi$ relies on the potential, that is to say on the
model. From the Klein-Gordon equation, we see that one can define two
regions in phase-space, depending on whether
$\left \vert \calP\right \vert <1$ or not. However, the situation is
in fact slightly more complicated since $\calP$ can be small for two
reasons: either $\epsilon_1\rightarrow 3$, namely
$\Gamma^2\rightarrow 6$, (of course, it is implicitly assumed that
$\epsilon_{1V}$ does not go to infinity in such a way that it
compensates for the smallness of $\epsilon_1-3$) or
$\epsilon_{1V}/\epsilon_1\ll 1$ with $\epsilon_1$ not necessarily
close to three. This leads \Ref{Goldwirth:1991rj} to define, not two,
but in fact three regions: region I is the region where
$\left \vert \calP\right \vert <1$ and $\epsilon_1>3/2$ (meaning
$2V/\Pi^2<1$), region II is the region where
$\left \vert {\mathcal P}\right \vert <1$ and $\epsilon_1<3/2$
(meaning $2V/\Pi^2>1$) and, finally, region III is where
$\left \vert \calP\right \vert \ge 1$.  In each of these regions, one
can find an approximate solution to the equations of motion. Notice
that one of the two boundaries between region II and region III,
namely $\calP=-1$, exactly corresponds to the slow-roll trajectory.

In region \RN{1}, the condition $\left \vert \calP\right \vert <1$
implies that \Eq{eq:kg-eqn} can be integrated once to obtain
\begin{equation}
\label{eq:phidot-N-reg12}
\Pi=\Pi_\uini\,\expo^{-3(N-\Nini)},
\end{equation}
where $\Pi_\uini$ is the initial value of the scalar field velocity at
the initial e-fold $\Nini$.  The corresponding evolution of the field
can be deduced from the equation
\begin{equation}
\label{eq:phi-N-diff-eqn}
\frac{\dd \phi}{\dd N}=\frac{\Pi_\uini}{H} \expo^{-3(N-\Nini)}.
\end{equation}
But since $\epsilon_1>3/2$ by definition of region \RN{1}, the kinetic
energy is dominant and the Friedmann equation, in this regime, is
given by $H^2\simeq \Pi^2/(6\Mp^2)$. Thereupon \Eq{eq:phi-N-diff-eqn}
can easily be integrated to obtain
\begin{eqnarray}
\label{eq:phi-N-reg1}
\phi=\phi_\uini \pm \Mp \sqrt{6}\left(N-\Nini \right).
\end{eqnarray}
The remarkable feature of those solutions is that they are model
independent, that is to say independent of the form of the
potential. It is also interesting to notice that, combining
\Eq{eq:phidot-N-reg12} and~$H^2\simeq \Pi^2/(6\Mp^2)$ exactly leads to
$\Gamma^2\simeq 6 $.

This regime is included within the phase-space trajectory of
\Eq{eq:GammaApprox}. This can be seen by expressing $H$ in terms of
$\Pi$ from \Eqs{eq:flefold} and \eqref{eq:PiofGamma}. One gets
\begin{equation}
H^2 = \dfrac{\Pi^2}{\Mp^2 \Gamma^2}\,,
\end{equation}
which immediately gives $H^2 \simeq \Pi^2 / (6\Mp^2)$ for
$\Gamma^2 \rightarrow 6$. The same limit implies \Eq{eq:phi-N-reg1}
from the very definition of $\Gamma$.

Let us now consider region \RN{2}. Since $\calP$ is small, it is clear
that \Eq{eq:phidot-N-reg12} is still valid. But, now, one has
$\epsilon_1<3/2$; as a consequence, we can write
$H^2 \simeq V/(3\Mp^2)$, and \Eq{eq:phi-N-diff-eqn} can be solved to
obtain
\begin{align}
\label{eq:phi-N-reg2}
N &= \Nini-\frac13 \ln \biggl[
\expo^{-3(N_{\utr}-\Nini)}
%\nonumber \\ &
 -\frac{\sqrt{3}}{\Mp \Pi_{\uini}}\int _{\phi_{\utr}}^{\phi}\kern -1em
\sqrt{V(\phi)}\,\dd \phi\biggr],
\end{align}
where $N_{\utr}$ is the number of e-folds at the transition between
regions \RN{1} and \RN{2} and $\phi_{\utr}$ is the value of the
scalar field at $N_{\utr}$.  Evidently, this time, the equation
describing the dependence of the scalar field amplitude on the initial
conditions would vary based on the specific inflationary potential
under consideration. However, for practical applications,
\Ref{Goldwirth:1991rj} assumes that the potential can be taken as
constant $V(\phi)\simeq M^4$ during phase II.  In that case,
\begin{align}
\Gamma & \simeq \frac{\sqrt{3}\Pi_\uini}{M^2}e^{-3(N-\Nini)}
\nonumber \\ &
\simeq 
\frac{\sqrt{3}\Pi_\uini}{M^2}e^{-3(N_{\utr}-\Nini)}
-3(\Phi-\Phi_{\utr}).
\end{align}
This approximation seems to differ from ours because it is made in
terms of the variable $\Pi$. As can be checked in \Eq{eq:GammaApprox},
$\Gamma(\phip)$ in the same regime ($\Gamma^2 < 3$, and $|\Gamma| >
|\Gammasr|$) depends only on $\Gammaini$ and $\phipini$. The term in
$\sqrt{V}$ appearing in \Eq{eq:phi-N-reg2} comes from the relation
between $\Pi$ and $\Gamma$, see \Eq{eq:PiofGamma}, and this regime is
again included within our \Eq{eq:GammaApprox}.

At the end of region \RN{2}, the quantity $\calP$ becomes larger than
one, and, a priori, the system enters the slow-roll regime (possibly
after a short transitory regime that cannot be described
analytically). In phase-space, the slow-roll trajectory reads
\begin{align}
\label{eq:kgsr}
\Pi &=-\frac{V_{\phi}}{3H}, \\
\label{eq:trajectorysr}
N-N_{\ustart} &=-\frac{1}{\Mp^2}\int _{\phi_{\ustart}}^\phi 
{\mathrm d}\psi \frac{V(\psi)}{V_{\psi}(\psi)},
\end{align}
where $\phi_{\ustart}$ is the value of the field at the start of the
slow-roll phase or the final value of the field at the end of region
\RN{2}. In fact, \Eq{eq:kgsr} can also be written as $\calP=-1$. As a
consequence, the behavior of the system depends whether one enters
region III through the boundary $\calP=-1$ or $\calP=1$.  In the first
case, the system directly goes from region II to the slow-roll
attractor while, in the second case, the system spends time in region
III before joining the attractor. Unfortunately, this evolution cannot
be described analytically. It usually corresponds to the regime where
the field changes direction. Such a problem does not occur in the
approximation scheme developed in the previous section.

\subsection{Application to well-known potentials}

\subsubsection{Exact numerical integration}
\label{sec:method}

To numerically integrate the system of
Eqs~\eqref{eq:hubble} to \eqref{eq:kg}, we have used the public
library
{\fieldinf}\footnote{\url{http://curl.irmp.ucl.ac.be/~chris/fieldinf.html}},
see \Refs{Ringeval:2007am,Ringeval:2013lea}. Starting from a grid of
initial conditions $(\phipini,\Gammaini)$, each trajectory in
phase-space is integrated in terms of the number of e-folds $N$, and
followed up to an ending point that we choose in such a way that
inflation would no longer be possible afterwards. This point is
numerically determined for each potential according to the following
method. A phase-space trajectory is numerically integrated along the
inflationary attractor to determine the field value, $\phip_1$, at
which inflation stops, {\ie}, the equation $\Gamma^2(\phip_1) = 2$ is
numerically solved. If the potential supports $n$ inflationary
separated regions, there are as many values of $\phip_{i}$, with
$i\in\{1,\dots,n\}$, which solve $\Gamma^2(\phip_i)=2$. Because of the
attractor nature of the inflationary domains, the values of $\phip_i$
do not depend on the initial conditions (see below). Let us define
\begin{equation}
\Hend^2 \equiv \min_{i=1,\dots,n}\left\{H^2(\phip_i)\right \} =
\min_{i=1,\dots,n} \left\{
\dfrac{V(\phi_i)}{2 \Mp^2} \right\} \,.
\end{equation}
Because $H$, and the energy density $\rho$, are monotonic decreasing
functions of $N$, we use $H(\Nend) = \Hend$ as the criterion to stop
the numerical integration. This condition ensures that we track all
the trajectories exploring the non-inflationary domains of the
potential with sufficient kinetic energy to climb up the potential and
inflate again later on. Finally, along each trajectory, we store the
number of e-folds spent in an inflationary regime, {\ie}, having
$\Gamma^2(\phip) < 2$. For various solutions ending up inflating, this
number can be very large, and for numerical efficiency, it has been
bounded to $10^3$. As can be seen in \Eq{eq:DGamma}, it is important
to stress that the trajectories in phase-space $(\phip,\Pi)$ do not
depend on the absolute normalization of the potential, say $M^4$.

Concerning the range of initial conditions, $\phipini$ is chosen to
encompass all the inflationary domains as well as regions much further
away in order to study their effects. The initial field velocities
fill the full range of mathematically allowed values $\Gammaini^2
<6$. Such a condition actually allows initial kinetic energies to be higher
than the Planck scale. Indeed, requiring the total energy density of
the field to be sub-Planckian, $\rho \le \Mp^4$, translates into the
constraint
\begin{equation}
\Gammaini^2 \le 6\left[1 - \dfrac{V(\phipini)}{\Mp^4} \right].
\label{eq:Planckbound}
\end{equation}
This limit depends on the overall normalization of the potential,
$M^4$, which in turn depends on the amplitude of the CMB
anisotropies. A precise determination of these numbers being outside
the scope of this work, we have chosen the worse case scenario in
which there is no bound on the initial kinetic energy.

In the following, we present our results for various specific
potentials.

\subsubsection{Large-field models}
\label{sec:lfi}

\begin{figure}
  \begin{center}
    \includegraphics[width=\figmaxw]{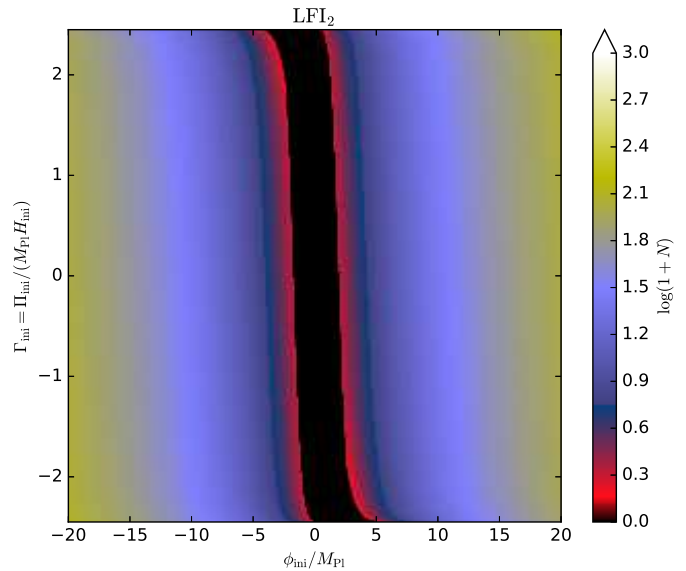}
    \caption{Number of e-folds of inflation for $\lfi_2$ achieved
      along phase-space trajectories starting from $2048^2$ initial
      conditions. No fine-tuning is required as almost all initial
      conditions lead to inflation, independently of the initial
      kinetic energy stored in the inflaton. At fixed value of
      $\phipini>0$, we notice that the number of e-folds is less if
      $\Gammaini<0$ than it is for $\Gammaini>0$. This effect is
      especially clear when $\Gammaini$ is close to its extremal value
      $\pm\sqrt{6}$ and $\phipini$ is not too large. This is because,
      if $\Gammaini<0$, then the field is initially pushed toward the
      minimum of the potential, while if $\Gammaini>0$, it initially
      climbs up the potential (the situation is symmetric for
      $\phipini < 0$).}
    \label{fig:lfiic}
  \end{center}
\end{figure}

Let us first examine the case of large-field inflation ($\lfi{}_p$)
with potential $V(\phi)=M^4(\phi/\Mp)^p$, where $p>0$. The slow-roll
solution for positive field values can be easily derived and
reads~\cite{Martin:2013tda}
\begin{equation}
\begin{aligned}
\phip(N) = \sqrt{\dfrac{p^2}{2} - 2 p \left(N-\Nend\right)}\,.
\end{aligned}
\label{eq:lfisr}
\end{equation}
The transitional phase described by \Eq{eq:GammaApprox} reaches this
solution after a field excursion $\Delta \phip$. As discussed in
\Sec{sec:matching}, in all regions where $|\Gammasr| \ll 1$, one has
\begin{equation}
  \Delta\phip \simeq \Delta\phipx \simeq \Delta\phipmax,
\end{equation}
where $\Delta \phipmax$ is given in \Eq{eq:deltaphipmax}. Here, as
described before, we use $\Gammax=\pm|\Gammasr|$. From the above
trajectory~(\ref{eq:lfisr}), a total number $\sixty$ of e-folds is
realized in the slow-roll regime if the field reaches the attractor at a
vacuum expectation value given by $\Mp\sqrt{p^2/2+2p\sixty}$. As a
consequence, requiring slow-roll inflation to last more than $\sixty$
e-folds means $\phipx> \sqrt{p^2/2+2p\sixty}$, or
\begin{equation}
\phipini +
\dfrac{1}{\sqrt{6}}\ln\left(\dfrac{1 + \dfrac{\Gammaini}{\sqrt{6}}}{1
  -\dfrac{\Gammaini}{\sqrt{6}}} \right) > \sqrt{\dfrac{p^2}{2} + 2 p
    \sixty} \simeq 2 \sqrt{30 p}\,,
\label{eq:lfidomain}
\end{equation}
where the last equality comes from choosing $\sixty = 60$ e-folds. For
even values of $p$, the potential is symmetric with respect to
$\phip=0$ and negative initial field values verify the opposite of the
above bound. From \Eq{eq:lfidomain}, we see that no fine-tuning of the
initial field values is required to start inflation. On the contrary,
only an extreme fine-tuning of $\Gammaini$ close to $-\sqrt{6}$ might
demand to push $\phipini$ to larger positive values to start
inflation.

\begin{figure}
  \begin{center}
    \includegraphics[width=\figw]{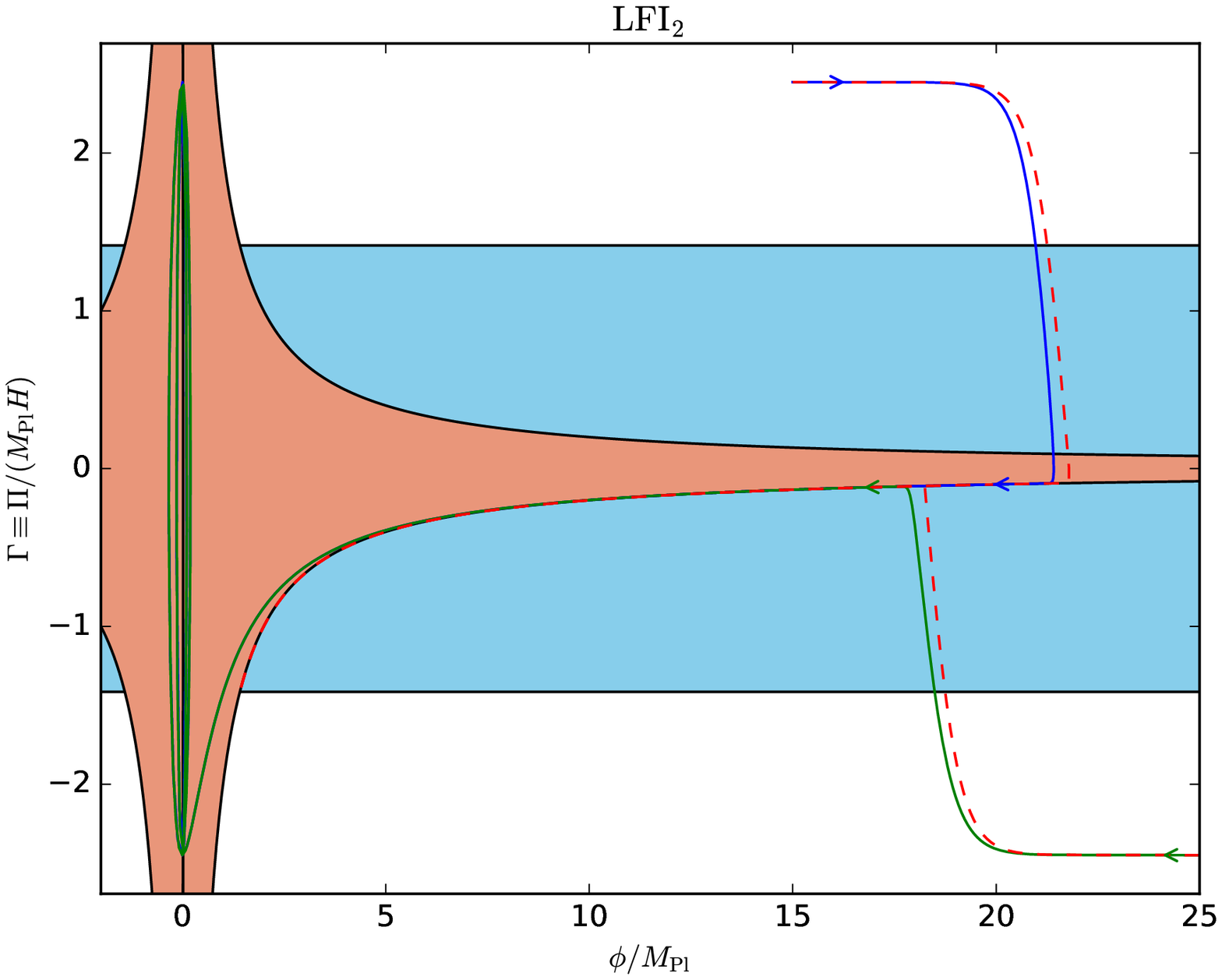}
    \includegraphics[width=\figw]{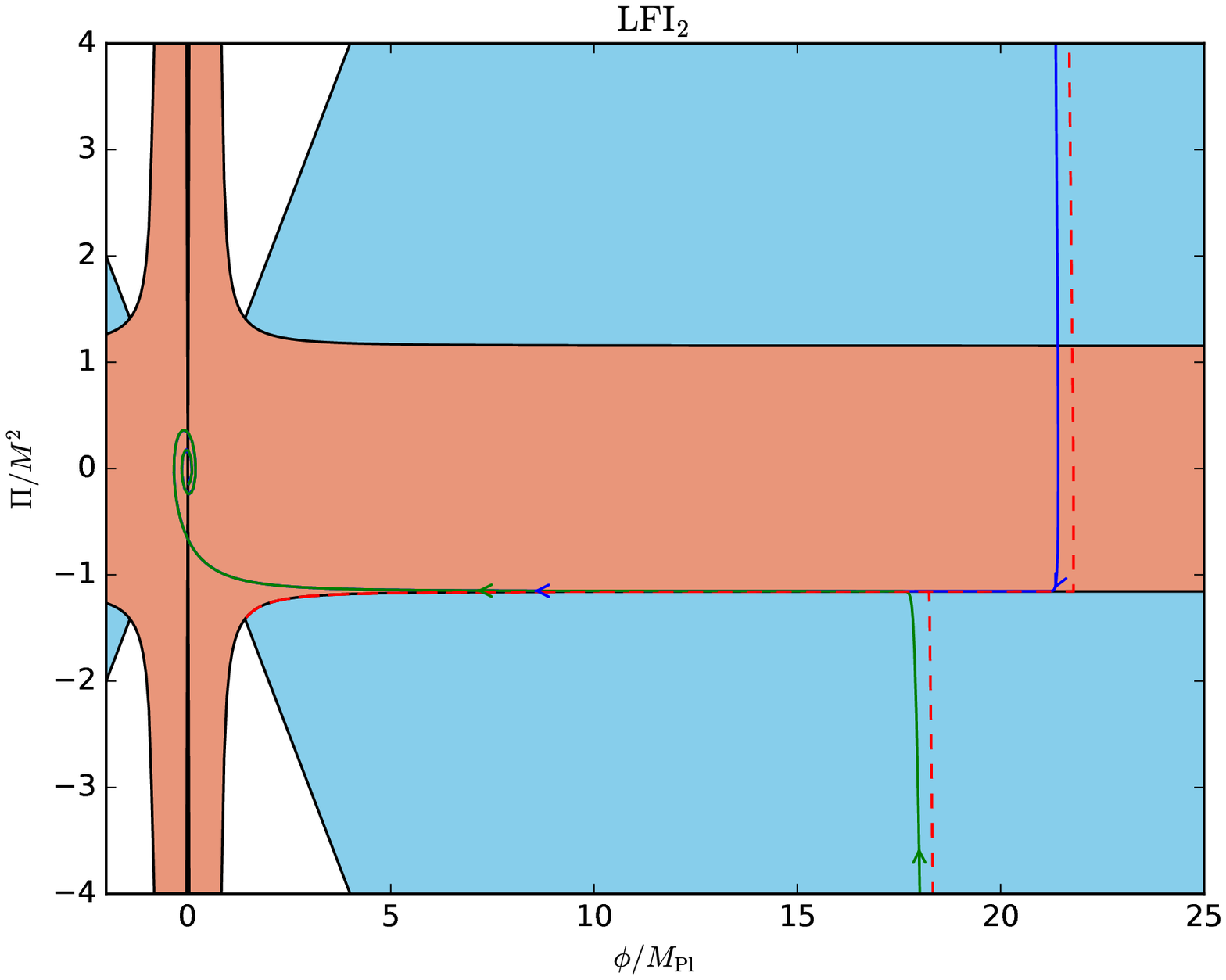}
    \caption{Phase-space trajectories for $\lfi_2$ (solid curves)
      starting deeply in the kination regime and relaxing toward slow
      roll. The blue line corresponds to a situation where the field
      initially climbs up its potential while the green line represents a
      case where it rolls it down. The upper panel displays the phase
      space $(\phip,\Gamma)$ while the lower panel corresponds to
      $(\phi,\Pi)$. The analytical approximations of \Eqs{eq:Gammasr}
      and \eqref{eq:GammaApprox} are represented as dashed red
      curves. In the upper figure, the horizontal blue region
      corresponds to $\Gamma^2 < 2$, {\ie}, inflation. The
      orange region narrowing at large-field values is the domain for which
      $|\Gamma| \le |\Gammasr|$. Both regions are also represented in
      the lower panel.}
    \label{fig:lfitraj}
  \end{center}
\end{figure}

This is confirmed in \Fig{fig:lfiic}, which shows the number of e-folds
of inflation achieved for $p=2$, starting from a grid of $2048^2$
initial conditions.  Almost the whole phase-space produces inflation,
without any fine-tuning of the initial conditions. The inverted ``S''
shape of the contours of equal e-foldings is well described by
\Eq{eq:lfidomain}. This may appear surprising as, strictly speaking,
\Eq{eq:lfidomain} is valid only for $|\Gamma| \gg |\Gammasr|$.
However, as explained in \Sec{sec:matching}, for small values of
$|\Gammasr|$, this inequality can only be violated (namely $\Gamma
<\Gammasr$) at small values of $|\Gamma|$. Since $\Gamma \propto
\dd \phip/\dd N$, this happens only in regions where the field value is
nearly constant, and \Eq{eq:lfidomain} is essentially valid
almost everywhere in phase-space.

In order to compare the analytical approximations presented in
\Sec{sec:asol} to the exact results, a few trajectories have been
represented in \Fig{fig:lfitraj}. The upper panel shows trajectories
in the phase-space $(\phip,\Gamma)$ while the lower panel is for
$(\phi,\Pi)$. The blue region encompasses values of $\Gamma^2 < 2$
corresponding to an accelerated expansion of the universe.  The orange
region (contained within the blue one at large-field values)
represents values of $|\Gamma| \le |\Gammasr|$. The solid curves are
two numerical integrated trajectories starting deeply in the kination
regime, {\ie}, with $\Gamma^2$ fine-tuned to $6$. They rapidly relax
toward the boundary between the blue and the orange regions where
they match the slow-roll attractor $\Gamma = \Gammasr$.

The dashed red curves represent the simplified analytical solution
discussed in \Sec{sec:matching}, made of \Eq{eq:GammaApprox}
extrapolated till it crosses $\Gammax = \pm |\Gammasr|$ and,
eventually, extrapolated again to $\Gammasr$ at constant $\phip$. In
other words, we have neglected both $\calC(N)$ and the relaxation
terms in \Eq{eq:NRsolSR}. Even with such a crude extrapolation, we
find good agreement with the exact numerical result almost
everywhere. For the purpose of illustration, we have extended by a
fraction of e-folds the exact trajectories in the regime $H < \Hend$
in order to show a few oscillations around the minimum of the
potential. Let us notice that, in the oscillatory domain,
$\Gammasr(\phip)$ blows up and none of the analytical approximations
presented before can be used. A more detailed investigation of these
regions is presented in \Sec{sec:uv}.

We conclude, as is well known, that no fine-tuning of initial
conditions is necessary for large-field inflation.

\subsubsection{Small-field inflation}
\label{sec:sfi}

\begin{figure}
  \begin{center}
    \includegraphics[width=\figmaxw]{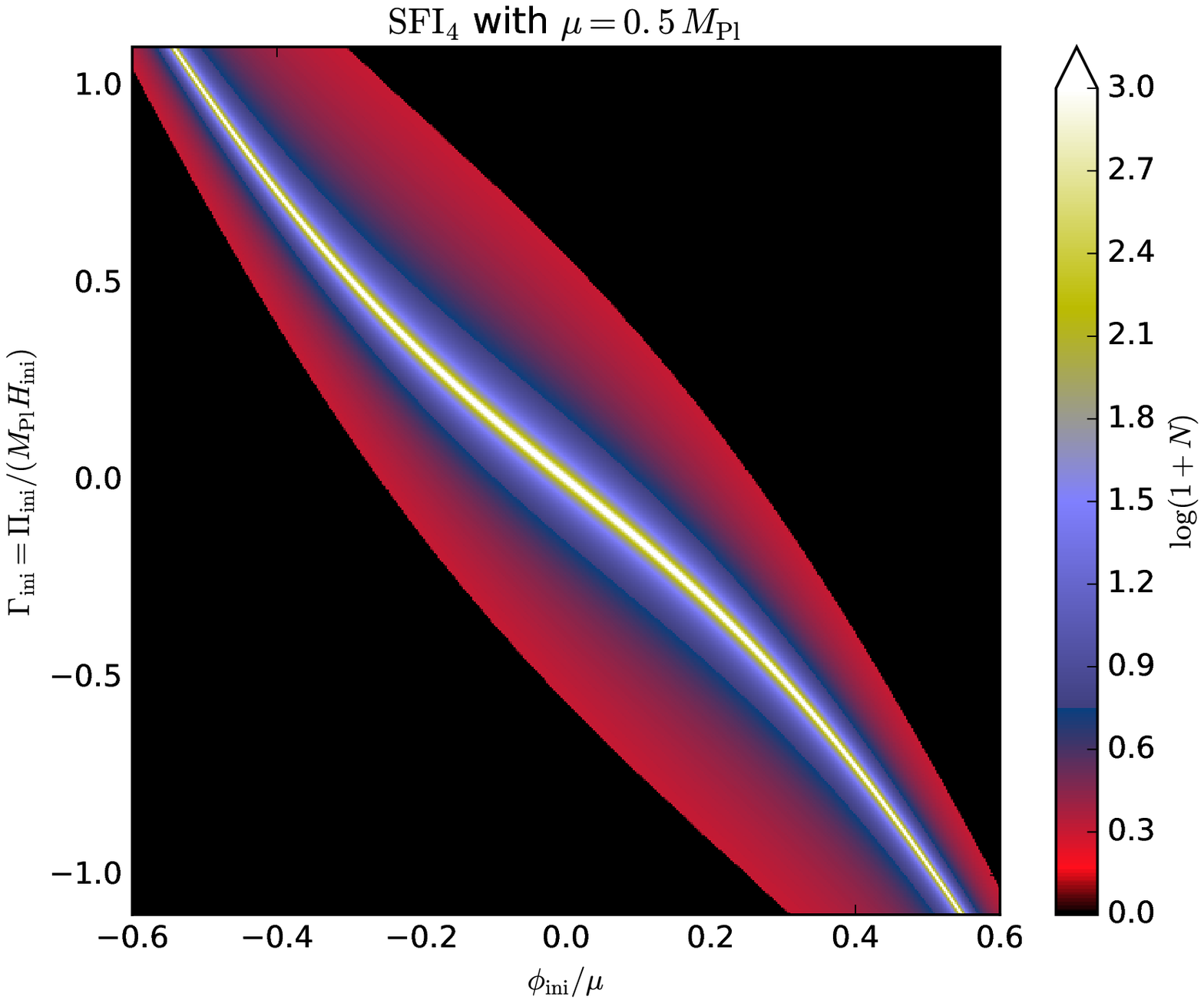}
    \includegraphics[width=\figmaxw]{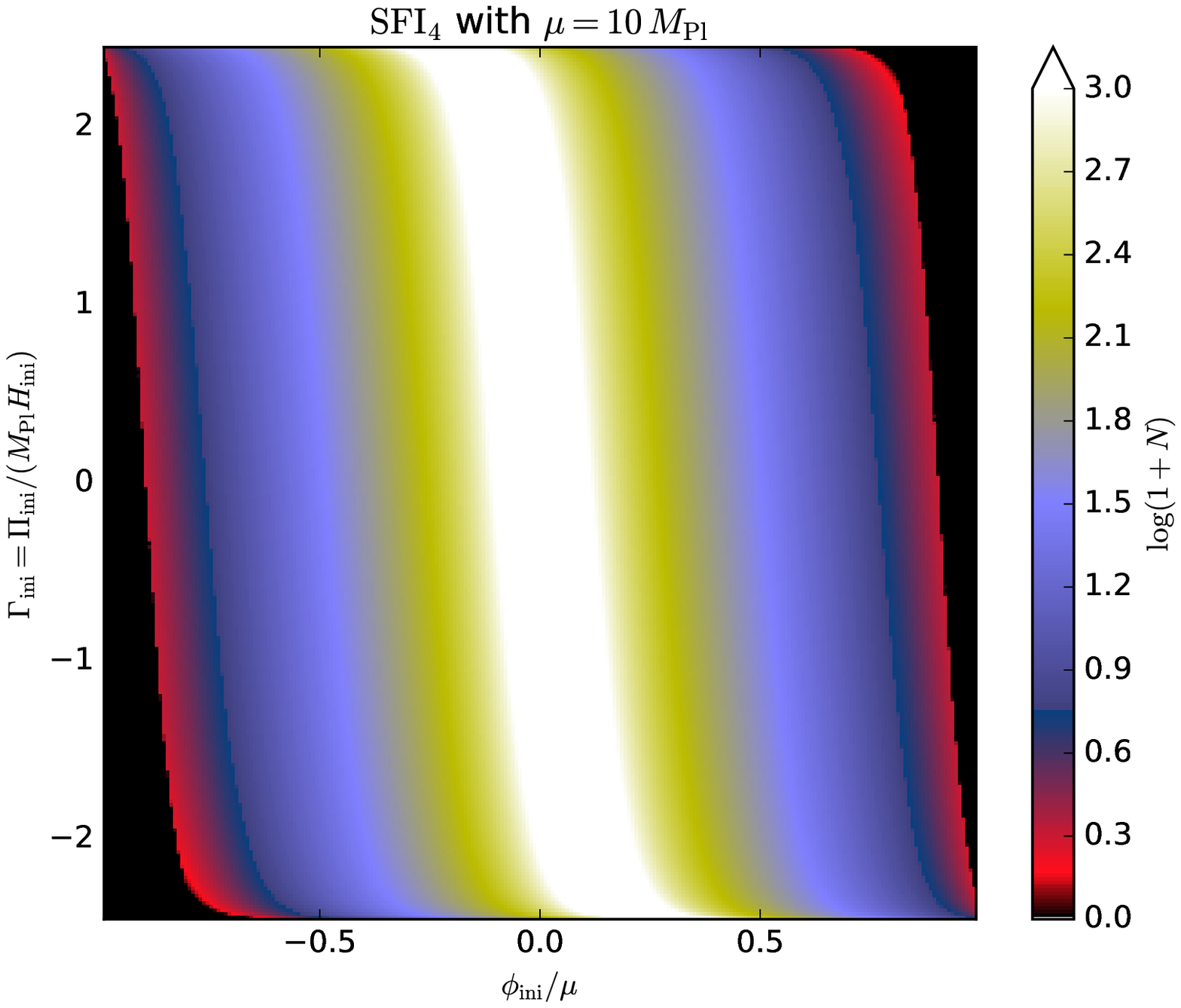}
    \caption{Number of e-folds of inflation for $\sfi_4$ achieved
      along phase-space trajectories starting from $2048^2$ initial
      conditions. Some fine-tuning is required for $\mu < \Mp$ (upper
      panel) while almost all initial conditions lead to inflation for
      $\mu > \Mp$ (lower panel). We see (upper panel) that if the
        field is not initially extremely close to the origin and
        starts at, say, positive values, then the only way to obtain
        a large number of e-folds of inflation is to tune its velocity
        such that it initially climbs up the potential (namely
        $\Gammaini<0$) and reaches the maximum with almost vanishing velocity.}
    \label{fig:sfiic}
  \end{center}
\end{figure}

\begin{figure}
  \begin{center}
    \hspace{9pt}\includegraphics[width=\figw]{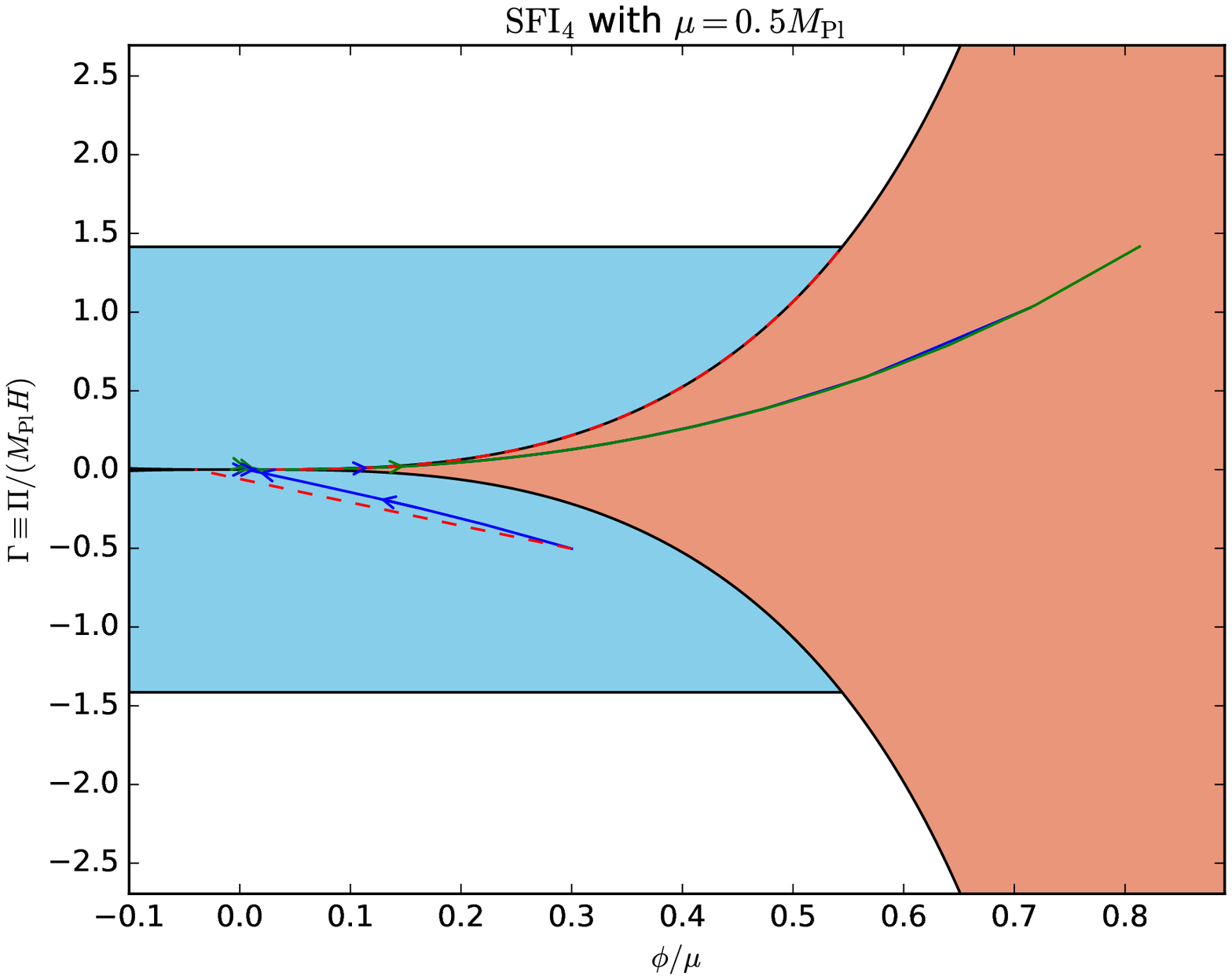}
    \includegraphics[width=1.05\figw]{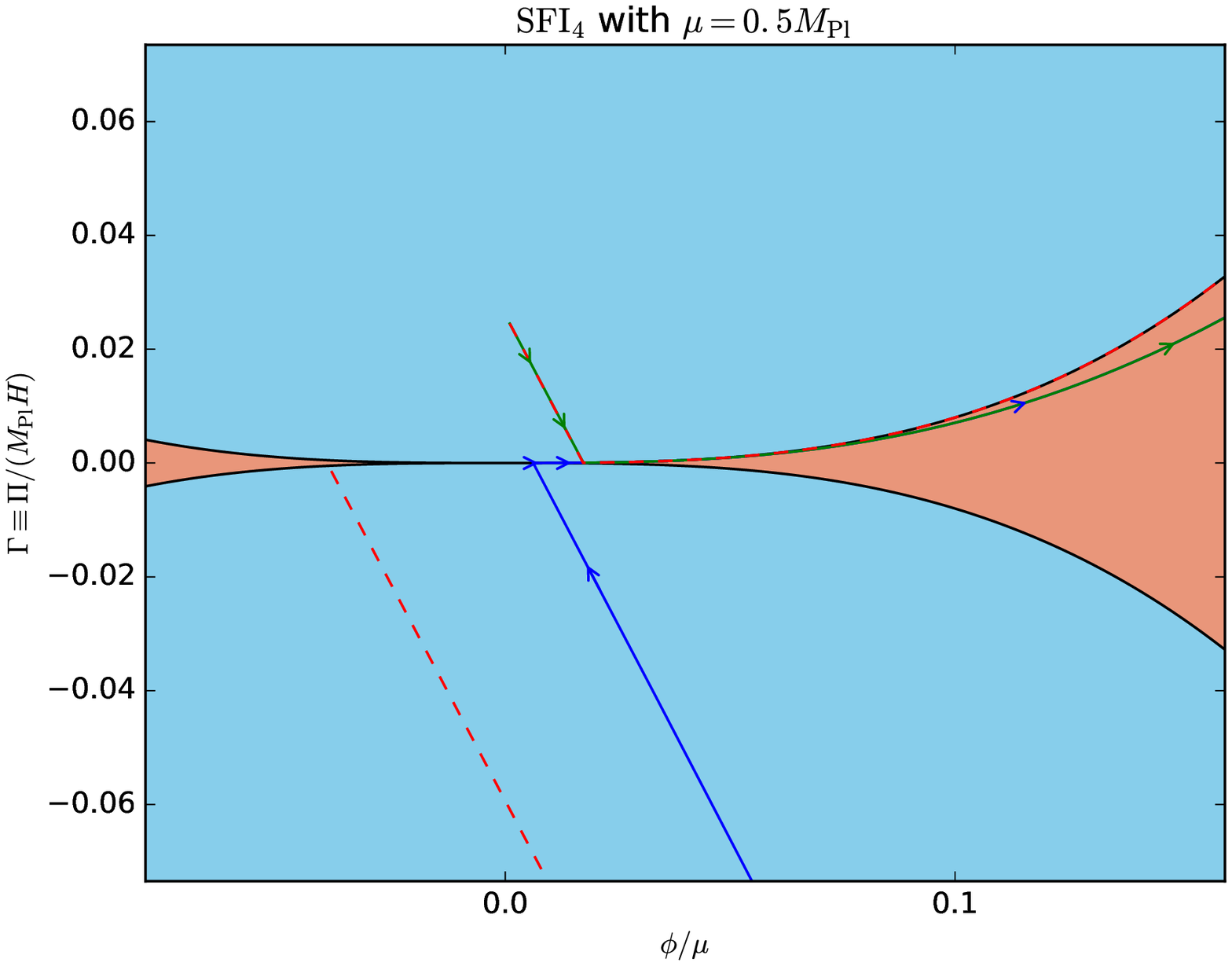}
    \caption{Phase-space trajectories for $\sfi_4$ (solid curve) in
      the fine-tuning regime for $\mu < \Mp$, in the space
      ($\phip,\Gamma$). The lower panel is a zoom in the region around
      $\phi=0$. The analytical approximations are represented with
      dashed-red curves and match well the exact result displayed with
      the green curve. The agreement with the blue curve is less
      compelling since the initial value of
      $\vert \Gamma/\Gammasr \vert$ is not so large in that case.  The
      analytical formula thus hits slow roll (the boundary of the
      orange region) on the wrong side of the potential, {\ie}, at
      $\phi<0$ instead of $\phi>0$, and hence cannot be used
      subsequently. Notice that in both cases, the inflationary exit
      is not well described by slow roll, precisely due to the
      steepness of the potential for $\phi$ approaching $\mu$. Because
      $V(\phi) \simeq M^4$ in the limit $\mu\ll \Mp$, trajectories in
      the space $(\phi,\Pi)$ are almost identical and have not been
      represented (see also \Fig{fig:sfimu10traj}).}
    \label{fig:sfimu05traj}
  \end{center}
\end{figure}

\begin{figure}
  \begin{center}
    \includegraphics[width=\figw]{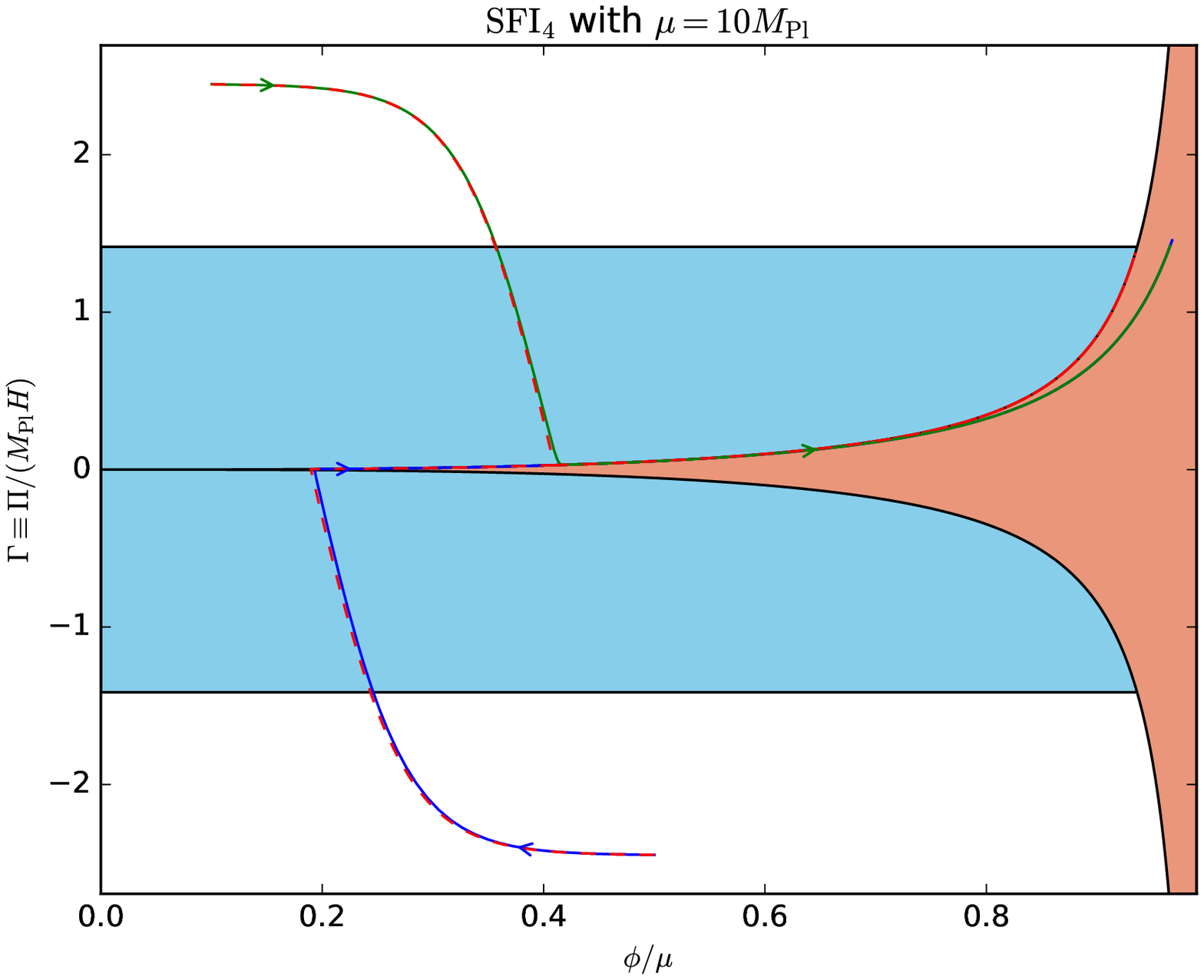}
    \includegraphics[width=\figw]{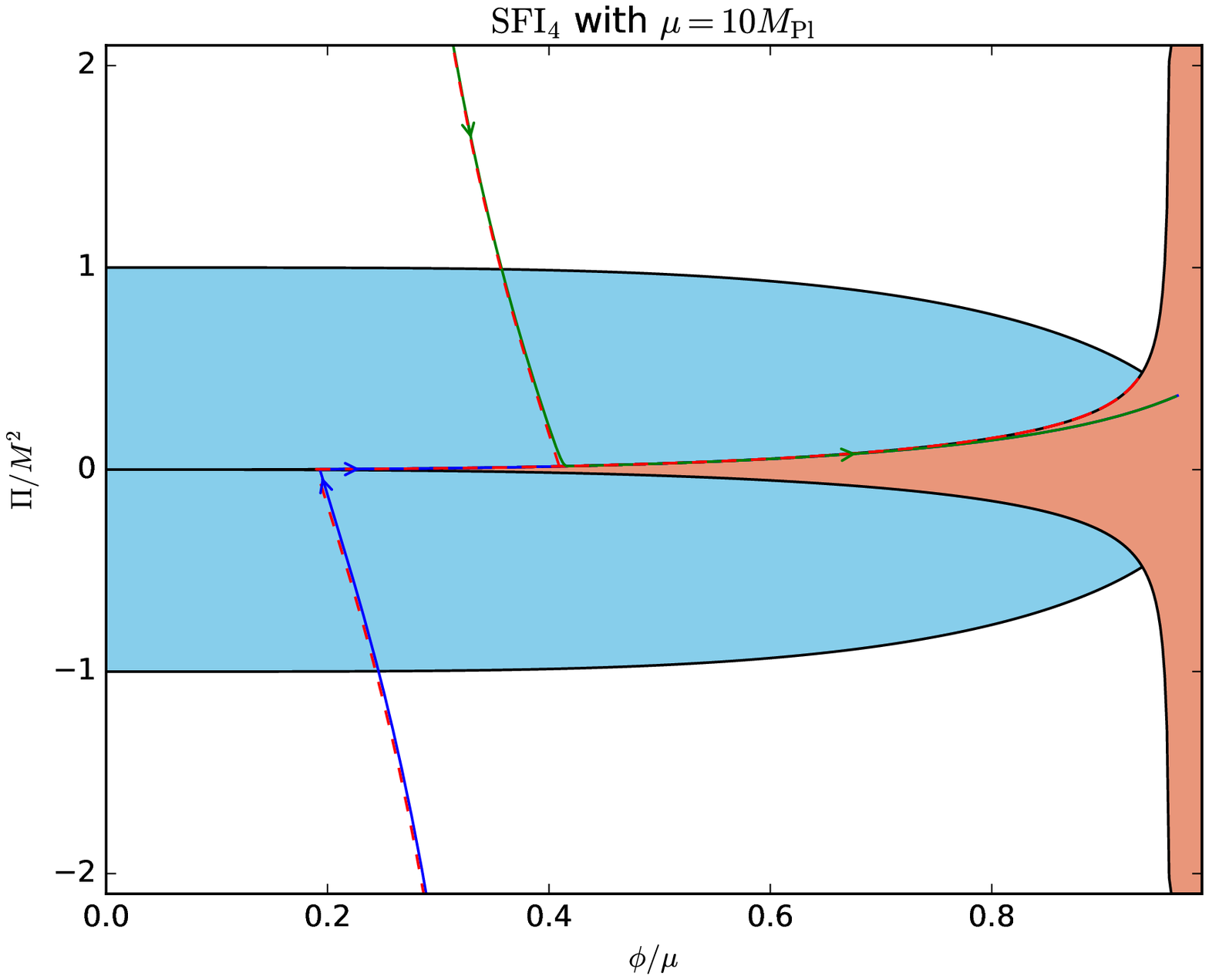}
    \caption{Phase-space trajectories for $\sfi_4$ (solid curve) with
      $\mu > \Mp$. As for the large-field models, almost all initial
      conditions relax toward slow roll and no fine-tuning is needed
      to start inflation. The upper panel displays the phase-space
      $(\phip,\Gamma)$ while the lower panel corresponds to
      $(\phi,\Pi)$. Our analytical approximations have been plotted in
      dashed red. The blue region corresponds to $\Gamma^2 < 2$, {\ie},
      inflation, while the orange region widening toward
      $\phi/\mu = 1$ is the domain for which
      $|\Gamma| \le |\Gammasr|$.}
    \label{fig:sfimu10traj}
  \end{center}
\end{figure}  

We now consider the case of the small-field inflation models $\sfi_p$,
with potential
\begin{align}
\label{eq:potsfi}
V(\phi)=M^4\left[1-\left(\frac{\phi}{\mu}\right)^p\right],
\end{align}
where $\mu$ is a mass scale and $p \ge 2 $ a power index. This
potential has no minimum and becomes negative if $\phi>\mu$. For this
reason $V(\phi)$ can be trusted only if $\phi < \mu$.

The slow-roll solution for $\sfi$ can be found in \Ref{Martin:2013tda}
and reads
\begin{equation}
\begin{aligned}
  2p \dfrac{\Mp^2}{\mu^2}\left(\Nend - N\right) & = \dfrac{2}{p-2}
  \left(x^{2-p}-\xend^{2-p} \right) 
  +  x^2 - \xend^2,
\end{aligned}
\label{eq:sfisr}
\end{equation}
where we have defined
\begin{equation}
x(N) \equiv \dfrac{\phi(N)}{\mu} = \dfrac{\Mp}{\mu} \phip(N),
\end{equation}
and $\xend$ is the slow-roll solution $\Gammasr^2(\xend) = 2$ for the
end of inflation,
\begin{equation}
\xend^p + \dfrac{p}{\sqrt{2}} \dfrac{\Mp}{\mu} \xend^{p-1} = 1.
\label{eq:sfixend}
\end{equation}
The above equations are also valid in the limit $p \to 2$ and can be
used to study $\sfi_2$. Let us, however, stress that consistency of
slow roll within $\sfi_2$ imposes the additional constraint
$\mu > \Mp$, see \Ref{Martin:2013tda}.

As above, using \Eq{eq:deltaphipmax} to approximate the transitional
regime before reaching slow roll, one has
\begin{equation}
  \xx \simeq \xini + \dfrac{1}{\sqrt{6}} \dfrac{\Mp}{\mu} \ln\left(\dfrac{1
    + \dfrac{\Gammaini}{\sqrt{6}}}{1 - \dfrac{\Gammaini}{\sqrt{6}}}
  \right).
\label{eq:sfixini}
\end{equation}
Slow roll produces $\Nend - \Nx > \sixty$ e-folds of inflation
provided $\xx$ satisfies
\begin{equation}
  \begin{aligned}
    \dfrac{2}{p-2}\xx^{2-p} + \xx^2 > 2 p
    \dfrac{\Mp^2}{\mu^2} \sixty + \dfrac{2}{p-2}\xend^{2-p} + \xend^2.
  \end{aligned}
\label{eq:sfixcross}
\end{equation}
Plugging \Eq{eq:sfixini} into \Eq{eq:sfixcross} gives the necessary
condition for a trajectory starting at $(\xini,\Gammaini)$ to produce
$\sixty$ e-folds of slow-roll inflation. As it is obvious from these
expressions, the amount of fine-tuning strongly depends on the ratio
$\mu/\Mp$.

Let us first assume that $\mu \ll \Mp$. As can be seen in
\Eq{eq:sfixini}, this regime amplifies the effect of $\Gammaini$ in the
actual value of $\xx$ at fixed $\xini$. Moreover, from
\Eq{eq:sfixend}, one has
\begin{equation}
\xend \simeq \left(\dfrac{\sqrt{2}}{p} \dfrac{\mu}{\Mp}
\right)^{1/(p-1)} \ll 1,
\end{equation}
and the whole slow-roll region $x \le \xend$ is confined in a small
domain at the top of the potential. The terms in $\xx^2$ and $\xend^2$
in \Eq{eq:sfixcross} can be neglected in this limit (recall that
$p > 2$). One finally gets, for $p > 2$,
\begin{equation}
  \begin{aligned}
    \xini & + \dfrac{1}{\sqrt{6}} \dfrac{\Mp}{\mu} \ln\left(\dfrac{1 +
      \dfrac{\Gammaini}{\sqrt{6}}}{1 - \dfrac{\Gammaini}{\sqrt{6}}}
    \right) \\ & < \left[p(p-2) \dfrac{\Mp^2}{\mu^2} \sixty +
        \left(\dfrac{p}{\sqrt{2}}\dfrac{\Mp}{\mu}
        \right)^{\frac{p-2}{p-1}} \right]^{-\frac{1}{p-2}}.
  \end{aligned}
\label{eq:sfimusmalldomain}
\end{equation}
The right-hand side of this expression is a very small number for $\mu
\ll \Mp$, showing that $\xini$ and $\Gammaini$ should be fine-tuned
along a narrow band in phase-space to produce a successful
inflationary era. Let us stress, however, that such a fine-tuning is not
only related to the presence of an initial kinetic energy. Setting
$\Gammaini=0$ in the previous equations does not solve the issue as
$\xini$ still has to be tuned at the top of the potential. The fine
tuning comes from the small field extension of the domain allowing for
long-enough inflation when $\mu \ll \Mp$.

Taking the opposite limit, namely $\mu \gg \Mp$, one gets
\begin{equation}
  \xend \simeq 1 - \dfrac{\Mp}{\sqrt{2} \mu},
\end{equation}
while, at leading order in $\Mp/\mu$, \Eq{eq:sfixcross} becomes
\begin{equation}
\xx < 1 - \dfrac{\Mp}{\mu} \sqrt{\dfrac{1}{2} + 2 \sixty}\,.
\end{equation}
Long enough inflation is therefore triggered for any initial conditions
satisfying
\begin{equation}
\begin{aligned}
    \xini & + \dfrac{1}{\sqrt{6}} \dfrac{\Mp}{\mu} \ln\left(\dfrac{1 +
      \dfrac{\Gammaini}{\sqrt{6}}}{1 - \dfrac{\Gammaini}{\sqrt{6}}}
    \right)  < 1 - \dfrac{\Mp}{\mu} \sqrt{\dfrac{1}{2} + 2 \sixty}\,,
  \end{aligned}
\label{eq:sfimulargedomain}
\end{equation}
and there is no fine-tuning for $\mu \gg \Mp$.

These findings are confirmed by the numerical results of
\Fig{fig:sfiic} where the case of $\sfi_4$ is presented. For
$\mu = 0.5 \Mp$ (top panel), one recovers the thin, fine-tuned band as
predicted by \Eq{eq:sfimusmalldomain}, but its shape is distorted for
$\phiini/\mu \gtrsim 0.1$. This is expected because $\Gammasr(\phip)$
is no longer a small quantity in these regions and the hypothesis
$|\Gamma| \gg |\Gammasr|$ ceases to be accurate. For $\mu = 10 \Mp$
(bottom panel), the contours of equal e-foldings are in perfect
agreement with the functional shape of \Eq{eq:sfimulargedomain} and no
fine-tuning is necessary to trigger inflation. The situation is at all
points similar to the large-field models discussed in
\Sec{sec:lfi}. Phase-space trajectories for $\sfi_4$  are
represented in Figs.~\ref{fig:sfimu05traj} and \ref{fig:sfimu10traj}.

\subsubsection{Initial conditions problem in hilltop models}
\label{sec:pbhilltop}

The small-field models discussed in the previous section are also
referred to as ``hilltop models'' in the
literature~\cite{Boubekeur:2005zm, Tzirakis:2007bf, Enckell:2018kkc}
and the previous results allow us to discuss various claims made in
\Ref{Ijjas:2013vea} about them.

Reference~\cite{Ijjas:2013vea} argues that, on general grounds, the Planck data
disfavors the large-field $\lfi_p$ models compared to $\sfi_p$,
postulated to be all fine-tuned. In other words, Planck would have
shown that inflation is in trouble since the data favor a class of
models for which the choice of initial conditions is
unnatural. However, we have just seen that this is not the case. All
$\sfi_p$ models with $\mu \gg \Mp$ are free of fine-tuning issues, at
least in the very same manner as the $\lfi_p$ models are.\footnote{If one
  is ready to accept the large-field models as theoretically
  ``viable'', then one cannot argue that letting the scale $\mu$ be
  super Planckian is problematic given that the field is also super
  Planckian in the $\lfi_p$ scenarios.} Moreover, as can be checked in
\Ref{Martin:2013nzq}, although the Planck data indeed make the
Bayesian evidence of $\lfi$ smaller than the one of $\sfi$, within the
$\sfi$ models, they slightly disfavor the $\sfi$ scenarios having
$\mu < \Mp$ compared to the ones with $\mu > \Mp$. Let us stress that
the CMB data are blind to the initial conditions of inflation and this
result only comes from the observable values of the tensor-to-scalar
ratio and the spectral index. It means that, even for $p \ne 2$, the
models $\sfi_{p>2}$ favored after Planck actually have fewer problems
with regards to the initial conditions than the models favored before
Planck.

Even if the previous argument is clearly in favor of inflation, it is,
in fact, anecdotal since the $\sfi$ models are not belonging to the
most favored models. In terms of Bayesian evidence, the most probable
models are the plateau models~\cite{Martin:2013nzq}. These plateau
models are very different (in particular, with respect to the initial
conditions problem) from hilltop/$\sfi$ models and should not be
confused. The terminology of \Ref{Ijjas:2013vea}, which includes them
in the same category, is therefore problematic from that
perspective. The typical example of plateau inflation is the
Starobinsky model that we discuss in the next section.

\subsection{Plateau is not hilltop!}
\label{sec:si}

\begin{figure}
  \begin{center}
    \includegraphics[width=\figmaxw]{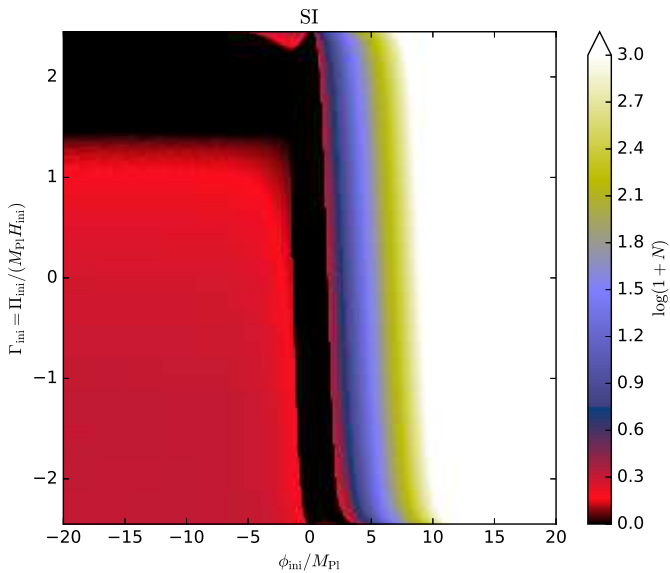}
    \caption{Number of e-folds of inflation for $\si$ achieved along
      phase-space trajectories starting from $2048^2$ initial
      conditions. Plateau-like models are very different from hilltop
      models. Almost all positive field values lead to slow-roll
      inflation, even for large kinetic energies.}
    \label{fig:siic}
  \end{center}
\end{figure}

\begin{figure}
  \begin{center}
    \includegraphics[width=\figw]{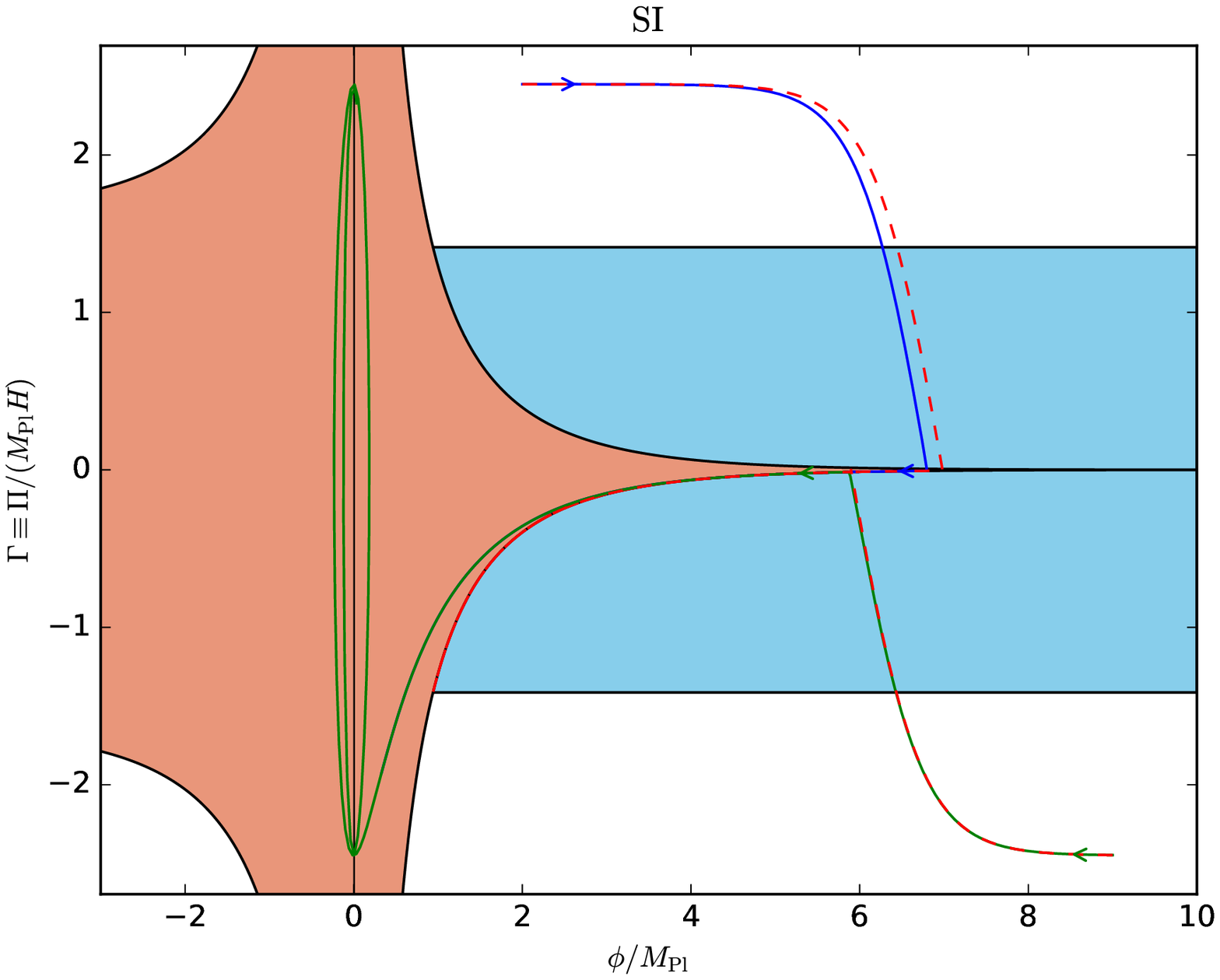}
    \includegraphics[width=\figw]{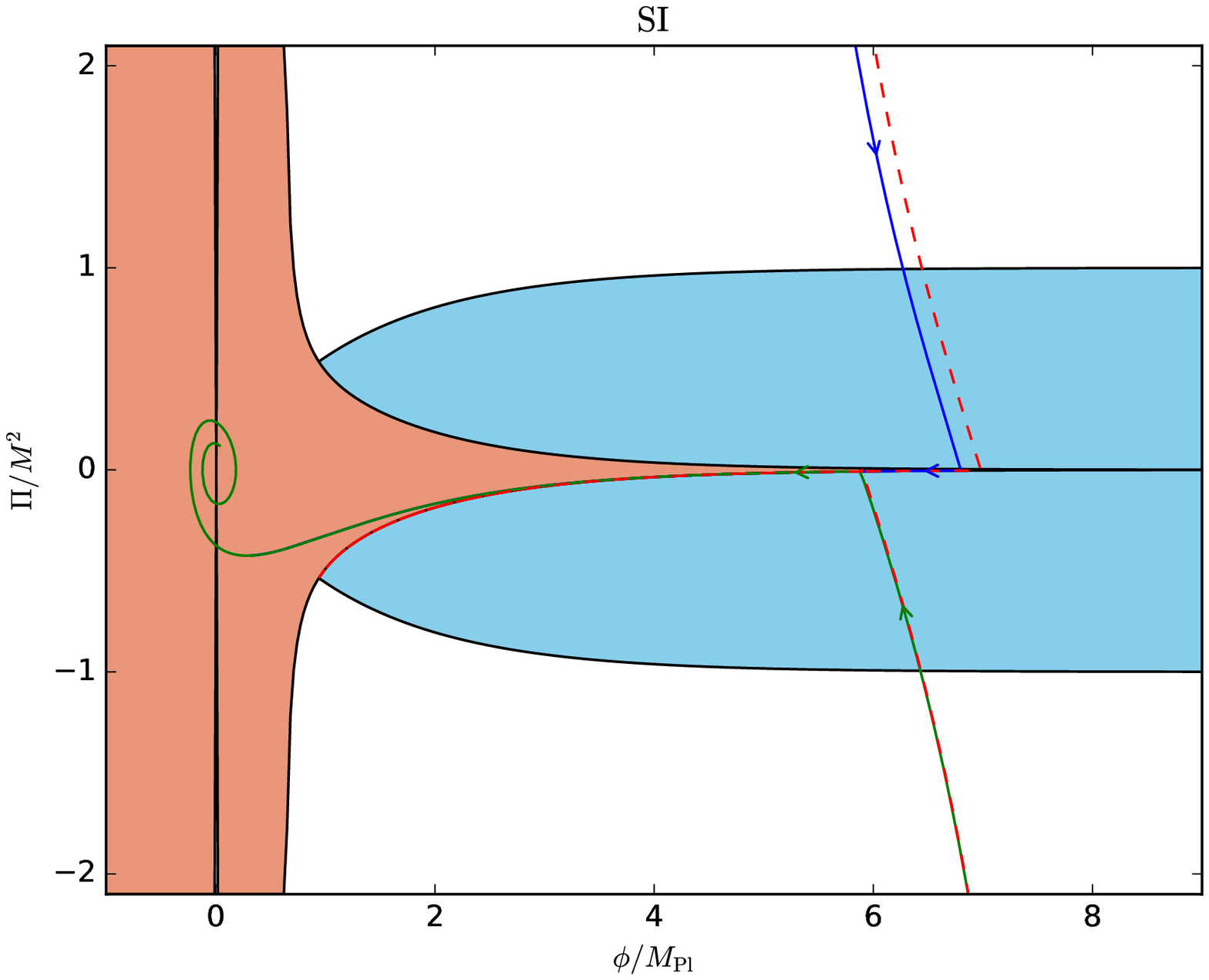}
    \caption{Phase-space trajectories for $\si$ (solid curve) starting
      deeply in the kination regime and relaxing toward slow
      roll. The upper panel corresponds to the phase-space
      $(\phip,\Gamma)$ while the lower panel is for $(\phi,\Pi)$. Our
      analytical approximations are represented as dashed red
      curves. The slight shift between the analytical and the exact
      upper trajectories comes from the fact that $\Gammasr$ (the
      boundary of the orange region) is not small compared to
      $\Gammaini$. The steepness of the potential toward its minimum
      slightly affects the kination regime described by
      \Eq{eq:GammaApprox}. Such an effect rapidly disappears at larger
      field values, as can be seen for the lower trajectory.}
    \label{fig:sitraj}
  \end{center}
\end{figure}

In this section, we consider the Starobinsky model ($\si$) which
exemplifies plateau inflation. As we will see, contrary to small-field
inflation, no fine-tuning is required to start inflation. The
potential is given by~\cite{Martin:2013tda}
\begin{align}
V(\phi)=M^4\left[1-\exp\left(-\sqrt{\frac23}
\frac{\phi}{\Mp}\right)\right]^2,
\label{eq:sipot}
\end{align}
and the slow-roll solution reads
\begin{equation}
\label{eq:si:SR:traj}
  \begin{aligned}
    \ee^{\sqrt{\frac{2}{3}} \phip(N)} - \sqrt{\dfrac{2}{3}} \phip(N) &
    =
    \dfrac{4}{3} \left(\Nend - N \right) \\ & + 1 + \dfrac{2}{\sqrt{3}} -
    \ln \left(1 + \dfrac{2}{\sqrt{3}} \right).
\end{aligned}
\end{equation}
As before, matching the kinetically driven regime to slow roll at
$\phipx$ gives the condition for $(\phipini,\Gammaini)$ to produce
$\Nend - \Nx > \sixty$ e-folds of slow-roll inflation. One gets
\begin{equation}
\begin{aligned}
   \ee^{\sqrt{\frac{2}{3}} \phipini} \left(\dfrac{1 + \dfrac{\Gammaini}{\sqrt{6}}}{1 -
  \dfrac{\Gammaini}{\sqrt{6}}} \right)^{\frac{1}{3}}
    - \sqrt{\dfrac{2}{3}} \phipini - \dfrac{1}{3} \ln \left(\dfrac{1 +
     \dfrac{\Gammaini}{\sqrt{6}}}{1 - \dfrac{\Gammaini}{\sqrt{6}}}
   \right)  \\  > \dfrac{4}{3} \sixty + 1 +\dfrac{2}{\sqrt{3}} -
   \ln\left(1+\dfrac{2}{\sqrt{3}} \right).
\end{aligned}
\label{eq:sidomain}
\end{equation}
No fine-tuning is necessary to satisfy such a condition. Compared to
$\lfi$, see \Eq{eq:lfidomain}, the presence of the exponential ensures
that slow-roll inflation \emph{always} occurs even for relatively low
values of $\phipini$. The numerical integration of the number of
e-folds is shown in \Fig{fig:siic} and matches well
\Eq{eq:sidomain}. The relaxation toward slow roll of a few
trajectories starting deeply in the kination regime is represented in
\Fig{fig:sitraj}.

The results of this section are probably the most important ones
regarding the initial conditions problem. In short, Planck favors
models, namely single-field plateau potentials, for which there is no
initial conditions problem at all. This is why inflation is not ``in
trouble'' after Planck but, on the contrary, is rather reinforced.

\subsection{The ``unlikeliness problem of inflation''}
\label{sec:uv}

Reference~\cite{Ijjas:2013vea} also argues that inflation is
``\emph{exponentially unlikely according to the inner logics of the
  inflationary paradigm itself}'' and that this problem is an
additional issue for inflation, independent of the initial conditions
problem previously discussed. The potential chosen by the authors to
exemplify this issue is $V(\phi) \propto [1 - (\phi/\phizero)^2]^2$
[the potential was written $V(\phi)=\lambda(\phi^2-\phi_0^2)^2$; see
Fig.~1(a) of that paper], which possesses a hilltop domain, at
$\phi<\phi_0$, and a large-field one, at $\phi>\phi_0$. The
``unlikeliness problem'' is the claim that inflation is more likely to
occur in the latter, whereas the data prefer the former.

The model is again referred to as a ``plateau-like model''. The
leading-order expansion of the potential in $\phi\ll\phizero$ is
\Eq{eq:potsfi} with $p=2$, $M^4=\lambda \phi_0^4$ and
$\mu=\phi_0/\sqrt{2}$. Equation~(3) of \Ref{Ijjas:2013vea} suggests that
only the regime $\mu < \Mp$ is considered, while it is stated that
$\phi \ll \phizero$ is required for inflation to occur. Their
terminology ``plateau-like'' is therefore unambiguously referring to
$\sfi_2$ models with sub-Planckian vev (which we have shown in
\Sec{sec:sfi} suffer from an initial-conditions fine-tuning issue).

As we have argued before, this terminology is inappropriate as
``plateau'' is not ``hilltop''. More importantly, the choice $p=2$ is
a very particular case. Because it is a hilltop model with a
non-vanishing mass, as discussed at length in \Ref{Martin:2013tda},
$\sfi_2$ with $\mu < 2 \Mp$ does not support slow-roll inflation at
all, the spectral index is very far from scale invariance and this
model is ruled out by any CMB data. Within all possible $\sfi_2$
models, only the ones having super-Planckian $\mu > 2 \Mp$ can be made
compatible with CMB measurements, and from \Eq{eq:sfimulargedomain},
super-Planckian $\sfi_2$ models do not suffer from any fine-tuning
issues.

In order to study the ``unlikeliness problem'', \Ref{Ijjas:2013vea}
needs, in fact, a model with a small-field part and a large-field
part, the latter being interpreted as a reasonable UV completion of
the former. Although the choice of $V(\phi) \propto [1 -
  (\phi/\phizero)^2]^2$ is quite unfortunate, consistent slow-roll
models having this property exist and, in the following, we study two
explicit examples. It is worth noticing again that these types of models
are not plateau-like and, therefore, are not among the best models
according to the Planck data. This implies, as discussed at the end of
\Sec{sec:pbhilltop}, that the ``unlikeliness problem'', if it exists,
can only affect models that are not favored by the
data. Nevertheless, let us be exhaustive in our discussion of
the criticisms raised against inflation.

\subsubsection{Generalized double-well inflation}
\label{sec:gdwi}

\begin{figure}
  \begin{center}
    \includegraphics[width=\figmaxw]{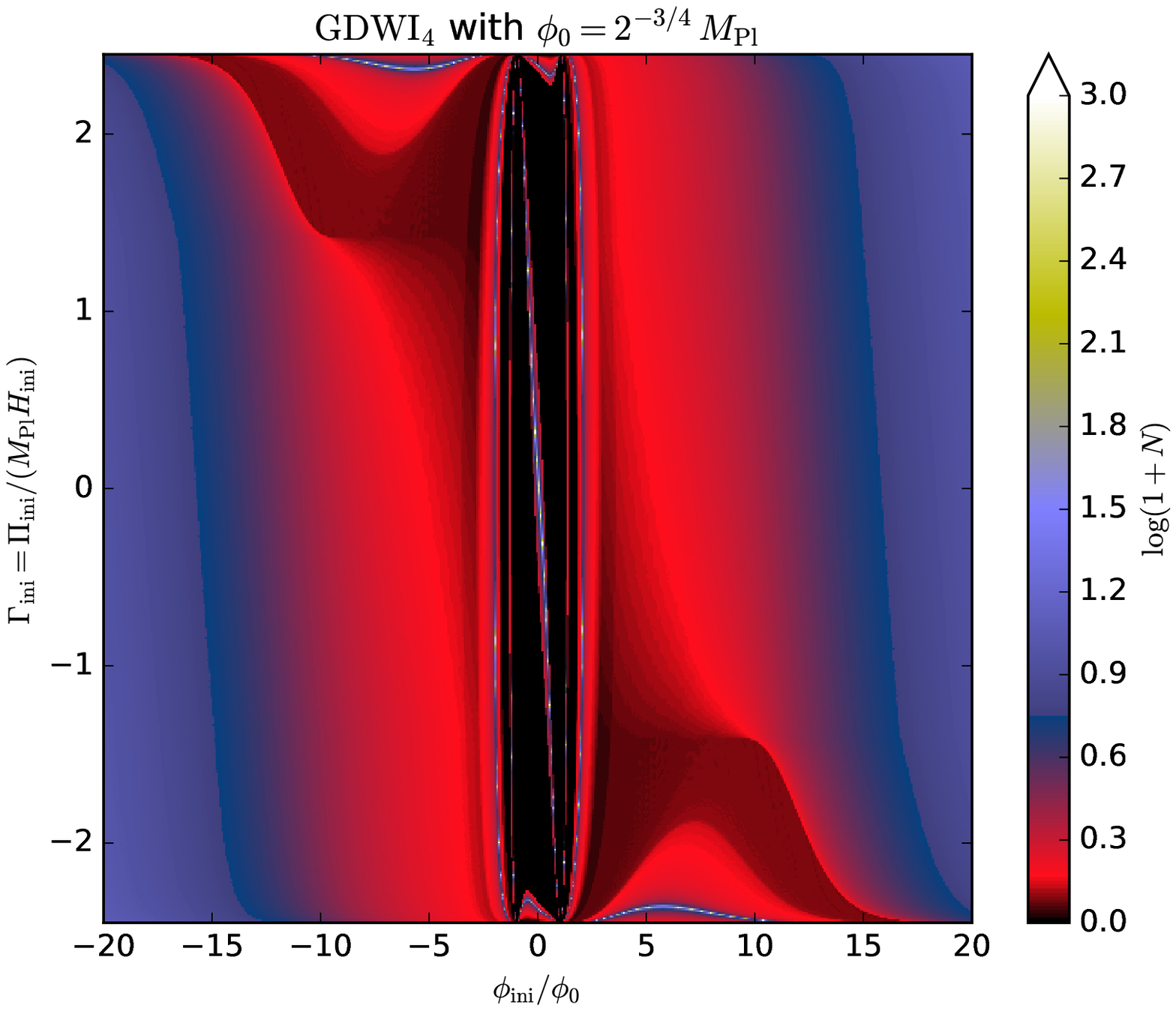}
    \includegraphics[width=\figmaxw]{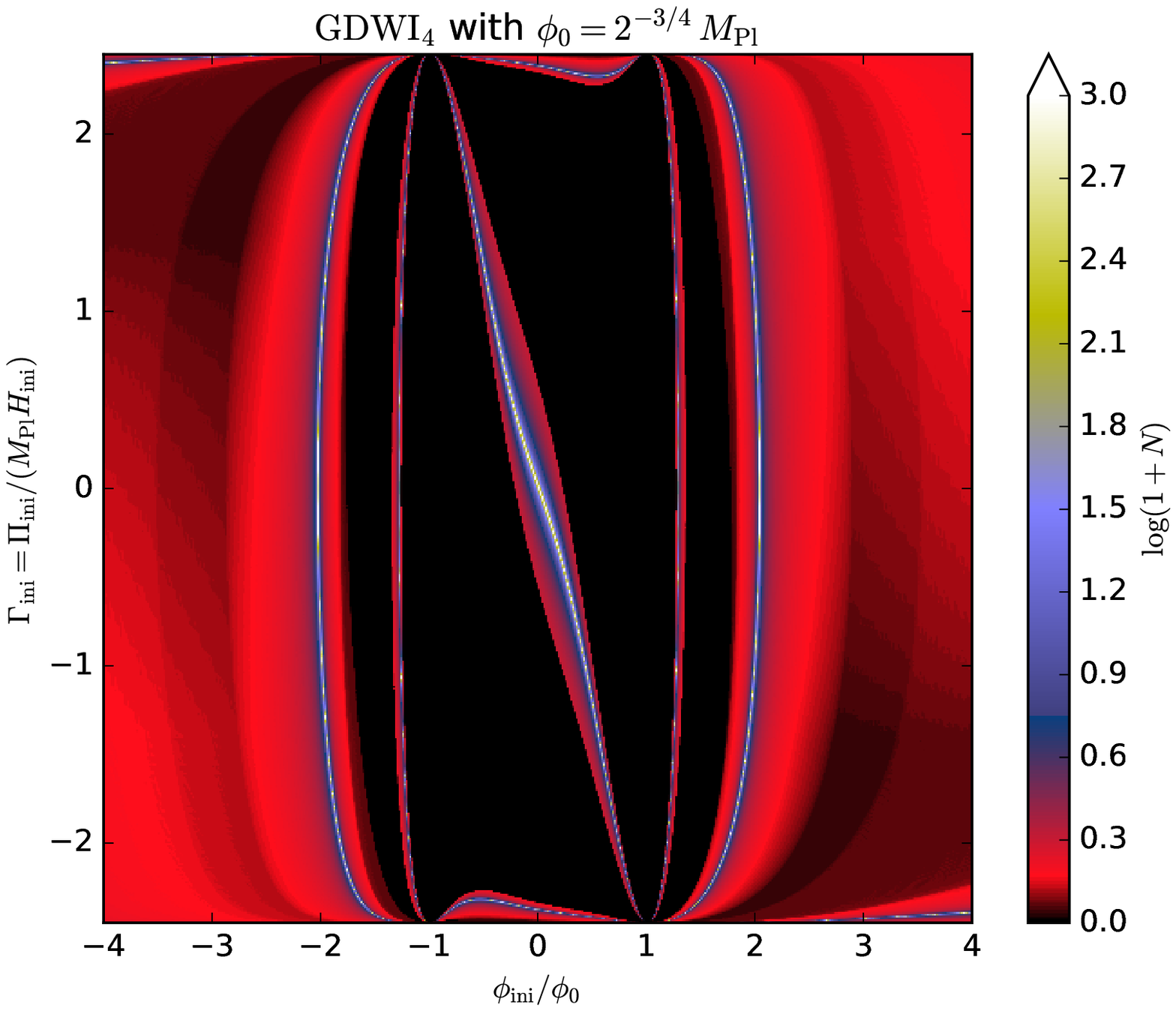}
    \caption{Number of e-folds of inflation for the fine-tuned
      $\gdwi_4$ model ($\phizero < \Mp$) achieved along phase-space
      trajectories starting from $2048^2$ initial conditions. The
      lower panel is a zoom into the central region for which
      $-4 \phizero <\phi < 4 \phizero$. The light narrow band in the
      middle is identical to the one obtained for $\sfi_4$ in
      \Fig{fig:sfiic}. However, the completed potential of $\gdwi_4$
      now allows for new successful initial conditions. The narrow
      band expands and spirals into steep regions of the
      potential.}
    \label{fig:gdwiic}
  \end{center}
\end{figure}

\begin{figure}
  \begin{center}
    \includegraphics[width=\figw]{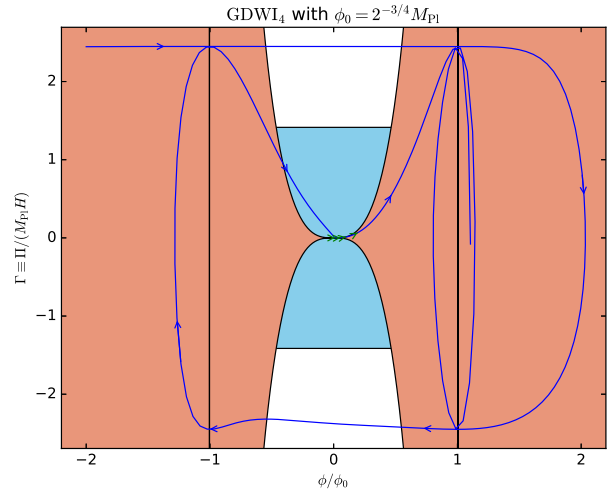}
    \caption{A phase-space trajectory for the fine-tuned $\gdwi_4$
      model ($\phizero < \Mp$), starting in kination and in a region
      of the potential that is steep, $\Gammasr(\phipini) > \Gammaini
      \simeq \sqrt{6}$. Compared to $\sfi_4$, the UV-completed
      potential allows for a complex bouncing dynamics that makes this
      trajectory producing hundreds of e-folds of hilltop inflation
      (around the green horizontal arrows at $\phi \simeq 0$). These
      trajectories are responsible for the multiplication of the
      successful inflationary regions in \Fig{fig:gdwiic}.}
    \label{fig:gdwitraj}
  \end{center}
\end{figure}

\begin{figure}
  \begin{center}
    \includegraphics[width=\figmaxw]{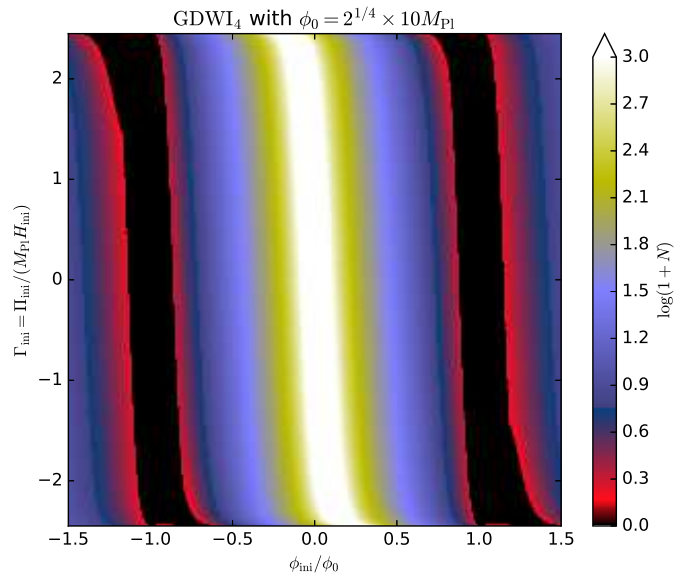}
    \caption{Number of e-folds of inflation for the $\gdwi_4$ model in
      the non-fine-tuned regime ($\phizero > \Mp$) along $2048^2$
      phase-space trajectories. The large inflationary band in the
      middle is identical to the one obtained for $\sfi_4$ in
      \Fig{fig:sfiic}. The structure of the successful domain is not
      affected by the UV completion.}
    \label{fig:gdwiiclarge}
  \end{center}
\end{figure}

In this section, we consider the generalized double-well model
($\gdwi_{2p}$), the potential of which can be written as
\begin{align}
\label{eq:potgdwi}
V(\phi)=M^4\left[\left(\frac{\phi}{\phizero}\right)^{2p}-1\right]^2,
\end{align}
where $\phizero$ is a vev and $p$ a positive power index. For
$\phi \ll \phizero$, \Eq{eq:potgdwi} can be expanded as
\begin{equation}
V(\phi) \simeq M^4\left[1 - 2 \left(\dfrac{\phi}{\phizero}
  \right)^{2p} \right].
\label{eq:dwiapprox}
\end{equation}
The case $p=1$ corresponds to the so-called double-well inflation
($\dwi$) studied in Sec.~4.14 of \Ref{Martin:2013tda}. As discussed in
this reference, $\dwi$ can be viewed as a UV completion of $\sfi_2$
with $\mu = \phizero/\sqrt{2}$. However, it is shown in this reference
that slow-roll inflation can only occur for
$\phizero > 2 \sqrt{2} \Mp$, for which there is no fine-tuning of the
initial conditions. This potential is therefore of limited interest to
discuss how the fine-tuning could be affected by the large-field
completion of the potential. As discussed previously, $\dwi$ is the
potential chosen in \Ref{Ijjas:2013vea} to supposedly illustrate the
fine-tuning of hilltop inflation.

The power index $p=2$ is, however, of immediate interest. As can be seen
from \Eq{eq:dwiapprox}, $\gdwi_4$ provides a large-field completion of
$\sfi_4$ with $\mu = 2^{-1/4} \phizero$. For $\phi \gg \phizero$,
\Eq{eq:potgdwi} behaves as the large-field model $\lfi_{4p}$. The
initial conditions to start inflation are given by \Eq{eq:lfidomain}
and are not fine-tuned. In the small-field regime, they are given by
\Eqs{eq:sfimusmalldomain} and \eqref{eq:sfimulargedomain} and are
either fine-tuned for $\phizero < \Mp$, or not fine-tuned for
$\phizero > \Mp$.

In \Fig{fig:gdwiic}, we have represented the number of inflationary
e-folds as a function of initial conditions for $p=2$ and
$\phizero = 2^{-3/4} \Mp$. This corresponds to the fine-tuning regime
of $\sfi_4$ with $\mu = 0.5 \Mp$. In the central region, for
$\phi \simeq 0$, we recover exactly the same structure as in
$\sfi_4$. The shape and position of the narrow band in which inflation
occurs are the same (see \Fig{fig:sfiic}). However, for
$\phi/\phizero \gtrsim 1$, new successful inflationary regions appear.
They are extensions of the central narrow band that are spiraling
many times into the steep parts of the UV-completed potential. Their
origin is evident from the example trajectory plotted in
\Fig{fig:gdwitraj}. Starting with a large kinetic energy in a steep
region of the potential, the field may cross the local maximum at
$\phi=0$ one or several times before falling into one of the two
minima. For some values of the initial kinetic energy, the last
crossing occurs with small enough velocity to enter a phase of
slow-roll inflation. This does not solve the fine-tuning problem of
$\sfi_4$ with $\mu < \Mp$, since the regions of successful initial
conditions in phase-space remain of small size. Still, the presence of
a large-field branch in the potential increases the size of the
successful regions of inflation rather than diminishing them.

In \Fig{fig:gdwiiclarge}, we have represented the number of
inflationary e-foldings in phase-space for $\gdwi_4$ in the regime
without fine-tuning, for $\phi/\phizero = 2^{1/4}\times 5 \Mp$. The
central region matches the one of $\sfi_4$ with $\mu = 10\Mp$ (see
\Fig{fig:sfiic}) and there is no fine-tuning. Interestingly, there are
no longer trajectories spiraling around the central region. This is
due to the fact that the potential is flat enough, namely $|\Gammasr|
\ll 1$, all over the region $\phi/\phizero \lesssim 1$. Therefore,
even if $|\Gammaini|$ is large,
\Eq{eq:sfimulargedomain} applies: the system relaxes toward slow roll
and cannot cross the whole hilltop region.

Let us now discuss the ``unlikeliness problem'' in the light of these
results. In a double-well potential with super-Planckian well
separation ($\phi_0>\Mp$), the data strongly support the hilltop
region of the potential against the large-field one, but there is no
fine-tuning issue of the initial conditions in either region, hence
there is no unlikeliness problem. If the well separation is
sub-Planckian ($\phi_0<\Mp$) and the effective mass does not vanish at
the top of the hill ($p=1$ or $\gdwi_{2}$), both inflating regions of
the potential are strongly disfavored by the data and the entire
potential is excluded. There is therefore no unlikeliness problem in
that case either. Only if the well separation is sub-Planckian and the
hill mass vanishes, the favored region of the potential, {\ie}, the
hilltop one, suffers from initial-conditions fine-tuning, though this
fine-tuning is not reinforced by the UV completion (it is rather the
contrary as discussed before). It is the only situation where one
could argue in favor of an ``unlikeliness problem''. This, however
corresponds to a very specific choice of the inflationary potential,
that is anyway disfavored compared to plateau
models~\cite{Martin:2013nzq}.

\subsubsection{Coleman-Weinberg inflation}
\label{sec:cwi}

\begin{figure}
  \begin{center}
    \includegraphics[width=\figmaxw]{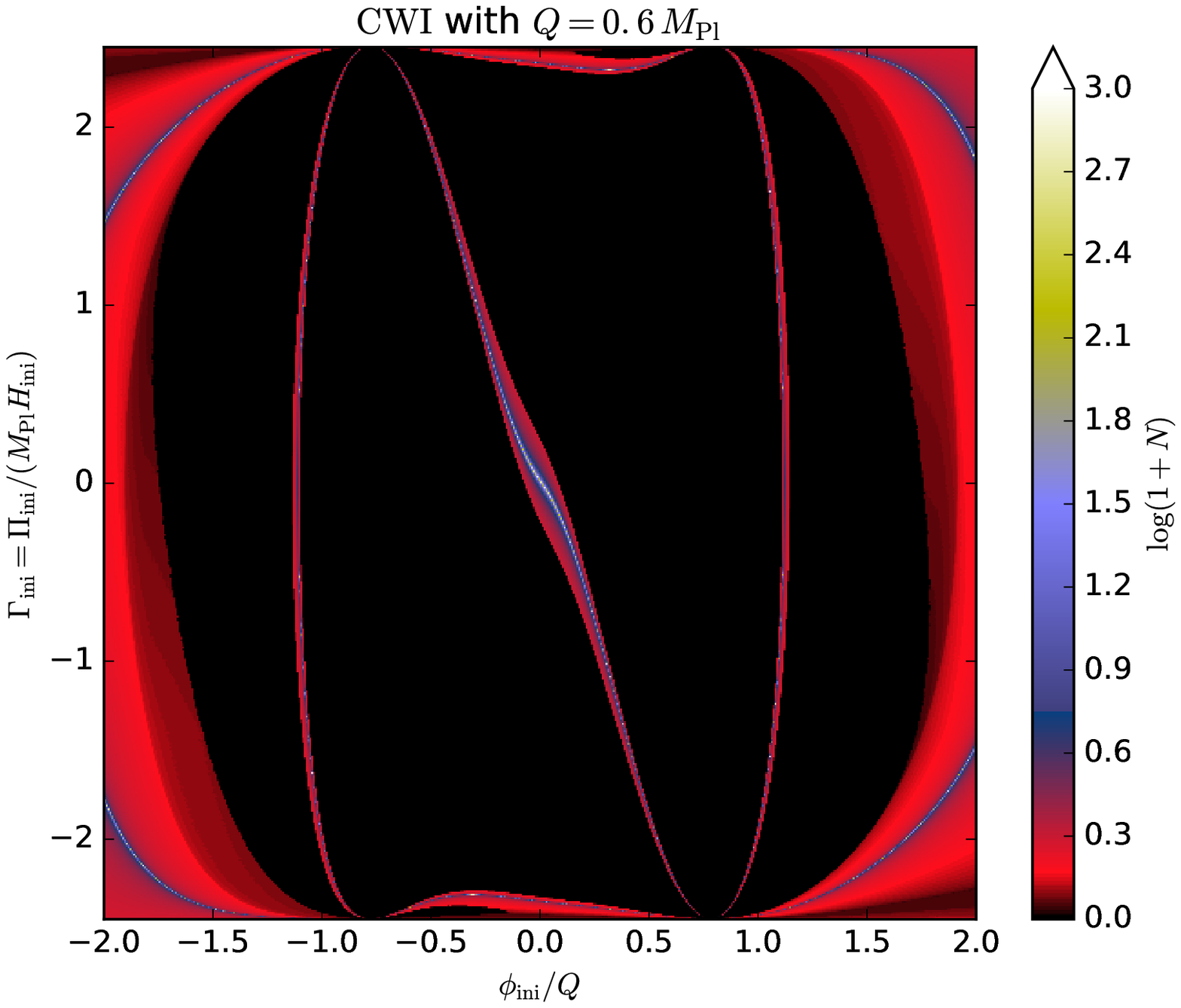}
    \includegraphics[width=\figmaxw]{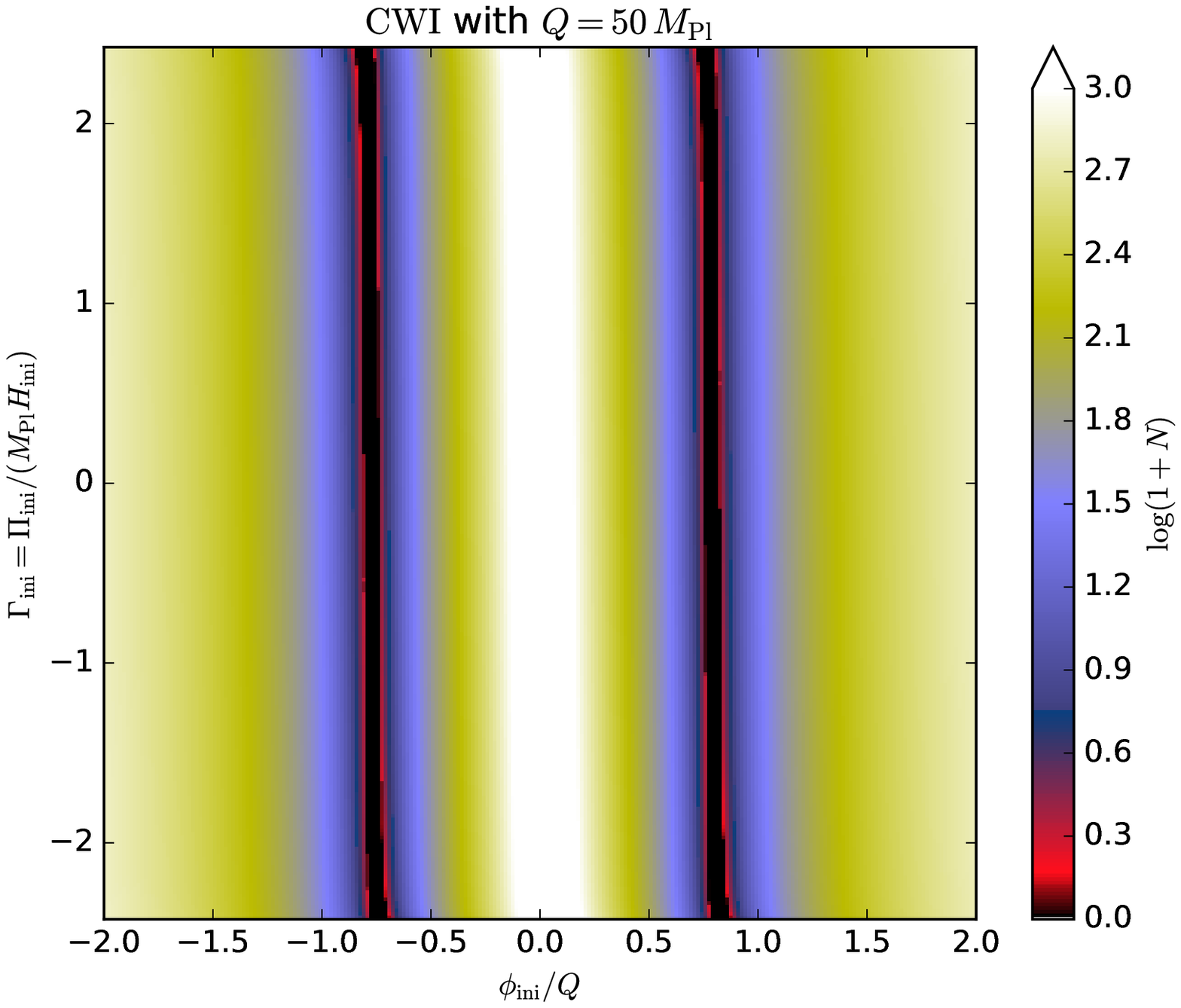}
    \caption{Number of e-folds of inflation in the $\cwi$ model for
      $2048^2$ initial conditions. The upper panel is in the
      fine-tuning regime $Q<\Mp$ while the bottom panel if for
      $Q > \Mp$. The situation is generic of hilltop inflation, see
      also \Figs{fig:gdwiic} and \ref{fig:gdwiiclarge}.}
    \label{fig:cwiic}
  \end{center}
\end{figure}

\begin{figure}
  \begin{center}
    \includegraphics[width=\figmaxw]{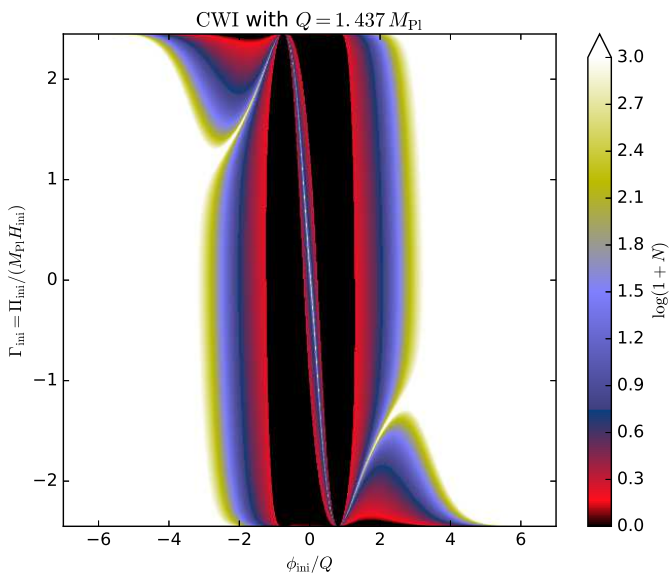}
    \caption{Number of e-folds of inflation in the $\cwi$ double
      inflationary regime, for $Q \simeq 1.4\,\Mp$. All initial
      conditions in the large-field regime, $\phiini/Q > 1$, produce
      trajectories climbing up into the central hilltop narrow
      band. There is no fine-tuning of the initial conditions anymore,
      and the observable window ({\ie}, the last $\sim60$ e-folds of
      inflation) is always in the hilltop domain of the potential,
      even though this regime would appear fine-tuned without the UV
      completion of the potential.}
    \label{fig:cwimagic}
  \end{center}
\end{figure}

\begin{figure}
  \begin{center}
    \includegraphics[width=\figw]{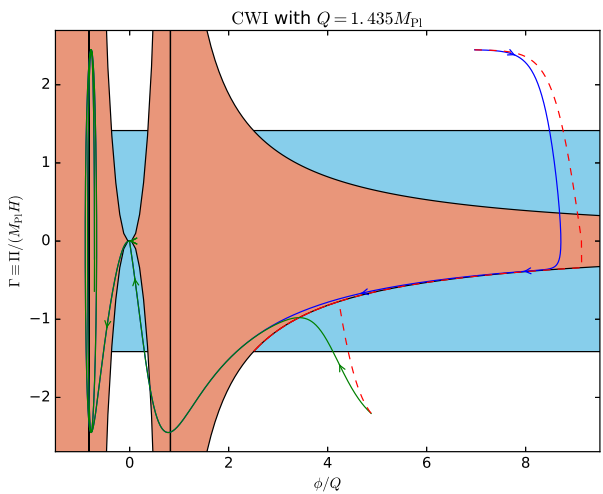}
    \caption{Example trajectories for $\cwi$ in the double inflation
      regime, for $Q = 1.435\,\Mp$. The two trajectories represented
      as solid lines start in the large-field inflationary region and
      end up producing about $500$ e-folds of hilltop inflation around
      $\phi \simeq 0$. The dashed red curves are the analytical
      approximations in the large-field regime and can be used only in
      that region.}
    \label{fig:cwitraj}
  \end{center}
\end{figure}

\begin{figure}
  \begin{center}
    \includegraphics[width=\figw]{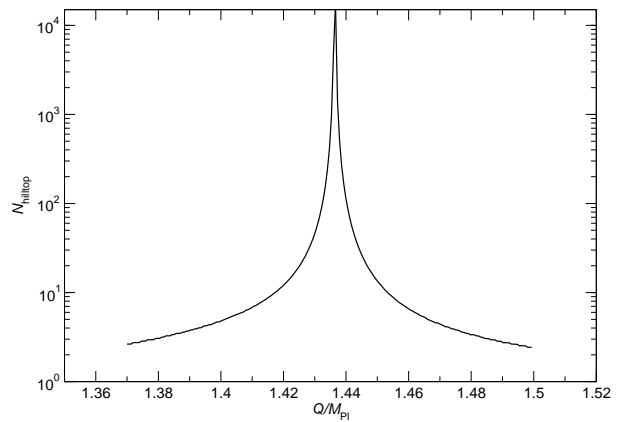}
    \caption{Number of e-folds of hilltop inflation realized within
      the $\cwi$ double-inflation regime. Because of the slow-roll
      attractor in the large-field domain, it depends only on the
      scale $Q$.}
    \label{fig:cwiefolds}
  \end{center}
\end{figure}

Although discussed for $\gdwi_4$ only, we expect the previous findings
to be a generic property of the hilltop models. In this section, we
indeed recover them in a particle-physics motivated model: the
Coleman-Weinberg potential (CWI)~\cite{Linde:1981mu}
\begin{equation}
V(\phi)= M^4 \left[1 + 2\ee \left(\dfrac{\phi}{Q}\right)^4 \ln
  \left(\dfrac{\phi^2}{Q^2}\right) \right].
\label{eq:potcw}
\end{equation}
The potential vanishes at its two minima for $\phi/Q = \pm \ee^{-1/4}$
and supports both hilltop inflation for $|\phi/Q| < \ee^{-1/4}$ and
large-field-like inflation for $|\phi/Q| > \ee^{-1/4}$. Slow-roll
solutions have been derived in \Ref{Martin:2013tda} and can be used
together with \Eq{eq:deltaphipmax} to derive the initial conditions
required to get enough e-folds of inflation in the hilltop region. The
situation is very similar to \Sec{sec:sfi} and we do not reproduce the
calculations here. The region $|\Gammasr| \ll 1$ is very confined
around $\phi=0$ when $Q < \Mp$ and starting inflation is fine-tuned in
that case.

The full numerical integration of $\cwi$ in the fine-tuning regime,
with $Q=0.6\Mp$, is presented in the upper panel of
\Fig{fig:cwiic}. The situation is in all points similar to $\gdwi_4$
(see \Fig{fig:gdwiic}). The narrow region of successful initial
conditions is extended into a spiraling band exploring the steep
parts of the potential that slightly alleviates the fine-tuning
problem. For completeness, we have also represented in the bottom
panel of \Fig{fig:cwiic} the non-fine-tuned case $Q=50\,\Mp$, where
the whole hilltop region $\phiini/Q < 1$ inflates.

A very interesting phenomenon appears for Planckian-like expectation
values $Q = \order{\Mp}$, the large-field region becomes connected to
the hilltop one. The kinetic energy acquired by the field when exiting
the large-field inflationary regime may become large enough to climb
into the hilltop domain, thereby triggering a second inflationary era
(see \Fig{fig:cwimagic}). Such an effect is not generic of hilltop
models as it clearly depends on how the hilltop domain is UV
completed. For the Coleman-Weinberg potential, we find that double
inflation appears for $Q\simeq 1.4\,\Mp$~\cite{Yokoyama:1998pt}. For
such a value of $Q$, even though the hilltop regime is still rather
fine-tuned, \emph{all large-field trajectories} end up in the narrow
inflationary band at the top of the potential. As a result, the fine
tuning of $(\Phi_\uini,\Gammaini)$ to start hilltop inflation is now
alleviated by a percent-like condition on the fundamental scale of the
theory, here $Q$. When this condition is satisfied, the universe may
spend a long time in the large-field inflationary regime, but
ultimately, the last $\sixty$ e-folds of inflation, the observable
ones, are realized in the hilltop domain. A typical trajectory in
phase-space has been represented in Fig.~\ref{fig:cwitraj}.  Let us
also notice that, in that case, the precise number of e-folds in the
hilltop regime becomes a function of $Q$ only and has been plotted in
\Fig{fig:cwiefolds}. Although not generic, this example illustrates
again that UV completion can only help in alleviating the fine-tuning
problem in hilltop models, when present.

In conclusion, the properties of the
Coleman-Weinberg potential are very similar to those of
$\gdwi_4$. The fact that there is no ``unlikeliness
problem'' in this type of scenarios therefore seems to be generic.

\subsection{UV-completion and initial conditions}
\label{sec:ccsi}

\begin{figure}
  \begin{center}
    \includegraphics[width=\figmaxw]{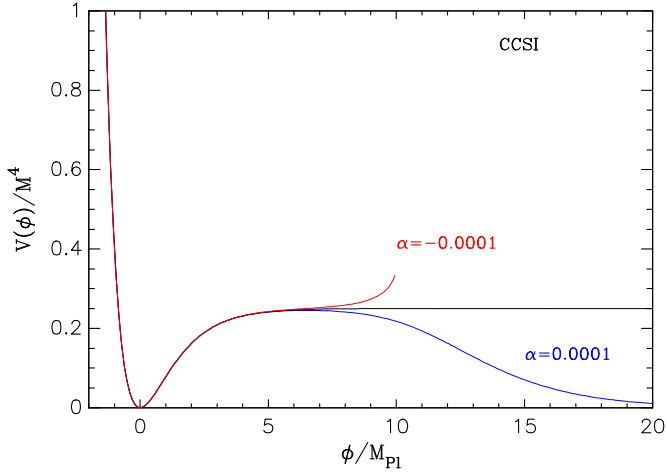}
    \caption{The potential~(\ref{eq:ccsipot}) of the cubicly corrected
      Starobinsky model, for $\alpha > 0$ ($\ccsi_1$) and $\alpha < 0$
      ($\ccsi_3$). The potential of Starobinsky inflation ($\si$) is
      the middle plateau ($M^4 \equiv \Mp^2 \mu^2/2$).}
    \label{fig:ccsipot}
  \end{center}
\end{figure}

In \Sec{sec:si}, we have established that there is no fine-tuning
issue for the initial conditions in the case of plateau inflation. The
reason is clear: the presence of a large plateau in the potential
ensures the relaxation of kination into slow roll according to
\Eq{eq:sidomain}. Physically, however, the plateau may not extend to
infinitely large-field values due to the presence of higher-order
corrections.  This is noticed in \Ref{Ijjas:2013vea}, which emphasizes
that in the Taylor expansion of $V(\phi)$, the desired flat behavior
can be obtained only if a precise cancellation order by order in
$\phi$ occurs.  Although there are mechanisms that automatically
produce such an ``exact'' plateau, see {\eg} \Ref{Kallosh:2013hoa}, a
legitimate question is then to determine whether UV-completed
plateau models suffer from a fine-tuning problem.

Let us first notice that if the correction is of the large-field type,
{\eg}, in a potential of the type
\begin{align}
\label{eq:phenocorrsi}
V(\phi) &=
V_{\si}(\phi) \, \heaviside{\frac{\phi_\mathrm{corr}}{\Mp}
-\frac{\phi}{\Mp}} \nonumber \\
&+V_{\si}(\phi_\mathrm{corr})
\left[1-\heaviside{\frac{\phi_\mathrm{corr}}{\Mp}
-\frac{\phi}{\Mp}} \right]\left(\frac{\phi}{\Mp}\right)^p,
\end{align}
where $V_\mathrm{SI}(\phi)$ is the Starobinsky potential of
\Eq{eq:sipot} and $\heaviside{x}$ is the Heaviside function, no
fine-tuning is required since neither the plateau branch at
$\phi<\phi_\mathrm{corr}$ nor the large-field branch at
$\phi>\phi_\mathrm{corr}$ suffers from a fine-tuning problem.

More generically otherwise, \Eq{eq:sidomain} shows that the plateau
needs only to cover a field range larger than $\Delta \phipmax$ for
relaxation to occur. Let us illustrate this property by considering
the cubicly corrected Starobinsky model ($\ccsi$), which is a type of
correction more physically motivated than the phenomenological
form~(\ref{eq:phenocorrsi}). It is a modified gravity $f(R)$ model
given by~\cite{Artymowski:2015mva, Artymowski:2015ida}
\begin{equation}
f(R) = R + \frac{R^2}{\mu^2} + \alpha\dfrac{R^3}{\mu^4}\,,
\label{eq:deff}
\end{equation}
where $\mu$ is a mass scale and $\alpha$ an expected small
dimensionless number. After a conformal transformation, any $f(R)$
theory can be cast into a scalar field theory, where the scalar field
$\phi$ is defined as~\cite{DeFelice:2010aj}
\begin{equation}
 \phi = \sqrt{\dfrac{3}{2}}\,\Mp \ln\left(\left|F\right|\right),
\end{equation}
where
\begin{equation}
  F(R) = \dfrac{\partial f(R)}{\partial R} =
  \exp{\left(\sqrt{\dfrac{2}{3}} \dfrac{\phi}{\Mp}\right)}.
\label{eq:FofR}
\end{equation}
The corresponding potential can be written as
\begin{equation}
\label{eq:f(R):pot}
V(\phi) = \frac{\Mp^2}{2}\,
\frac{\left|F\right|}{F}\,\frac{R\,F - f}{F^2}\,.
\end{equation}

Let us first consider the case where $\alpha=0$. Defining
\begin{equation}
y \equiv \sqrt{\dfrac{2}{3}} \dfrac{\phi}{\Mp}\,,
\end{equation}
and solving \Eq{eq:FofR} for $R$, one gets
\begin{equation}
  R = \dfrac{\mu^2}{2}\left(\ee^y - 1\right).
\label{eq:R-staro}
\end{equation}
The potential~(\ref{eq:f(R):pot}) is then given by
\begin{equation}
 V(\phi) = \frac{\Mp^2\,\mu^2}{8}\left(1 - \ee^{-y}\right)^2,
\label{eq:staropot}
\end{equation}
which, as expected, corresponds to the standard Starobinsky model of
\Sec{sec:si}. Therefore, the higher-order terms in \Eq{eq:deff} are
natural gravity-motivated corrections to $\si$. Notice that the
potential of Starobinsky inflation matches the one of Higgs
inflation~\cite{Bezrukov:2007ep}, where one assumes that the inflaton
field is the Higgs boson non minimally coupled to gravity. This
picture could also motivate the form of other possible
corrections~\cite{Barvinsky:2008ia, DeSimone:2008ei, Bezrukov:2010jz}.

\begin{figure}
  \begin{center}
    \includegraphics[width=\figmaxw]{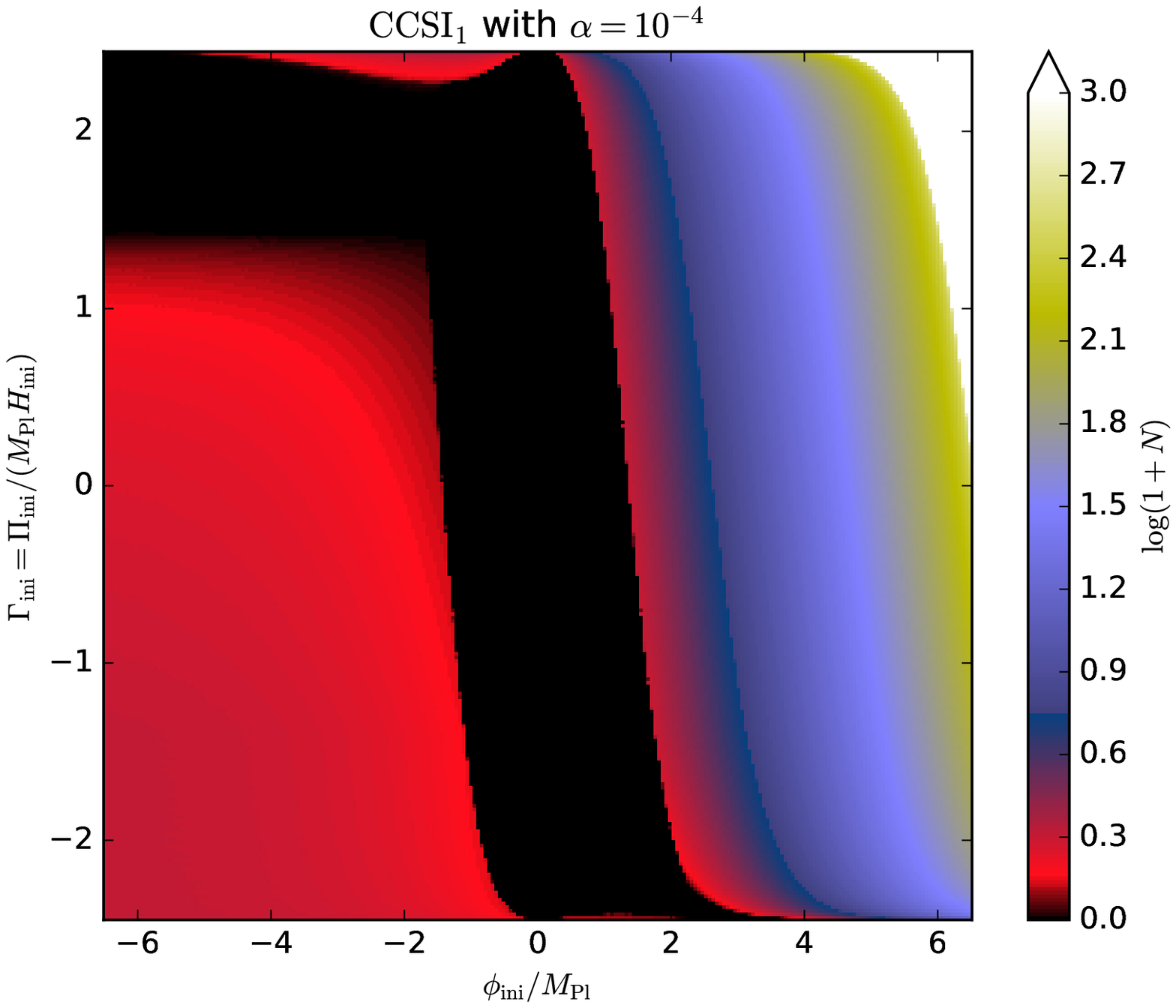}
    \includegraphics[width=\figmaxw]{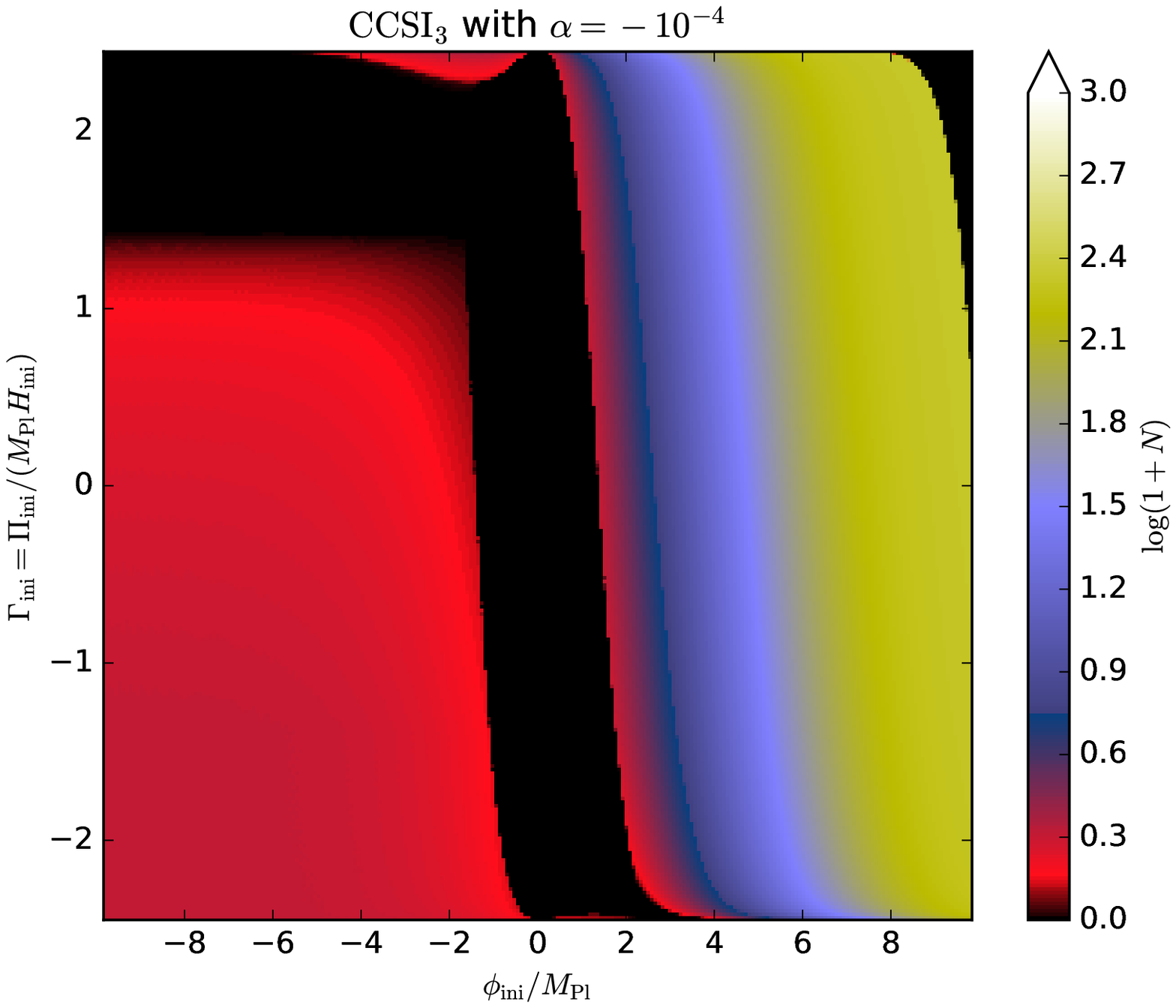}
    \caption{Number of e-folds of inflation for the $\ccsi_1$
      ($\alpha >0$, upper panel) and $\ccsi_3$ ($\alpha < 0$, lower
      panel) models along $2048^2$ phase-space trajectories. Even
      though the plateau is reduced compared to $\si$, starting
      inflation does not require fine-tuning of the initial
      conditions. At large-field values, the number of e-folds is
      unbounded in $\ccsi_1$ while the UV cutoff ensures that it
      remains finite in $\ccsi_3$. The upper-right black corner in the
      lower panel corresponds to cases where the field climbs up the
      potential to $\phiuv$ where it stops being defined. In this
      case, the integration is stopped and the corresponding forbidden
      region is simply displayed in black.}
    \label{fig:ccsiic}
  \end{center}
\end{figure}

Let us now consider the case where $\alpha$ is non-zero. Solving
\Eq{eq:FofR} for $R$ gives
\begin{equation}
 1 + 2 \dfrac{R}{\mu^2} + 3 \alpha \dfrac{R^2}{\mu^4} = \ee^y,
\end{equation}
whose roots are
\begin{equation}
 R = \frac{\mu^2}{3 \alpha}\left[-1 \pm 
\sqrt{1 + 3\alpha \left(\ee^y - 1\right)}\right].
\end{equation}
We choose the positive sign in the above equation so that it reduces
to the expression~\eqref{eq:R-staro} for the standard Starobinsky
model in the limit $\alpha\to 0$.  With this solution, the potential
reads
\begin{equation}
\begin{aligned}
  V(\phi) &= \dfrac{\Mp^2 \mu^2}{2} \left(1 - \ee^{-y} \right)^2 \\ & \times
  \dfrac{1 + \sqrt{1 + 3\alpha \left(\ee^y - 1\right)} + 2\alpha
      \left(\ee^y - 1 \right)}{\left[1 + \sqrt{1+3 \alpha \left(\ee^y
        - 1\right)} \right]^3}\,,
\end{aligned}
\label{eq:ccsipot}
\end{equation}
which matches \Eq{eq:sipot} for $\alpha = 0$. For $\alpha > 0$
($\ccsi_1$), the potential is always well defined at large-field
values. The plateau is distorted and there is a maximum at
\begin{equation}
\phiVmax = \sqrt{\dfrac{3}{2}} \Mp \ln \left(\dfrac{2 
+ 4 \sqrt{\alpha}}{\sqrt{\alpha}} \right),
\end{equation}
above which the potential asymptotically goes to zero. For
$\alpha < 0$ ($\ccsi_3$), the potential is only defined for
$\phi < \phiuv$, where
\begin{equation}
\label{eq:ccsi:phiuv}
 \phiuv = \sqrt{\frac{3}{2}} \Mp \ln\left(1 - \dfrac{1}{3\alpha}\right).
\end{equation}
At $\phiuv$, the potential is finite and its value is the one on the
asymptotic plateau (when $\alpha=0)$ multiplied by $4/[3(1-3\alpha)^2]$.

The potentials of $\ccsi_1$ and $\ccsi_3$ have been represented in
\Fig{fig:ccsipot}. Clearly, the correction breaks the infinitely wide
plateau that was present in $\si$. In \Fig{fig:ccsiic}, we have
plotted the number of e-folds of inflation in phase-space for these
two potentials. As expected, inflation still occurs in a large part of
phase-space and no fine-tuning is required. These plots can be
compared to \Fig{fig:siic}. For $\ccsi_1$, there is almost no change
compared to $\si$, the whole large-field domain produces
inflation. However, a crucial change is that the number of e-folds in
the region $\phi > \phiVmax$ is infinite and inflation never ends. For
$\ccsi_3$ the situation is reversed. The field being bounded by
$\phi < \phiuv$, the total number of e-folds of inflation can be large
but never exceeds $10^3$.\footnote{This highlights the sharp
  difference between UV-corrected plateau potentials and hilltop
  models, since an arbitrarily large number of e-folds can always be
  realized in the latter, regardless of the width of the hill. This
  further shows why plateau models, even with UV corrections, cannot
  be categorized jointly with hilltop models, see \Sec{sec:si}.}

Let us estimate how small the parameter $\alpha$ should be in order to
have $\sixty$ e-folds of slow-roll plateau inflation. Approximating
\Eq{eq:si:SR:traj} in the large $\sixty$ regime, one has
$\phipx\simeq \sqrt{3/2}\ln(4\sixty/3)$. Requiring that
$\phiVmax>\phipx \Mp$ in $\ccsi_1$ then leads to
$\alpha<(2\sixty/3-2)^{-2}\simeq 10^{-3}$ for $\sixty\simeq 50$, and
in $\ccsi_3$, the condition $\phiuv>\phipx\Mp$ leads to
$\vert \alpha\vert < 1/(4\sixty-3)\simeq 5\times 10^{-3}$. No extreme
fine-tuning is thus required on the small expansion parameter
$\alpha$, and plateau inflation is therefore rather robust to small
corrections that may eventually break the plateau.

\subsection{Model simplicity}
\label{sec:discuss}

How natural the inflationary paradigm is can also be discussed by
considering whether the data can be explained with simple models of
inflation or if it forces us to consider more complicated and
contrived scenarios. For instance, in \Ref{Ijjas:2013vea}, it is
argued that the models ruled out by Planck, like $\lfi_4$, are among
the simplest ones since they require only one parameter, while ``the
plateau-like models require three or more parameters and must be
fine-tuned'', hence are much less simple. In this section, we analyze
this question.

Let us first notice that due to the non-detection of
non-Gaussianities, isocurvature perturbations or departures from
scale invariance, the minimal models of inflation relying on a single
scalar field in the slow-roll regime, remain in excellent agreement
with the data.

The main issue is then obviously in the definition of ``simplicity''
for a model.  Even if one accepts the naive definition in terms of the
number of free parameters, the Starobinsky model $\si$ is as simple
as, say $\lfi_4$, since both contain a single parameter and
require only $\phi \gtrsim 5 \Mp$ in order to have hundreds or more e-folds
of inflation.

A more objective meaning to the concept of simplicity can be given in
the Bayesian approach, which penalizes wasted parameter space and
rewards models that achieve a good compromise between quality of fit
and lack of fine-tuning\footnote{The number of unconstrained
  parameters can also be accounted for with the Bayesian
  complexity~\cite{Kunz:2006mc} or by counting effective $\chi^2$
  degrees of freedom~\cite{Lewis:2013hha}.}. In this sense,
plateau/hilltop inflation is no more complicated than large-field
models, as shown in \Ref{Martin:2013nzq}.

Let us also notice that parameters counting can be ambiguous.  For
instance, the same potential as in $\si$ can be obtained from the
Higgs field action non minimally coupled to gravity, the so-called
Higgs-inflation model $\hi$~\cite{Bezrukov:2007ep}. In $\hi$, one
might argue that more than one parameter is present: $\xi$, the
non-minimal coupling to gravity and $\lambda$ the self-interacting
Higgs coupling constant. They nonetheless combine into a single
quantity $M^4=\Mp^4\lambda /(4\xi^2)$ to produce the one-parameter
potential of \Eq{eq:sipot}. The situation is exactly the same for
$\lfi_4$: the potential is $V(\phi)=\lambda \phi^4$ but it can be
justified from more fundamental theories containing several parameters
that all combine to give $\lambda$. This is exactly what happens if
one constructs supergravity (SUGRA) models of $\lfi$ as discussed in
Sec.~4.2.1 of \Ref{Martin:2013tda}, see Eq.~(4.33). So a one-parameter
model in this context means a one-parameter model as far as CMB
predictions are concerned.

We conclude that, at this stage, the data are perfectly compatible
with the minimal and simplest implementations of inflation.

\section{Initial conditions beyond isotropy and 
homogeneity}
\label{sec:beyondFL}

The results discussed above assume that the universe is homogeneous
and isotropic. This is evidently not satisfactory since inflation is
precisely supposed to homogenize and isotropize the universe. This
issue is clearly crucial for inflation and is part of the general
problem of initial conditions. Technically, however, it is much more
complicated than the problem treated in \Sec{sec:beyondI}.

\subsection{Beyond isotropy}
\label{sec:beyondiso}

A first step toward a more complete investigation of this question is
to maintain homogeneity and relax isotropy only and see whether
inflation isotropizes the universe, see
\Refs{Steigman:1983hr,Anninos:1991ma,Turner:1986gj}. This strategy can be
exemplified by considering the Bianchi I metric which reads
\begin{align}
\label{eq:Bianchi1:metric}
\ud s^2=-\ud t^2+a_i^2(t)\left(\ud x^i\right)^2,
\end{align}
where each direction in space now has its own scale factor. The same
metric can also be expressed as
$\ud s^2=-\ud t^2 +a^2(t)\gamma_{ij}\ud x^i
\ud x^j$ with
\begin{align}
a(t)\equiv \left[a_1(t)a_2(t)a_3(t)\right]^{1/3}, 
\end{align}
and 
\begin{align}
\gamma_{ij}=
\begin{pmatrix}
e^{2\beta_1(t)} & 0 & 0 \\
0 & e^{2\beta_2(t)} & 0 \\
0 & 0 & e^{2\beta_3(t)} 
\end{pmatrix},
\end{align}
with $\sum_{i=1}^{i=3}\beta_i=0$. As before, we assume the matter
content of the early universe to be dominated by a scalar field
$\phi(t)$, with a potential $V(\phi)$.  Then, the Einstein equations
lead to
\begin{align}
\label{eq:friedani}
3\frac{\calH ^2}{a^2}&=
\frac{1}{\Mp^2}\left[\frac{\phi'^2}{2a^2}
+V(\phi)\right]+\frac{\sigma^2}{2a^2}, \\
\label{eq:pani}
-\frac{1}{a^2}\left(\calH ^2+2\calH '\right)&=
\frac{1}{\Mp^2}\left[\frac{\phi'^2}{2a^2}
-V(\phi)\right]+\frac{\sigma^2}{2a^2},\\
\label{eq:shear}
\left(\sigma^i_j\right)'+2\calH \sigma^i_j &=0,
\end{align}
where $\calH =a'/a$, a prime denoting a derivative with respect to
conformal time. In the above equation $\sigma_{ij}$ is the shear,
defined as
\begin{align}
\sigma_{ij}=\frac12 \gamma_{ij}'=
\begin{pmatrix}
\beta_1'e^{2\beta_1} & 0 & 0 \\
0 & \beta_2'e^{2\beta_2} & 0 \\
0 & 0 & \beta_3'e^{2\beta_3} 
\end{pmatrix},
\end{align}
and $\sigma ^2=\sigma_{ij}\sigma^{ij}=\sum_{i=1}^{i=3}\beta_i'^2$ with
$\sigma ^i_j=\gamma ^{ik}\sigma_{kj}$. An isotropic universe
corresponds to a vanishing shear. Indeed, if the $\beta_i$'s are
constant, one can always redefine the spatial coordinates $x_i$ such
that \Eq{eq:Bianchi1:metric} reduces to the FLRW metric.

In order to study the dynamics of the system, one has to solve the
Einstein equations~(\ref{eq:friedani}), (\ref{eq:pani})
and~(\ref{eq:shear}). This is especially easy for \Eq{eq:shear}, which
does not directly depend on $\phi(t)$. The corresponding solution is
given by $\sigma ^i_j=S^i_j/a^2$, where $S^i_j$ is a time-independent
tensor. As a consequence, one has $\sigma^2=S^2/a^4$ where
$S^2=S^i_jS^j_i$ is a constant.  This implies that the shear is, in
fact, equivalent to a stiff fluid with an equation-of-state parameter
$w_{\sigma}\equiv p_{\sigma}/\rho_{\sigma}=1$ and energy density
$\rho_{\sigma}=\Mp^2S^2/(2a^6)$.

Two situations must then be considered. If
$\rho_\phi \gg \rho_{\sigma}$ initially, then the universe inflates
and quickly isotropizes since $\rho_{\sigma}\propto e^{-6N}$. If, on
the contrary, the shear initially dominates,
$\rho_{\sigma} \gg \rho_\phi$, then the universe expands as
$a\propto t^{1/3}$ and the expansion is not accelerated. In that case,
initially the field is slowly rolling, $\rho_\phi$ is
approximately constant, and, since $\rho_{\sigma}\propto a^{-6}$,
after a transitory period inflation starts and isotropizes the
universe; or, the kinetic energy of the scalar field dominates over
its potential energy, both $\rho_\phi$ and $\rho_\sigma$ decay as
$\propto a^{-6}$, and once the potential energy becomes larger than the
kinetic energy, inflation starts and also isotropizes the universe.

We conclude that, generically, inflation makes the universe isotropic,
and the presence of initial shear is not a threat for inflation.

\subsection{Beyond homogeneity}
\label{sec:beyondhomo}

Despite the previous analysis, which is clearly a good point for
inflation, the most difficult question remains to be addressed, namely
whether inflation can homogenize the universe. Technically,
this is a complex problem since one must now consider a situation that
is initially inhomogeneous (and also anisotropic).

An analytical approach that has been used in the literature to
investigate this problem is the so-called ``effective-density
approximation'', which was studied in
\Refs{Goldwirth:1989pr,Goldwirth:1991rj} (for different methods and/or
arguments, see also
\Refs{Albrecht:1984qt,Albrecht:1985yf,Albrecht:1986pi,Brandenberger:2016uzh}
and \Refs{Vachaspati:1998dy,Berera:2000xz, Easther:2014zga}). The idea
is to consider an inhomogeneous scalar field on an isotropic and
homogeneous FLRW background, assuming that the backreaction of the
field inhomogeneities does not modify too much the FLRW metric, and
manifests itself only via a new term in the Friedmann-Lema\^itre
equation that simply changes the value of the Hubble
parameter. Concretely, one takes
\begin{align}
  \phi(t,\bmx)=\phi_0(t)+\Re \left[\delta \phi(t)
e^{i \bmk\cdot \bmx /a(t)}\right],
\end{align}
and assumes that the corresponding Klein-Gordon equation can be split
into two equations for the zero mode and for the inhomogeneous
mode. This leads to
\begin{align}
\label{eq:kginhomo1}
\ddot{\phi}_0+3H\dot{\phi}_0+V_{\phi}(\phi_0) &=0, \\
\label{eq:kginhomo2}
\ddot{\delta \phi}+3H\dot{\delta \phi}+\frac{k^2}{a^2}\delta \phi &=0.
\end{align}
Let us notice that, despite the notation, $\delta \phi(t)$ needs not
be small compared to $\phi_0(t)$. The crucial ingredient of this
approximation scheme is that, in \Eq{eq:kginhomo2}, the potential does
not appear. We therefore assume that the length scale of the
inhomogeneities is small enough for the potential energy to be
negligible compared to the gradients. As mentioned above, the
Friedmann-Lema\^itre equation is then expressed as
\begin{equation}
\label{eq:friedinhomo}
H^2=\frac{1}{3\Mp^2}\left[\frac12 \dot{\phi}_0^2+V(\phi_0)
+\frac12 \dot{\delta \phi}^2+\frac12 \frac{k^2}{a^2}\delta \phi^2\right]
-\frac{{\mathcal K}}{a^2}.
\end{equation}
This approximation should be valid if the wavenumber $\bmk$ is such
that the wavelength of the inhomogeneous part is much smaller than the
Hubble radius, namely $k \gg a H$, see
Refs.~\cite{Goldwirth:1989pr,Goldwirth:1991rj}. If, on the contrary,
it is much larger than the Hubble radius, then this should just amount
to a normalization of the homogeneous field in our local Hubble
volume. In this framework, the energy density of the inhomogeneities
is defined by
$\rho_{\delta \phi}=\rho_{\dot{\delta \phi}}+\rho_{\nabla}$, with
$\rho_{\dot{\delta \phi}}=\dot{\delta \phi}^2/2$ and
$\rho_{\nabla}=k^2\delta \phi^2/(2a^2)$ while the energy density of
the homogeneous mode is, as usual, given by
$\rho_{\phi_0}=\dot{\phi}_0^2/2+V(\phi_0)$. Then, the problem can be
reformulated in the following way: if, initially,
$\rho_{\delta \phi}\gg \rho_{\phi_0}$, namely if initially the
universe is strongly inhomogeneous, then can $\rho_{\delta \phi}$
decrease such that $\rho_{\phi_0}$ takes over and inflation starts,
thus making the universe homogeneous?

\begin{figure}
\begin{center}
\includegraphics[width=\figw]{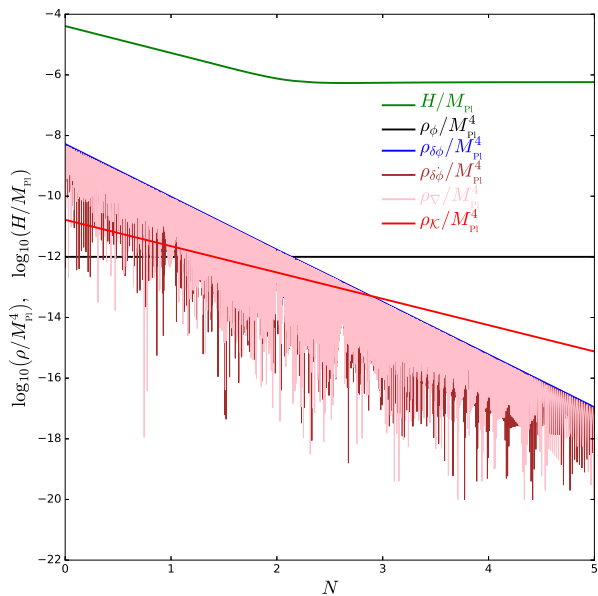}
\caption{Evolution of the Hubble parameter and of the various energy
  densities obtained by numerical integration of
  Eqs.~(\ref{eq:kginhomo1}),~(\ref{eq:kginhomo2})
  and~(\ref{eq:friedinhomo}). The potential is chosen to be the
  Starobinsky one, see \Eq{eq:sipot} with a scale $M=0.001\Mp$, which,
  roughly speaking, matches the CMB normalization. The initial value
  of the field is $\phi_0=6\Mp$ and the initial velocity is taken to
  be $\dot{\phi}_0=-V_{\phi}(\phi_0)/[3V(\phi_0)]$ (which is the
  slow-roll velocity). In the absence of inhomogeneities, with these
  initial conditions, inflation would start and would lead to more
  than $\simeq 100$ e-folds. The initial value of $\delta \phi$ is
  taken to be $0.01\Mp$ (and is therefore less than the Planck mass as
  required, see the main text) while the initial velocity of $\delta
  \phi(t)$ is given by $\dot{\delta \phi}_{\uini}=0$.  The scale $k$
  is chosen to be $k/a_{\uini}=10^{-2}\Mp$ and the initial curvature
  has been set to $\rho_{\calK} = -5 \times 10^{-11} \,\Mp^4$. This
  implies that $H_{\uini}/\Mp\simeq 4 \times 10^{-5}$, $\rho_{\phi,
    {\uini}}\simeq 10^{-12}\Mp^4$, and $\rho_{\delta \phi,
    {\uini}}\simeq 5 \times 10^{-9}\Mp^4$.  Those initial conditions
  are such that $H_{\uini}^2a_{\uini}^2/k^2\simeq 1.66\times
  10^{-5}\ll 1$ and $\rho_{\phi, {\uini}}/\rho_{\delta
    \phi,{\uini}}\simeq 2\times 10^{-4}$, namely the inhomogeneities
  largely dominate initially.}
\label{fig:energydensitystaro}
\end{center}
\end{figure}

Initially, we can choose $\dot{\phi}_0$ and $\phi_0$ such that, in
the absence of inhomogeneities, slow-roll inflation would start
(therefore, those values depend on the potential that we assume). If,
initially, the inhomogeneities dominate, then \Eq{eq:friedinhomo} can
be expressed as $H^2\simeq \rho_{\delta \phi}/(3\Mp^2)$, or
\begin{align}
\label{eq:Friedinihomo}
\frac{3a^2H^2}{k^2}\simeq \frac12 \frac{\dot{\delta \phi}^2}{\Mp^2}
\frac{a^2}{k^2}+\frac 12\frac{\delta \phi^2}{\Mp^2}.
\end{align}
One can also include a non-vanishing initial curvature, but as long as
$\calK$ is not large enough to make the universe collapse, its effects
quickly disappear (see the red line in
\Fig{fig:energydensitystaro}). However, we stress that, if the initial
curvature is much larger, this picture could be drastically
modified. As a matter of fact, we have checked that, if one increases
its contribution by one order of magnitude, the values of the other
parameters used in \Fig{fig:energydensitystaro} being otherwise the
same, then the universe recollapses. Therefore, if it is not necessary
to assume that curvature is initially tiny, it is nevertheless true
that it should be subdominant. This is the hypothesis that we make in
the following and this is the reason why it is ignored in the above
equation. Then, the condition that the wavelength of the
inhomogeneities is smaller than the Hubble radius implies that the
left-hand side of \Eq{eq:Friedinihomo} is small. This results in the
following initial conditions: $\delta \phi \ll \Mp$ and $\dot{\delta
  \phi}^2/\Mp^2\ll k^2/a^2$.

In \Fig{fig:energydensitystaro}, we have numerically integrated
\Eqs{eq:kginhomo1},~(\ref{eq:kginhomo2})
and~(\ref{eq:friedinhomo}). Initially, we see that
$\rho_{\delta \phi}\gg \rho_{\phi}$, namely the universe is strongly
inhomogeneous. Then, $\rho_{\dot{\delta \phi}}$ (brown line) and
$\rho_\nabla$ (pink line) and, therefore, $\rho_{\delta \phi}$,
decrease (blue line) as $a^{-4}$ while $\rho_{\phi}$ is
approximately constant. During this phase, the Hubble parameter is
not constant and we do not have inflation.

The fact that $\rho_{\delta \phi}$ behaves as radiation can be
understood from noticing that
\begin{align}
\label{eq:deltaphisol}
\delta \phi(t)\simeq 
\Re\left[\delta \phi_{\uini}\frac{a_\uini}{a(t)}
e^{i\frac{k}{a_\uini H_\uini}\frac{a(t)}{a_\uini}}\right],
\end{align}
provides a solution to \Eqs{eq:kginhomo2} and~(\ref{eq:Friedinihomo})
in that case.\footnote{This solution is valid at next-to-leading order
  in $Ha/k$, and generalizes the formula found in
  \Ref{Goldwirth:1991rj}, see Eq.~(7.10) of that reference, which is
  valid at leading order in $Ha/k$ only. At that order unfortunately,
  one cannot derive the overall scaling in $a$ in \Eq{eq:deltaphisol},
  which, however, determines the damping rate of inhomogeneities. This
  is why one needs to work at next-to-leading order.} This implies
that, very quickly, $\rho_{\phi}$ takes over, the universe becomes
homogeneous, the Hubble parameter settles to a constant and inflation
starts. In this regime, \Eqs{eq:kginhomo2} still possesses an
analytical solution, given by
\begin{align}
\delta \phi(t)\simeq 
\Re\left[\delta \phi_{\uini}\frac{a_\uini}{a(t)}
e^{i\frac{k}{a_\uini H_\uini}\frac{a_\uini}{a(t)}}\right],
\end{align}
at leading order in $aH/k$. At that order, this implies that
$\rho_{\delta \phi}$ still decays as $1/a^4$ during inflation, as can
be checked in \Fig{fig:energydensitystaro}. This is the case until $k$
crosses out the Hubble radius, at which point the entire
effective-density approximation scheme breaks down.

We have reproduced the above analysis for the Coleman-Weinberg
potential, see \Eq{eq:potcw} for $Q=10^{-3}\Mp$ and found the same
qualitative behavior (the same result was also found for LFI models),
which seems to indicate that it is independent of the potential
chosen, in agreement with
Refs.~\cite{Goldwirth:1989pr,Goldwirth:1991rj}.

We conclude that, in the previous setting, the presence of large
inhomogeneities cannot prevent inflation. However, as also stressed in
Refs.~\cite{Goldwirth:1989pr,Goldwirth:1991rj}, this conclusion is
obtained assuming the size of the inhomogeneities to be much smaller than the
Hubble radius. This means that the previous results are, in fact,
limited and have a small impact on our general understanding of how
the universe becomes homogeneous during inflation. As a matter of
fact, in the most general situation (in particular if the size of the
inhomogeneous mode is of the order of the Hubble radius), only a
numerical integration of the full Einstein equations can provide a
correct answer.

This has been carried out by several authors. The first numerical
solutions~\cite{Goldwirth:1989pr,Goldwirth:1989vz,
  Goldwirth:1991rj,Goldwirth:1990pm} were obtained under the
assumption that spacetime is spherically symmetric. This simplifies
the calculations since then the problem only depends on time and on
one radial coordinate. Nevertheless, the Einstein equations remains
partial (as opposed to ordinary as in the situation treated before)
non-linear differential equations.  This analysis was improved in
\Refs{KurkiSuonio:1987pq,Laguna:1991zs,KurkiSuonio:1993fg} in which
the spherical symmetry assumption was relaxed. More recently,
\Refs{East:2015ggf,Clough:2016ymm, Bloomfield:2019rbs} have run new
simulations (and seem to confirm the validity of the behavior
$\rho_{\delta\phi}\propto a^{-4}$ found above, even when $k\sim a H$).
All these works have technical restrictions and, at this stage, it is
difficult to draw a completely general conclusion. However, it seems
that LFI and plateau models work better than SFI, and that although
the size of the initial homogeneous patch is an important parameter of
the problem, strong gradients may also help in starting inflation (see
also \Refs{Calzetta:1992gv,Perez:2012pn}).

It is also worth mentioning that, speculating on what Quantum Gravity
could be, it does not seem unreasonable to assume that a patch of Planck
size should be homogeneous. If this patch can be stretched to the
observable universe today, then the homogeneity problem would be
solved. However, without inflation, this is not possible. Indeed, in
the hot big bang model, the Planck energy density is reached at
redshift $z_{\uPl}\simeq 10^{30}$. The Planck length, $\ell_{\uPl}
\simeq 10^{-35}\, \meter$, at this initial redshift, is thus stretched
to $\simeq 10^{-5} \,\meter$ today, to be compared to the Hubble
radius today, $r_\uH \simeq 10^{26}\, \meter$. This is a generic
feature of decelerated expansion, which redshifts length scales by an
amount always smaller than the increase of the Hubble radius. But, on
the contrary, with a sufficient number of inflationary e-folds, the
initial Planck patch becomes larger than the Hubble patch today and,
within the above-mentioned hypothesis, the homogeneity problem is
solved. Notice that this does not address, within Quantum Gravity, the
problem of starting inflation, which is a topic by
itself~\cite{Linde:1983cm,Linde:1985ub, Coule:1999wg}.

We conclude that the fundamental issue as to whether inflation
homogenizes the universe is still open, although most recent numerical
works on this topic seem to be suggesting it does. Among all the
potential problems that have been raised against inflation, it is
clearly the most serious one.

\section{The Trans-Planckian Problem}
\label{sec:tpl}

In the previous sections, we studied how inflation depends on initial
conditions. A similar question exists for the perturbations. In fact,
if one traces backwards in time the length scales of cosmological
interest today, they are generically smaller than the Planck length at
the onset of inflation. It is in this regime that the initial
conditions (the adiabatic vacuum) are chosen. But one can wonder
whether this is legitimate and whether quantum field theory in curved
spacetime is valid in this case. Notice that the energy density of
the background remains much less than the Planck energy density, so
that the use of a classical background is well justified, and it is
only the wavelengths of the perturbations that can be smaller than the
Planck length. This issue is known as the trans-Planckian problem of
inflation~\cite{Martin:2000xs, Brandenberger:2000wr,
  Brandenberger:2012aj, Brandenberger:2004kx}.

In absence of a final theory of quantum gravity, it is difficult to
calculate what would be the modifications to the behavior of the
perturbations if physics beyond the Planck scale were taken into
account. But what can be done is to introduce several ad hoc but
reasonable modifications and test whether the inflationary predictions
are robust~\cite{Martin:2000xs, Brandenberger:2000wr} under those.

It has been shown that the inflationary power spectrum can be modified
if physics is not adiabatic beyond the Planck scale. So, to a certain
extent, the predictions are not robust. However, if one uses the most
conservative way of modeling the modifications originating from the
spacetime foam, one finds that the corrections scale as
$(H/M_\uc)^p$, where $H$ is the Hubble scale during inflation and
$M_\uc$ the energy scale at which new physical effects pop up
(typically the Planck scale or, possibly, the string scale); $p$ is an
index which, in some cases, can simply be
one~\cite{Martin:2003kp}. Those corrections are therefore typically
small. In some sense, this result can be viewed as decoupling between
the Planck and inflationary scales. However, there are other ways of
modeling the new physics (for instance, for some choices of modified
dispersion relations~\cite{Martin:2002kt}) that could lead to more
drastic modifications.

Notice that the form of these corrections is somehow generic. Choosing
the adiabatic vacuum consists in singling out a specific
Wentzel-Kramers-Brillouin (WKB) branch in
the evolution of the cosmological perturbations, which is a second
order differential equation. Any deviation from this will necessarily
introduce interferences with the other branch and, as a result,
superimposed oscillations will appear in the inflationary correlation
functions~\cite{Brandenberger:2002hs}. Although the amplitude and the
frequency of those oscillations are model dependent, their presence
have been searched for in the CMB data but no conclusive signal has been
found so far~\cite{Martin:2003sg, Martin:2004yi, Martin:2004iv,
  Martin:2006rs}.

In conclusion, it is fair to say that trans-Planckian effects are, at
least in their most conservative formulations, not a threat for
inflation. On the contrary, they should be viewed as a window of
opportunity~\cite{Martin:2000bv,Easther:2001fi}: if we are lucky
enough, one might use them to probe the Planck scale, something that
would clearly be impossible with other means~\cite{Martin:2000xs,
  Brandenberger:2000wr, Easther:2002xe, Armendariz-Picon:2003gd,
  Brandenberger:2004kx, Easther:2004vq, Brandenberger:2012aj}.

\section{Inflation and the Quantum Measurement Problem}
\label{sec:qminflation}

The inflationary mechanism for structure formation is based on General
Relativity and Quantum Mechanics. As a consequence, the behavior of
inflationary perturbations is described by the Schr\"odinger equation
that controls the evolution of their wavefunction. Initially, the
system is placed in its ground state, which is a coherent state, and
then, due to the expansion of spacetime, it evolves into a very
peculiar state, namely a two-mode squeezed state. This state is
sometimes described as ``classical'' since most of the corresponding
quantum correlation functions can be obtained using a classical
distribution in phase-space~\cite{Polarski:1995jg, Kiefer:1998qe,
  Martin:2015qta}. However, it also possesses properties usually
considered as highly non classical. It is indeed an entangled state,
very similar to the Einstein-Podolsky-Rosen (EPR) state, with a large
quantum discord~\cite{Martin:2015qta}, which allows one to construct
observables for which the Bell inequality is
violated~\cite{Martin:2016tbd,Martin:2017zxs}.

The above picture, however, raises an issue~\cite{Sudarsky:2009za,
  Goldstein:2015mha}: the quantum state of the perturbations is not an
eigenstate of the temperature fluctuation operator. A non-unitary
process needs therefore to be invoked, during which the state evolves
from the two-mode squeezed state into an eigenstate of
$\widehat{\delta T}/T$.  In other words, the quantum state of the
perturbations is homogeneous and something is needed to project it
onto a state that contains inhomogeneities.

This problem is no more than the celebrated measurement problem of
Quantum Mechanics, which, in the Copenhagen approach, is ``solved'' by
the collapse of the wavefunction. In the context of Cosmology,
however, the use of the Copenhagen interpretation appears to be
problematic~\cite{Hartle:2019hae}. Indeed, it requires the existence
of a classical domain, exterior to the system, which performs a
measurement on it.  In quantum cosmology for instance, one calculates
the wavefunction of the entire Universe and there is, by definition,
no classical exterior domain at all. In the context of inflation, one
could argue that the perturbations do not represent all degrees of
freedom and that some other classical degrees of freedom could
constitute the exterior domain, but they do not qualify as
``observers'' in the Copenhagen sense. The transition to an eigenstate
of the temperature fluctuation operator, which necessarily occurred in
the early Universe (structure formation started in the early
Universe), thus proceeded in the absence of any observer, something at
odds with the Copenhagen interpretation.

How is this problem usually addressed? One possibility is to resort to
the many-world interpretation together with
decoherence~\cite{Kiefer:2006je}. It can also be understood if one
uses alternatives to the Copenhagen interpretation such as ``collapse
models''; see \Refs{Perez:2005gh, DeUnanue:2008fw, Leon:2010fi,
  Martin:2012pea, Canate:2012ua, Das:2013qwa}. In this case, one
obtains different predictions that can be confronted with CMB
measurements. Other solutions involve the Bohmian interpretation of
Quantum Mechanics~\cite{Peter:2006hx, PintoNeto:2011ui,
  Peter:2016kan}.

In conclusion, let us stress that the quantum measurement problem is
present in quantum mechanics itself and is not specific to inflation:
any mechanism where cosmological structures originate from quantum
fluctuations would have to face it. As for the trans-Planckian
problem, inflation can, however, be viewed as a window of opportunity
that could shed light on fundamental issues of Quantum Mechanics using
astrophysical measurements.

\section{The Likelihood of Inflation}
\label{sec:probainf}

Another class of criticisms against inflation is based on the idea that
there exists a natural measure on the space of classical universes and
that, according to this measure, the probability of having a
sufficient number of e-folds of slow-roll inflation is tiny. In this
section, we examine these arguments.

The main idea is the following. Let us consider a system having $n$
degrees of freedom $q_i$ and described by the Hamiltonian
$\Ham(q_i,p_i)$ where $p_i=-\partial \Ham/\partial q_i$ is the
conjugate momentum of $q_i$. The evolution of the system can be
followed in the $2n$-dimensional phase-space endowed with the
coordinates $(q_i,p_i)$. Then, there exists a natural symplectic form
given by $\omega=\sum_{i=1}^n\dd p_i\wedge \dd q_i$ which leads to the
Liouville measure, namely $\Omega=(-1)^{n(n-1)/2}/n!\, \omega^n$. This
measure is commonly and successfully used in statistical physics.

In the context of inflation, where gravity is relevant, one can also
describe the system in terms of a Hamiltonian and, therefore, attempt
to define a natural measure. Indeed, the Einstein-Hilbert action, in
the homogeneous case, leads to the mini superspace Lagrangian
\begin{equation}
\mathcal{L}=-3\frac{\Mp^2}{\calN}a\dot{a}^2+3\calN \Mp^2\calK a
+\frac{a^3\dot{\phi}^2}{2
  \calN}-\calN a^3 V(\phi).
\label{eq:Leff}
\end{equation}
Here, $\calN$ is the lapse and plays the role of a Lagrange
multiplier. The conjugate momenta read
\begin{equation}
\label{eq:momenta}
  p_{\calN}=0, \qquad p_a=-6 \Mp^2 \dfrac{a \dot{a}}{\calN}, \qquad
  p_\phi = a^3\dfrac{\Pi}{\calN}\,.
\end{equation}
Performing a Legendre transform, one obtains the Hamiltonian
\begin{equation}
\label{eq:Hamiltonian:minisuperspace}
\Ham = \calN\left[-\frac{p_a^2}{12\Mp^2 a}+\frac{p_\phi^2}{2 a^3}
-3\Mp^2 \calK a+a^3
  V\left(\phi\right)\right].
\end{equation}
The equation of motion for $\calN$ sets it to be a constant, and the
variation of the Lagrangian with respect to $\calN$ gives the
Friedmann-Lema\^itre equation~(\ref{eq:hubble}) provided $\calN=1$,
which we choose to be the case in what follows. The Hamiltonian
equation of motion for $\phi$ is nothing but the Klein-Gordon
equation~(\ref{eq:kg}), while the Hamiltonian equation of motion for
$a$, combined with the Friedmann-Lema\^itre equation, gives the
Raychaudhuri equation~(\ref{eq:hubbledot}). In practice, the
Friedmann-Lema\^itre equation can be deduced from the Klein-Gordon and
Raychaudhuri equations up to an integration constant (which
corresponds to fixing $\calN$).

As a result, the dynamics is effectively Hamiltonian on a
four-dimensional phase-space $( \phi, p_\phi , a, p_a )$ and
this can motivate the choice of the simplectic form
\begin{align}
\label{eq:ghs}
\omega_{_\uGHS}=\dd p_a\wedge \dd a+\dd p_\phi \wedge \dd\phi.
\end{align}
The associated measure is known as the Gibbons-Hawking-Stewart
(GHS) measure~\cite{Gibbons:1986xk}.

At this point however, a first difficulty arises. Since General
Relativity is a constrained setup, the physical system lives in fact
on the surface $\Ham=0$ and not in the entire four-dimensional phase
space. One can nevertheless consider the measure induced by the GHS
form when pullbacked on this surface, which reads
\begin{equation}
\begin{aligned}
\omega_{\uGHS,\Ham=0}& =-6\Mp^2a^2 \dd H\wedge \dd a 
\\ &
+
\frac{6\Mp^2a^3H}{\sqrt{6H^2\Mp^2-2V(\phi)+ \dfrac{6\Mp^2\calK}{a^2}}}
\dd H\wedge \dd \phi 
\\ &
+
\frac{3a^2\left(6H^2\Mp^2-2V(\phi)+ \dfrac{4 \Mp^2\calK}{a^2}\right)}
{\sqrt{6H^2\Mp^2-2V(\phi)
+\dfrac{6\Mp^2\calK}{a^2}}}
\dd a\wedge \dd \phi.
\end{aligned}
\end{equation}

As detailed in \Ref{Schiffrin:2012zf}, this measure is still not
satisfactory because it is degenerate, namely the determinant of
$\omega_{\uGHS,\Ham=0}$ vanishes, and, as a consequence, it cannot
lead to a correct volume form. In order to fix this second problem,
the usual procedure consists in restricting the natural measure, not
to the constraint surface itself, but to a surface within the
constraint surface, $\Surf(N,H,\phi)=K_*$, which intersects each
trajectory only once. By considering different values of $K_*$, one
describes a succession of surfaces in the constraint surface, which
can be viewed as describing the ``passage of time''. The standard
choice is to define $\Surf$ by $H=H_*$, which leads to the form
introduced by Gibbons and Turok in \Ref{Gibbons:2006pa}, namely
\begin{equation}
\begin{aligned}
\label{eq:gt}
\omega_{\uGHS,\Ham=0,H=H_*} &=
3a^2 \\ &  \kern-1em  \times
\dfrac{6H_*^2\Mp^2-2V(\phi)+\dfrac{4 \Mp^2\calK}{a^2}}{\sqrt{6H_*^2\Mp^2
-2V(\phi)+\dfrac{6\Mp^2\calK}{a^2}}} \dd a\wedge \dd \phi.
\end{aligned}
\end{equation}

This measure, however, suffers from a number of drawbacks as underlined
in \Ref{Schiffrin:2012zf}. As discussed in great detail in Sec.~III of
this reference, in statistical physics, the use of the natural
symplectic form is justified only if some conditions are
satisfied. Reference~\cite{Schiffrin:2012zf} shows that, in the context of
inflation, none of these conditions is actually fulfilled. The
physical motivations behind the GHS measure seem therefore elusive.
Moreover, the measure~\eqref{eq:gt} is, in fact, infinite and needs to
be regularized. As discussed in \Ref{Schiffrin:2012zf}, the physical
conclusions that one may draw unfortunately depend on the
regularization scheme, and no universal scheme has been found so
far. A common procedure is to introduce a cutoff $a_\uc$ in $a$ and to
take the limit $a_\uc \to \infty$. Then, the probability of having
slow-roll inflation lasting at least $\sixty$ e-folds is given by
\begin{align}
P(\sixty)= \dfrac{\displaystyle\int _{\calD(\phipini,\sixty)}
\sqrt{6H_*^2\Mp^2-V(\phi)}\, \dd \phi}
{\displaystyle \int 
\sqrt{6H_*^2\Mp^2-V(\phi)}\, \dd \phi },
\end{align}
in agreement with Eq.~(62) of \Ref{Schiffrin:2012zf}. In the
numerator, the $\phi$ integral is taken for initial values of the
inflaton leading to at least $\sixty$ e-folds and for values of $\phi$
such that the square root is defined, the corresponding integration
domain being denoted by $\calD(\phipini,\sixty)$. In the denominator,
the range of integration comprises all values such that the square
root is defined.  Finally, we have also assumed flat spacelike
sections, namely $\calK=0$.

Then comes the question of which value of $H_*$ should be taken. The
choice made in \Ref{Gibbons:2006pa} is to choose $H_*$ as the Hubble
parameter at the end of inflation from which one obtains [see also
  \Ref{Schiffrin:2012zf}, Eq.~(69)]
\begin{align}
\label{eq:unlikeinf}
P(\sixty)\propto e^{-3\sixty}.
\end{align}
This result is at the origin of the claim that inflation is extremely
unlikely. It is, however, tightly related to the choice of taking $H_*$
at the end of inflation, see \Refs{Schiffrin:2012zf,Corichi:2013kua}:
since slow roll is an attractor in phase-space, it is clear that,
measured at the end of inflation, the volume occupied by the
inflationary trajectories is very small.

The presence of a dynamical attractor is, in fact, a positive feature of
inflation, as revealed by the result obtained when taking $H_*$
``initially'', say, at the Planck scale (or at scales slightly below if
one wants to avoid quantum gravity effects). This choice seems better
justified and leads to a completely different result. If one takes
$V(\phi)=m^2\phi^2/2$, then \Refs{Carroll:2010aj,Schiffrin:2012zf}
have shown that
\begin{align}
P(\sixty)\simeq 1-\frac{4m}{\pi H_*}\sqrt{\dfrac{2\sixty}{3}},
\end{align}
which leads to $P(60)\simeq 0.99996$, namely sufficient inflation is
almost certain. This number, however, depends on the potential. For
instance, it has also been evaluated for Natural Inflation in
\Ref{Carroll:2010aj} which finds $P(60)\simeq 0.171$. Given the Planck
CMB data, what should be done is to carry out this calculation for the
best inflationary scenarios, namely the plateau single-field
potentials such as the Starobinsky model. Using \Eq{eq:sipot} together
with \Eq{eq:si:SR:traj} expanded in the large-$\Phi$ limit, one
arrives at
\begin{align}
\label{eq:psixtysi}
P(\sixty)=\frac{\displaystyle
\int_{\ln(4\sixty/3)}^{\infty}\dd x \sqrt{1-\adim \left(1-e^{-x}\right)^2}}
{\displaystyle \int_{0}^{\infty}\dd x \sqrt{1-\adim \left(1-e^{-x}\right)^2}},
\end{align}
where $\adim$ is a dimensionless quantity given by
\begin{align}
\adim \equiv \frac{M^4}{6H_*^2\Mp^2}.
\end{align}
The upper limit of the integral is infinite because, in a plateau
model, $6H_*^2\Mp^2-V(\phi)$ is, in the slow-roll approximation,
always positive regardless of $\phi$. The two integrals present in the
ratio~(\ref{eq:psixtysi}) are infinite, but they can be regularized by
replacing the upper infinite limit with a finite cutoff
$x_\mathrm{sup}$, upon which they behave as
$\sim \sqrt{1-\adim}\, x_\mathrm{sup}$. Letting
$x_\mathrm{sup}\rightarrow \infty$, one then simply obtains
$P(\sixty)=1$. Therefore, if one were ready to accept the previous
considerations (with all the caveats mentioned), the conclusion would
be that Planck has precisely singled out the models for which the
probability of getting enough e-folds of inflation is unity, and that
would rather reinforce the status of inflation. Let us also notice
that \Ref{Ijjas:2015hcc} has remarked that the use of the Liouville
measure in the cosmological context is often disputed but has argued
that any other approaches lead the same conclusions. The previous
considerations, however, show that it is sufficient to take a
reasonable value for $H_*$ to completely change the consequences.

Let us quickly review the results obtained in this section. The claim
that inflation is unlikely, which is expressed mathematically by
\Eq{eq:unlikeinf}, is based on the use of the GHS measure which,
contrary to the case of statistical physics, does not satisfy any of
the conditions required for the validity of the Liouville
measure~\cite{Schiffrin:2012zf}. This measure therefore lacks physical
justifications, as other proposed measures in the
literature~\cite{Finn:2018krt}. If one were to use it anyway, the
result would strongly depend on the regularization scheme. If one
further insists and employs one such scheme where a cutoff on the
scale factor is introduced, if one chooses $H_*$ at the beginning of
inflation rather than at the end, one finds that sufficient inflation
is very likely (and even certain for the plateau models that seem to
emerge from the data) rather than unlikely. This is due to the
presence of a dynamical attractor, namely slow roll, and leads us to
conclude that the claim that inflation is unlikely can safely be
discarded.

\section{Initial Conditions, Attractors and Measures}
\label{sec:measure}

The discussion in \Sec{sec:beyondI} about fine-tuning in the space of
initial conditions is crucially based on the existence of an
attractor, which appears to be a fundamental property of
inflation. However, as was emphasized in \Ref{Remmen:2013eja}, the
presence of an attractor during inflation can be challenged. First,
there is the Liouville theorem, which states that volumes are
conserved in phase-space and goes against the intuitive meaning of
what an attractor is. Second, an attractor is a coordinate-dependent
notion.

Concerning this second point, it was shown in \Sec{sec:beyondI} that
there is a convergence of inflationary trajectories toward the
slow-roll trajectory when plotted in the coordinates $(\phi,\Pi)$ or
$(\phip,\Gamma)$. However, as noticed in \Ref{Remmen:2013eja}, if the
same trajectories are represented in terms of the coordinates
$(\phi,p_\phi)$, the attractor behavior is lost (see Fig.~2 of
\Ref{Remmen:2013eja}). The reason is that $p_\phi$ is related to
$\dot{\phi}$ by a time-dependent function, namely $a^3$ as can be seen
in \Eqs{eq:momenta}, and that, as we will see, the rate at which
trajectories approach the slow-roll attractor is precisely given by
$1/a^3$.  In fact, the attractor behavior of any system (not
necessarily in Cosmology) obtained with given coordinates can always
be erased by changing these coordinates, in particular by suitably
multiplying them by some time-dependent function. The presence of an
attractor, and, hence, the sensitivity to initial conditions, is
therefore intimately related to the choice of a measure in phase
space.

In fact, among the measures that can be proposed, two categories can
be distinguished. The first category corresponds to measures that can
be written as $f(x,y)\dd x\wedge \dd y$, where $x$ and $y$
parametrize phase-space and obey $\dd x/\dd \tau =X(x,y)$ and
$\dd y/\dd \tau=Y(x,y)$ (here $\tau$ denotes the time label used to
formulate the equations of motion), namely the equation of motion are
autonomous, that is to say they do not explicitly depend on
$\tau$. The second category contains measures that are not of type I,
either because $f$ explicitly depends on the time label $\tau$, or
because the equations of motion are not autonomous, or both. Measures
of category II are a priori not well-defined unless, using some additional
prescription, one can reexpress the time label in terms
of the dynamical variables of the system.

In this section, we discuss alternatives to the flat measure in
$(\phip,\Gamma)$ implicitly assumed in \Sec{sec:beyondI}, their
motivations, and their relevance for characterizing dynamical
attractors and the fine-tuning of initial conditions. We will specify
the category (I or II) of each measure and show that the presence of
an attractor is always found with measures of category I.

\subsection{Alternative measures}

Let us introduce several measures that can be viewed as
``natural''. Our goal is by no means to be exhaustive (one probably
could define other measures) or to argue that one measure is better
than the others. The aim is rather to show that the notion of an
inflationary attractor is robust in the sense that, given reasonable
measures, and for various attractor criteria, it is almost always
present.

\subsubsection{Field-space $t$-measure}
\label{sec:tm}

Since \Eqs{eq:hubble} and~(\ref{eq:kg}) give rise to a closed,
time-independent differential system in the space $( \phi,\Pi)$, where
we recall that $\Pi=\dot{\phi}$, a first natural choice for a measure
is one that is flat in this space, $\dd\phi\wedge\dd\Pi$. Because this
comes from parametrizing trajectories with cosmic time $t$ (in other
words the time label $\tau$ is $\tau=t$), we will refer to this as the
$t$-measure. This measure clearly belongs to category I because the
function $f$ does not depend on time (it is one) and the equations of
motion are, as just noticed above, autonomous since $\dot{\phi}=\Pi$
and
$\dot{\Pi}=-3[\Pi^2+2V(\phi)]^{1/2}\Pi/(\sqrt{6}\Mp)-V_{\phi}$. This
has to be contrasted with the case where we choose to work with
another time coordinate, say conformal time $\tau=\eta$, and consider
the $\eta$-measure $\dd\phi\wedge\dd\tilde{\Pi}$ that is flat if phase
space is parametrized by $\phi$ and $\tilde{\Pi}=\phi'=a\Pi$. This
measure is of category II because, while the function
$f(\phi,\tilde{\Pi})$ is still one, the scale factor explicitly
appears in the differential system given by \Eqs{eq:hubble}
and~(\ref{eq:kg}), which is not time independent anymore:
$\phi'=\tilde{\Pi}$,
$\tilde{\Pi}'=-2[\phi'^2+2a^2V(\phi)]^{1/2}/(\sqrt{6}\Mp^2)-a^2V_\phi$. Another
way to see it is to work in the space $( \phi,\Pi)$ with the measure
$\dd\phi\wedge\dd\tilde{\Pi}=a\dd\phi\wedge\dd\Pi$. As mentioned
above, the equations for $\phi$ and $\Pi$ are autonomous but now the
function $f(\phi,\Pi)=a$ becomes explicitly time dependent. Since $a$
is not a function of $\phi$ and $\Pi$ it does not define a proper
form. A workaround is to remark that $a$ can be integrated along a
given phase-space trajectory. As such, a function $a(\phi,\Pi)$ can be
defined along each trajectory, but the relative values of the scale
factor between different trajectories is ambiguous, unless one defines
an initial time slice in phase space where, say, one imposes
$a=1$. This illustrates the statement made above, namely that measures
of category II need an additional prescription to be fully defined,
and further shows that not all choices of time labels lead to measures
of category I.

\subsubsection{Hamiltonian measure}
\label{sec:hm}

The $t$-measure may not seem very natural since the coordinates $\phi$
and $\Pi$ are not canonical variables, in the sense that they satisfy
equations of motion that do not derive from a Hamiltonian. However,
the dynamics of the full four-dimensional phase-space, made of $\phi$,
$\Pi$, the scale factor $a$ and its conjugate momentum $p_a$, is
Hamiltonian since it can be obtained from the Einstein-Hilbert action,
as discussed in \Sec{sec:measure}, where it is shown to lead to GHS
measure~\cite{Gibbons:1986xk} (or its pulled-back version). This
measure is of category I but nonetheless suffers from the drawbacks
highlighted in \Sec{sec:measure}.

\subsubsection{Hamiltonian induced measure}
\label{sec:him}

Even though $\phi$ and $\Pi$ are not canonical variables, the
projection of the dynamics from the four-dimensional space
$( \phi,p_\phi,a,p_a) $ onto $( \phi,\Pi)$ has the remarkable property
of being well defined (trajectories do not cross) and second order. In
this sense one can perfectly consider the dynamics in the plane
$( \phi,\Pi)$ endowed with the measure induced from the one in the
four-dimensional ``canonical'' space, that is to say the form
$\dd\phi\wedge\dd p_\phi = a^3 \dd\phi\wedge \dd\Pi$ that is flat in
the space $( \phi,p_\phi)$. However, $f(\phi,\Pi)=a^3$ and, as a
consequence, the measure belongs to category II.

In what follows, we will refer to this choice as the Hamiltonian
induced metric. Let us notice that it suffers from the same flaw as
the $\eta$-measure, see \Sec{sec:tm}: one needs to specify an initial
time slicing in the field phase-space.

Let us also note that even though the dynamics in the two-dimensional
space $( \phi,\dot{\phi} )$ is well defined and second order, there is
no guarantee a priori that it can be obtained from a Hamiltonian. If
the scalar field is a test field, it is the case and the Hamiltonian
is simply given by \Eq{eq:Hamiltonian:minisuperspace} where $a(t)$ is
a fixed, {\ie},non-dynamical, function. If $\phi$ dominates the energy
budget of the universe and has a quadratic potential, in
\Ref{Remmen:2013eja} it is shown that this is also the case, by virtue
of Douglas' theorem~\cite{Douglas:1939}.  Otherwise the question
remains open.

\subsubsection{Field-space $N$-measure}
 
The Hamiltonian measure of \Sec{sec:hm} may at first seem more natural
as stemming directly from the Einstein-Hilbert action. However, one
may question the relevance of promoting the scale factor $a$ into a
dynamical variable as important as $\phi$, since $a$, alone, is a
non-measurable quantity. If the spacelike sections are flat, only
scale factor ratios are relevant for astrophysics and cosmology, such
as in redshift measurements, or as in the Hubble parameter.

The Einstein-Hilbert action in the FLRW metric is obtained by
integrating \Eq{eq:Leff} with $\calN=1$, {\ie},
\begin{equation}
S = \int a^3 \left[-3 \Mp^2 \dfrac{\dot{a}^2}{a^2} + \dfrac{1}{2}
  \dot{\phi}^2 - V(\phi) \right] \ud t .
\end{equation}
This expression makes clear that, up to the $a^3$ term, $a$ and its
derivative $\dot{a}$ only appear within the ratio precisely given by
the Hubble parameter $H = \dot{a}/{a}$. This remark suggests to change
the time coordinate from $t$ to $N\equiv \ln a$ and one gets
\begin{equation}
S = \int e^{3 N}\left[ - 3 \Mp^2 H + \dfrac{H}{2} \Mp^2 \Gamma^2 -
  \dfrac{V(\phip)}{H} \right] \ud N,
\label{eq:SN}
\end{equation}
where $\Gamma=\ud \phip/\ud N$ and $\phip \equiv \phi/\Mp$, as in
\Sec{sec:beyondI}. Therefore, the Lagrangian reads
\begin{equation}
\calL\left(N,H,\Gamma,\phip\right) \equiv e^{3N} \left[ - 3 \Mp^2 H +
  \dfrac{H}{2} \Mp^2 \Gamma^2 - \dfrac{V(\phip)}{H} \right].
\label{eq:LN}
\end{equation}
This Lagrangian is an explicit function of the time integration
variable, $N$, and describes a non-conservative dynamical system. The
Hubble parameter in this Lagrangian has no dynamics and acts as an
auxiliary field. The only dynamical degree of freedom is $\phip$ (and
its derivative $\Gamma$) such that phase-space is two-dimensional.

Let us check that \Eq{eq:SN} gives back the Friedmann-Lema\^itre and
Klein-Gordon equations for a self-gravitating scalar field. The
Euler-Lagrange equation with respect to $\Phi$ gives
\begin{equation}
\dfrac{\ud \Gamma}{\ud N} + \left(3 + \dfrac{1}{H} \dfrac{\ud H}{\ud
  N} \right) \Gamma + \dfrac{1}{H^2 \Mp^2} \dfrac{\ud V}{\ud \phip} = 0\,,
\label{eq:Sdphi}
\end{equation}
while $\delta S/\delta H=0$ yields
\begin{equation}
H^2 = \dfrac{2}{\Mp^2} \dfrac{V}{6 - \Gamma^2}\,,
\label{eq:SdH}
\end{equation}
which matches the first of \Eq{eq:flefold}. Taking the logarithm of
\Eq{eq:SdH} and differentiating with respect to $N$ gives a first
order differential equation for $\Gamma$ which, combined with
\Eq{eq:Sdphi}, gives back the first Hubble flow function
\begin{equation}
-\dfrac{1}{H} \dfrac{\ud H}{\ud N} = \dfrac{1}{2} \Gamma^2,
\end{equation}
which matches the second of \Eq{eq:flefold}. 

Starting from the Einstein-Hilbert action, and having the prejudice of
considering only observable quantities in the dynamical system, we
reach the conclusion that the Hubble parameter is an auxiliary field
while the dynamics is dissipative and two-dimensional in the phase
space $( \phip,\Gamma)$. This is in contrast with the
Hamiltonian measure where, starting from the same action, the choice
of the dynamical variables was made to have a conserved Hamiltonian at
the expense of having a four-dimensional phase-space.

Recalling that $\Gamma=\dd\Phi/\dd N$, we refer to the measure induced
by $\ud \phip \wedge \ud \Gamma$ as the field-space $N$-measure. This
measure belongs to category I since $f(\Phi,\Gamma)=1$ and the
corresponding equations of motion are autonomous:
$\dd \Phi/\dd N=\Gamma$,
$\dd \Gamma/\dd N=-(3-\Gamma^2/2)\Gamma-(6-\Gamma^2)V_\phi/(2V)$.

\subsection{Attractors as volume shrinkers}

Dynamical attractors play an important role in cosmology since they
have the ability to erase the dependence on initial conditions from
the predictions of a given model. The idea is that, if one starts from
a set of initial points in phase-space, enclosed within a certain domain
of phase-space volume $\vol$, the trajectories stemming from these
points will all merge toward the attractor trajectory. One may expect
that the volume of the region they encompass thus goes to zero as time
proceeds, since the volume of a one-dimensional line in a
more-than-two-dimensional space vanishes. For this reason, a first
definition of a ``dynamical attractor'' one may propose is a
trajectory around which phase-space volume decreases, $\ud \vol<0$.

\subsubsection{Field-space $t$-measure}
\label{sec:VolumeShrinker:dphi_wedge_ddotphi}

Let us consider two vectors $\bmu_1$ and $\bmu_2$ both attached at
time $t$ to the point $(\phi,\dot{\phi})$ in phase-space and with
components $(\delta\phi_i,\delta\Pi_i)$ for $i=1,2$
respectively. The volume spanned by these two vectors is given by
\begin{equation}
\label{eq:V(t)}
\vol\left(t\right)=\bmu_1\left(t\right)\wedge \bmu_2\left(t\right) 
= \delta\phi_1 \delta\Pi_2 - \delta\Pi_1\delta\phi_2.
\end{equation}
In \App{app:PhaseSpaceVolume_t-Measure}, it is shown that this
infinitesimal volume evolves according to
\begin{equation}
\label{eq:Vdot:tmeasure}
\dfrac{\dd \ln \vol}{\dd t} =  -3H\left(1+\frac{ \Pi^2}{6\Mp^2H^2}\right).
\end{equation}
In an expanding universe, $H>0$, the entire phase-space has attractive
properties, while in a contracting universe, $H<0$, the entire phase
space is repulsive. Let us also notice that the typical timescale
associated with the contraction (respectively expansion) of the phase
space volume is one e-fold.

\subsubsection{Hamiltonian measure}

If one uses the Hamiltonian measure to compute volumes in the
four-dimensional phase-space $( \phi,p_\phi,a,p_a ) $, one
finds that the volume is always preserved, as a consequence of
the Liouville theorem (see \App{app:liouville}). This proves that
the volume is preserved according to this measure, and that there is
no phase-space attractor or repeller in this sense.

\subsubsection{Hamiltonian induced measure}

If one considers an arbitrary metric $g_{ij}$, where $i$ and $j$
  are either 1 or 2, \Eq{eq:V(t)} needs to be replaced by
$\vol\left(t\right)=\bmu_1\left(t\right)\wedge \bmu_2\left(t\right)
\det(g_{ij})$, and a new term appears in \Eq{eq:Vdot:tmeasure}, namely
\begin{equation}
\label{eq:Vdot:withmetric:phi_wedge_phidot}
\dfrac{\dd \ln \vol}{\dd t} = -3H-\frac{\Pi^2}{2\Mp^2 H}
+\frac{\dot{\det}\left(g_{ij}\right)}{\det\left(g_{ij}\right)}.
\end{equation}
For the induced Hamiltonian measure, one has $\det(g_{ij}) = a^3$ and
as explained in \Sec{sec:him}, an initial prescription for the scale
factor is mandatory. One can take $a$ to be uniform inside the initial
infinitesimal volume (neglecting corrections suppressed by higher
powers of $\vol$), and one obtains
\begin{align}
\label{eq:dVdt:InducedHamiltonian}
\frac{\dd\ln \vol}{\dd t} = -\frac{\dot{\phi}^2}{2\Mp^2H}\,.
\end{align}
The same conclusions as the ones reached in
\Sec{sec:VolumeShrinker:dphi_wedge_ddotphi} thus apply here, namely
phase-space is attractive (repulsive) in an expanding (contracting)
universe.  Let us, however, notice that the typical timescale
associated with the phase-space contraction (expansion) is different,
since \Eq{eq:dVdt:InducedHamiltonian} gives rise to
$\dd\vol/\dd N = -\epsilon_1 \vol$. The typical timescale is
therefore $1/\epsilon_1$ e-folds and can be very long during inflation
if $\epsilon_1 \ll 1$. In this case, the Hamiltonian induced measure
induces quasi conservation of the phase-space volume.

\subsubsection{Field-space $N$-measure}

For the $N$-measure, the calculation closely follows the one presented
for the $t$-measure, and is performed in
\App{app:PhaseSpaceVolume_N-Measure}. One obtains
\begin{equation}
\label{eq:Vdot:Nmeasure}
\dfrac{\ud \ln \vol}{\ud N} = \frac{3}{2}
\left[\Gamma-\Gamma_+(\phip)\right] \left[\Gamma-\Gamma_-(\phip)\right],
\end{equation}
where
\begin{equation}
\Gamma_{\pm}(\phip) = \pm \sqrt{2 + \dfrac{\Gammasr^2(\phip)}{9}} 
+ \dfrac{\Gammasr(\phip)}{3}\,.
\label{eq:Gammapm}
\end{equation}
The right-hand side of Eq.~\eqref{eq:Vdot:Nmeasure} is negative only
for $\Gamma$ in the range $\Gamma_{-} < \Gamma < \Gamma_{+}$.

In the usual situation where $\Gammasr(\phip) \ll 1$, $\vol$ decreases
as soon as $|\Gamma| \lesssim \sqrt{2}$, {\ie}, as soon as inflation
takes place (and independently of its nature: slow roll, transitional,
ultra-slow roll). In the slow-roll regime, $\Gamma\simeq\Gammasr$, one
obtains $\dd\ln\vol/\dd N\simeq -3$, {\ie}, a behavior similar to
\Eq{eq:Vdot:tmeasure}.

In the kination limit, $\Gamma^2 \to 6$, $\vol$ decreases only if
$\Gammasr^2(\phip) > 6$, {\ie}, the potential should be very steep,
and $\Gammasr(\phip) \Gamma >0$ implies that the field follows the
gradient of its potential. This is consistent with the stability
analysis of \Sec{sec:asol}.

\subsection{Attractors as flow compressors}

The fact that the Hamiltonian volume is conserved simply means that if
two flow lines get closer, the dispersion of points \emph{along} these
lines must increase, such that the volume is squeezed along the
directions orthogonal to the flow and stretched along the direction
parallel to the flow. This is not incompatible with the intuitive idea
of having an attractor, which simply involves compression of flow
lines irrespective of the way they are labeled by time.

This is why one could argue that what matters most is
not the distance between points in phase-space per se, but the
distance between flow lines. One way to characterize the flow
evolution is by evaluating the Lyapunov exponents, but their values
are usually location and direction dependent in phase-space and one
has to consider a spectrum of Lyapunov exponents~\cite{Clesse:2009ur}.

Here we follow a different approach. Let $\orb(M)$ denote the orbit of
$M$ in phase-space, {\ie}, the set of points in either the past or the
future of $M$ under dynamical evolution. This is the flow line that
$M$ belongs to, or said differently, the equivalence class of $M$
under the dynamical relation. Starting from a distance measure $d$ in
phase-space, we then construct the quantity $\dist$, defined as
\begin{equation}
\begin{aligned}
\label{eq:OrbitDistance:def}
\dist\left(M_1,M_2\right) & 
= \max\left\lbrace d\left[M_1,\orb\left(M_2\right)\right],
d\left[M_2,\orb\left(M_1\right)\right] \right\rbrace,
\end{aligned}
\end{equation}
where
$d\left[M_1,\orb\left(M_2\right)\right]
=\min_{M\in\orb\left(M_2\right)}\left[d\left(M_1,M\right)\right] $.
This is a symmetric, non-negative function, that vanishes if and only
if $M_1$ and $M_2$ are along the same trajectory. However it does not
satisfy the subadditivity inequality,
{\ie}, $\dist(M_1,M_2)+\dist(M_2,M_3)>\dist(M_1,M_3)$, in general. This
is therefore not a proper distance as mathematically defined but this
does not matter for our purposes. One can define an attractor as being
a region where
\begin{align}
\dd\dist <0 .
\end{align}
This means that, considering a reference phase-space trajectory (the
attractor), starting from an initial condition away from the
attractor, one gets closer and closer to the attractor as time
proceeds, according to the phase-space distance $d$. For the sake of
clarity, we restrict the study of $\dist$ for the two-dimensional
measures only.

\subsubsection{Field-space $t$-measure}

Let us first notice that in order to bring the two field coordinates
to the same dimension, a mass parameter $\mu$ must be introduced, so
that $d^2=\mu^2\dd\phi^2+\dd\Pi^2$. When we had to calculate phase
space volumes above, this parameter was irrelevant (if constant in
time) but it does play a role when computing distances a priori. One
can take $\mu$ to the Planck mass, the mass of the inflaton field, or
maybe the Hubble parameter evaluated at a given time.

Let us now consider a point of coordinates $\phi$ and $\Pi$ in phase
space, together with another point of coordinates $\phi+\delta\phi$
and $\Pi+\delta\Pi$. In \App{app:flowCompression:naive}, it is shown
that the distance $\dist$ between these two points obeys
\begin{align}
\label{eq:d_dist_d_t_t_measure}
\dfrac{\dd \ln \dist}{\dd t} =
\dfrac{\left(\dfrac{V'}{\Pi}-\dfrac{\Pi^2}{2\Mp^2 H}\right)\delta\Pi
-\left(\dfrac{V'\Pi}{2\Mp^2H}+V''\right)\delta\phi}{\delta\Pi
+\left(\dfrac{V'}{\Pi}+3H\right)\delta\phi}\,.
\end{align}

Let us notice that the mass scale $\mu$ has dropped off from this
result. However, the sign of the right-hand side in this expression
depends on the initial displacements $\delta\phi$ and
$\delta\Pi$.\footnote{One can check that in the case where the flow
  lines are straight lines in field space, this dependence cancels
  out. For instance, if all flow lines are parallel straight lines,
  $\ddot{\phi}(\phi,\dot{\phi})=A\dot{\phi}$,
  \Eq{eq:d_dist_d_t_t_measure:gen} gives a vanishing result, in
  agreement with the fact that the flow map is neither attractive nor
  anti-attractive in that case. As another example, in the case where
  the flow lines are straight lines intersecting at the origin,
  $\ddot{\phi}(\phi,\dot{\phi})=\dot{\phi}^2/\phi$, one finds from
  \Eq{eq:d_dist_d_t_t_measure:gen} that
  $\dd\ln\dist/\dd t = \Pi/\phi$, in which, as announced, the
  dependence on $\delta\phi$ and $\delta\Pi$ has canceled out, and
  which is agreement with the intuition that the flow map is
  attractive in the anti-diagonal quadrants and repelling in the
  diagonal quadrants.\label{footnote:dist:straightlines}} We consider
three possibilities.

The first one corresponds to a fluctuation in the velocity direction
only, $\delta{\phi}=0$. One obtains
\begin{align}
\label{eq:d_dist_dt:t_measure:deltaphi_eq_0}
\left. \frac{\dd \ln \dist}{\dd t}\right\vert_{\delta\phi=0} =
-3H\left(1+\frac{\epsilon_2}{6}\right),
\end{align}
where the second Hubble flow function $\epsilon_{n+1} = \ud
\ln{\epsilon_{n}}/\ud N$ for $n=1$ appears.  An attractor behavior is
then obtained when $\epsilon_2>-6$. The slow-roll regime, for which
$\vert \epsilon_2\vert\ll 1$, is therefore in the attractive
region. In the ultra-slow-roll regime, $\epsilon_2=-6$, the distance
is preserved, in agreement with
footnote~\ref{footnote:dist:straightlines}.

Let us now consider the case where the initial displacement is along
$\phi$ only, and $\delta\Pi=0$. This gives rise to
\begin{align}
\label{eq:d_dist_dt:t_measure:deltaPi_eq_0}
\left.\frac{\dd \ln \dist}{\dd t}\right\vert_{\delta\dot\phi=0}=-3H
\left(1+\dfrac{\epsilon_1^2-2\epsilon_1\epsilon_2 +
  \dfrac{\epsilon_2^2}{4} +
  \dfrac{\epsilon_2\epsilon_3}{2}}{\dfrac{3\epsilon_2}{2}-3\epsilon_1}\right).
\end{align}
Let us stress that, as for \Eq{eq:d_dist_dt:t_measure:deltaphi_eq_0},
this expression is exact and does not assume anything about the Hubble
flow functions. In this case the situation is more complicated, but in
the slow-roll regime, one has
$\dd\ln\dist/\dd t = -3H[1+\order{\epsilon}]$ so the same conclusions
as above apply, even though subtleties could arise in situations when
$\epsilon_2\simeq 2\epsilon_1$ (see below).  In the ultra-slow-roll
inflation limit, $\epsilon_2=-6$ and $\vert\epsilon_1\vert \ll 1$,
$\vert\epsilon_3\vert \ll 1$, the right-hand side of
\Eq{eq:d_dist_dt:t_measure:deltaPi_eq_0} again vanishes.

Finally, let us consider the case where the initial field displacement
is orthogonal to the field-space trajectory,
$\mu^2\Pi \delta\phi+\dot{\Pi}\delta\Pi=0$, since it is the direction
along which the reduction in the distance is sought. One obtains
\begin{equation}
\label{eq:d_dist_dt:t_measure:gen}
\begin{aligned}
&\left. \frac{\dd \ln \dist}{\dd t}\right|_{\perp}  =
-\dfrac{3H}{1+\dfrac{H^2}{\mu^2}
\left(\epsilon_1-\dfrac{\epsilon_2}{2}\right)^2}
% \\ & \times 
 \left[1+\frac{\epsilon_2}{6}
\right.  \\ & \left.
  -
  \dfrac{H^2}{\mu^2}\dfrac{2\epsilon_1-\epsilon_2}{2}
 % \right.\\ & \times \left.  
  \left(-\epsilon_1+\frac{\epsilon_2}{2}+\frac{\epsilon_1^2}{3} -
  \frac{2}{3}\epsilon_1\epsilon_2+\frac{\epsilon_2^2}{12} +
  \frac{\epsilon_2\epsilon_3}{6}\right)\right].
\end{aligned}
\end{equation}
The limits $\mu\rightarrow \infty$ and $\mu\rightarrow 0$ allow one to
recover \Eqs{eq:d_dist_dt:t_measure:deltaphi_eq_0}
and~(\ref{eq:d_dist_dt:t_measure:deltaPi_eq_0}) respectively, and this
result is, in fact, the generic formula. For the three choices of values
for $\mu$ mentioned above, $\mu=H_*$, $\sqrt{V_*''}$ or $\Mp$, one can see
that the structure $\dd\ln\dist/\dd t = -3H[1+\order{\epsilon}]$ is
preserved, so the slow-roll regime is in the attractive domain, with a
relaxation timescale of order one e-fold. The above expression also
makes the limit $\epsilon_2\rightarrow 2 \epsilon_1$ regular.  In the
ultra-slow-roll inflation limit, the right-hand side of
\Eq{eq:d_dist_dt:t_measure:gen} vanishes, and the distance is
preserved.

\subsubsection{Hamiltonian induced measure}

In \App{app:flowCompression:Hamiltonian} it is shown that, compared to
the $t$-measure, one gets an additional term for the distance
evolution, which does not depend on the initial displacements $\delta\phi$ and
$\delta\Pi$,
\begin{equation}
\begin{aligned}
\label{eq:d_dist_dt:Hamiltonian_measure:gen}
\dfrac{\dd \ln \dist}{\dd t} & = \left. \dfrac{\dd \ln \dist}{\dd t}\right\vert_{t-\mathrm{measure}}
+\dfrac{3H}{1+\dfrac{H^2}{\mu^2} \left(\epsilon_1
-\dfrac{\epsilon_2}{2}\right)^2},
\end{aligned}
\end{equation}
where the first term in the right-hand side is given by
\Eq{eq:d_dist_d_t_t_measure}.  If the initial displacement is in the
velocity direction only, or in the field direction only, an attractor
behavior is obtained only if $H/\mu$ is bounded by a certain
combination of the slow-roll parameters that can be obtained from
\Eqs{eq:d_dist_dt:t_measure:deltaphi_eq_0}
and~(\ref{eq:d_dist_dt:t_measure:deltaPi_eq_0}) respectively. In
particular, if one chooses the scale $\mu$ to be much larger than the
Hubble scale, the same results as for the $t$-measure are recovered.

However, if the initial displacement is orthogonal to the field-space
trajectory, one can see that the leading-order term in
\Eq{eq:d_dist_dt:t_measure:gen} is exactly canceled, so that the
attractor behavior in the $\mathcal{O}(1)$ e-fold in the slow-roll
regime is lost, in agreement with Fig.~2 of
\Ref{Remmen:2013eja}. Whether the dynamics is attractive then depends
on the choice of the scale $\mu$ with respect to the combination of
the slow-roll parameters appearing in \Eq{eq:d_dist_dt:t_measure:gen}.

\subsubsection{Field space $N$-measure}

The calculations are formally identical to the ones for the field
space $t$-measure, except that since $\Phi$ and $\Gamma$ have same
mass dimension, there is no need to introduce an additional mass
parameter. One gets
%\begin{equation}
%\dfrac{\ud \ln \dist}{\ud N} = \dfrac{\Gamma_0 \left( \dfrac{\partial
 %   \DGamma}{\partial \phip} \delta \phip + \dfrac{\partial
  %  \DGamma}{\partial \Gamma} \delta \Gamma \right) - \DGammazero
 % \delta \Gamma}{\Gamma_0 \delta \Gamma - \DGammazero \delta \phip}\,,
%\end{equation}
\begin{equation}
\dfrac{\ud \ln \dist}{\ud N} = \dfrac{\left(3-\dfrac{\Gamma^2}{2}\right)
{\Gammasr}_{,\phip}\,\delta\Phi+\left(\Gamma^2-\dfrac{\Gammasr\Gamma}{2}
-3\dfrac{\Gammasr}{\Gamma}\right)\delta\Gamma}{\delta\Gamma
+\left(3-\dfrac{\Gamma^2}{2}\right)
\left(1-\dfrac{\Gammasr}{\Gamma}\right)\delta\Phi}.
\end{equation}

For initial displacements $\delta \phip = 0$, one obtains
\begin{equation}
\left. \dfrac{\ud \ln \dist}{\ud N} \right|_{\delta \phip = 0} 
= \Gamma^2-\frac{\Gammasr\Gamma}{2}-3\frac{\Gammasr}{\Gamma}.
\label{eq:deltaphipnull}
\end{equation}
The sign analysis of the right-hand side can be performed exactly but
ends up being not particularly illuminating. Let us focus instead on
the relevant physical situations. In the slow-roll regime, for which
$\Gamma \simeq \Gammasr$ and $\vert\Gammasr\vert\ll 1$, we have
\begin{equation}
\label{eq:d_dist_d_N_Nmeasure_delta_Phi_eq_0}
\left. \dfrac{\ud \ln \dist}{\ud N}\right|_{\delta \phip=0, \Gamma
  \simeq \Gammasr} \simeq
-\left(3 - \dfrac{1}{2} \Gamma^2 \right),
\end{equation}
and flow lines are always compressed. Interestingly, the
ultra-slow-roll limit, $\Gammasr \to 0$ with $\Gamma \gg \Gammasr$,
gives
\begin{equation}
\left. \dfrac{\ud \ln \dist}{\ud N}\right|_{\delta \phip=0, \Gamma \gg
  \Gammasr} \simeq \Gamma^2 > 0.
\end{equation}
As a result, ultra-slow roll induces flow lines expansion when
starting from $\delta \phip=0$.

For initial displacements having $\delta \Gamma=0$, one gets
\begin{equation}
\left. \dfrac{\ud \ln \dist}{\ud N}\right|_{\delta \Gamma=0} = 
\dfrac{\Gamma {\Gammasr}_{,\Phi}}{\Gamma - \Gammasr} \,.
\end{equation}
For slow roll, this expression becomes singular as $\Gamma \to
\Gammasr$. This means that $\Gamma$ must be calculated at
next-to-leading order in slow roll to evaluate the above
equation. Solving \Eq{eq:kgefold} perturbatively around
$\Gamma\simeq\Gammasr$, one obtains $\Gamma\simeq
\Gammasr(1-{\Gammasr}_{,\phip}/3)$ [in agreement with Eq.~(22) of
  \Ref{Vennin:2014xta}]. This leads to
\begin{equation}
\label{eq:d_dist_d_N_Nmeasure_delta_Gamma_eq_0}
\left. \dfrac{\ud \ln \dist}{\ud N}\right|_{\delta \Gamma=0, \Gamma
  \simeq \Gammasr} \simeq -3,
\end{equation}
and flow lines are compressed as in
\Eq{eq:d_dist_d_N_Nmeasure_delta_Phi_eq_0}. For ultra-slow roll, one
gets
\begin{equation}
\left. \dfrac{\ud \ln \dist}{\ud N}\right|_{\delta \Gamma=0, \Gammasr
  \ll \Gamma \ll 1} \simeq - \dfrac{\ud^2 \ln V}{\ud \phip^2}\,,
\end{equation}
which means that only convex potentials lead to flow lines
compression, a result that is compatible with the stability analysis
performed in \Ref{Pattison:2018bct}.

Finally, for initial displacements orthogonal to the trajectory, one
gets
\begin{widetext}
\begin{equation}
  \begin{aligned}
    \left. \dfrac{\ud \ln \dist}{\ud N} \right|_{\perp} & = 
    \dfrac{ \left(3 -
      \dfrac{\Gamma^2}{2}  \right)^2 \left(1 -
      \dfrac{\Gammasr}{\Gamma} \right) {\Gammasr}_{,\phip}+\Gamma^2-   
\dfrac{3\Gammasr}{\Gamma} - \dfrac{\Gamma \Gammasr}{2} }
{1 + \left(3 - \dfrac{\Gamma^2}{2}  \right)^2 \left(1 -
      \dfrac{\Gammasr}{\Gamma} \right)^2}\,.
    \end{aligned}
\end{equation}
\end{widetext}
In the slow-roll limit, $\Gamma \to \Gammasr$, we get the same
behavior as in \Eqs{eq:d_dist_d_N_Nmeasure_delta_Phi_eq_0}
and~(\ref{eq:d_dist_d_N_Nmeasure_delta_Gamma_eq_0}) and flow lines are
always compressed. In ultra-slow roll, $\Gamma \gg \Gammasr$ and
$\Gamma \ll 1$, one gets
\begin{equation}
\left. \dfrac{\ud \ln \dist}{\ud N} \right|_{\perp,\Gammasr \ll \Gamma
  \ll 1}
\simeq \frac{\Gamma^2}{10} - \frac{9}{10} \dfrac{\ud^2 \ln V}{\ud \phip^2}\,.
\end{equation}
As before, flow lines can only be compressed for a convex potential
and provided that the field velocity in e-folds remains low enough,
$\Gamma^2< 9(\ud^2 \ln V/\ud \phip^2)$.

\subsection{Slow-roll attractor} 

The above considerations lead to the conclusion that slow roll is in
the flow lines compressing region of phase-space with a relaxation
timescale of order one e-fold. Let us now show that slow roll is the
actual attractor of that region 
(see \Refs{Liddle:1994dx, Pattison:2018bct} for
other demonstrations).

In terms of $(\phip,\Gamma)$, the equation of motion is given by
\Eq{eq:kgefold}. Since $\Gamma^2 < 6$, this leads to
$\dd\Gamma/\dd N\propto \Gammasr-\Gamma$.  This is why, if
$\Gamma > \Gammasr$, then $\ud \Gamma/\ud N < 0$ and $\Gamma$
decreases, while if $\Gamma < \Gammasr$, $\ud \Gamma/\ud N > 0$ and
$\Gamma$ increases. As a result, within the whole phase-space,
$\Gamma(N)$ is attracted toward $\Gammasr$. Notice, however, that
$\Gammasr$ is also varying with $N$ such that this attractor behavior
does not necessarily imply that the slow-roll solution
$\Gamma \simeq \Gammasr$ is adiabatically reached. In fact, stable
ultra-slow roll~\cite{Pattison:2018bct} is precisely an example in
which slow roll is never attained.

The stability of the slow-roll solution itself can be studied by
considering a small deviation from $\Gamma = \Gammasr$, parametrized
by
\begin{equation}
\Gamma(N) = \Gammasr(N) + \epsilon(N).
\label{eq:GammaEpsilon}
\end{equation}
Plugging this expression into \Eq{eq:kgefold} gives
\begin{equation}
\label{eq:SR_stab_deps_dN}
\dfrac{\ud \ln \epsilon}{\ud N} = -3 + \left(\epsilon +
\Gammasr\right)\left( \dfrac{\epsilon + \Gammasr}{2} 
-\dfrac{{\Gammasr}_{,\phip}}{\epsilon}\right).
\end{equation}
Since it was shown above \Eq{eq:d_dist_d_N_Nmeasure_delta_Gamma_eq_0}
that $\Gamma-\Gammasr$ receives a correction of order
$\Gammasr {\Gammasr}_{,\phip}$ in the slow-roll regime, the deviation
$\epsilon$ should be thought of as being much larger than this
correction [working at leading order in slow roll, otherwise the
expansion~(\ref{eq:GammaEpsilon}) would have to be carried out around
the corrected value of $\Gamma$], while being much smaller than
$\Gammasr$. Provided $\Gammasr$ and ${\Gammasr}_{,\Phi}$ remain much
smaller than one, one therefore obtains
\begin{equation}
\epsilon(N) \simeq \epsilon(\Nini) \exp\left[-3(N - \Nini)\right],
\end{equation}
and slow roll is stable, with a relaxation time of the order one
e-fold.

Equation~\eqref{eq:SR_stab_deps_dN} also shows that the second derivative of the
potential (the term ${\Gammasr}_{,\phip}$) can play a role but only at
the expense of having $\Gamma \gg \Gammasr$, which is precisely the
condition to land in the ultra-slow-roll regime.

\subsection{Information conservation}

Finally, let us mention another possible way to characterize
attractors: their ability to erase information about initial
conditions. A natural framework to define such a property is the one
of information theory. Consider two probability distributions
$P_1(\bmx)$ and $P_2(\bmx)$ in phase-space (here described by the
vector $\bmx$). The ``information distance'' between these two
probability distributions can be measured according to the
Kullback-Leibler divergence
\begin{align}
\label{eq:DKL:def}
\DKL\left(P_1\vert\vert P_2\right) = \int \dd\bmx P_1\left(\bmx\right)
\ln \dfrac{ P_1\left(\bmx\right)}{ P_2\left(\bmx\right)}\,.
\end{align}
Starting from either of two initial conditions in phase-space
described by two probability distributions, the ability to reconstruct
which of the two initial distributions one started from depends on the
information distance between these distributions. Therefore, if the
information distance decreases as time proceeds, the ability to
reconstruct initial conditions is getting washed away and one has an
``eraser''. In the opposite case, the sensitivity on the initial
conditions becomes more severe as time proceeds and the dynamics is
repulsive, if not chaotic.

For the case of a scalar field, the classical evolution of the system
is continuous and because $\DKL$ is measure invariant, this implies
that the Kullback-Leibler divergence is always conserved during the
flow, for any continuous measure.

For instance, assuming that the classical evolution of the system maps
$\bmx \to \bmy = \bmF(\bmx)$ uniquely (which is the case for a
dynamical system), one has
\begin{equation}
\begin{aligned}
  \DKL & = \int \ud \bmy  P_1(\bmy) \ln\left[\dfrac{P_1(\bmy)}
{P_2(\bmy)}\right]
  \\ & = \int \ud \bmx \left|\det(\bmF') \right| P_1(\bmy)
  \ln\left[\dfrac{P_1(\bmy) |\det(\bmF')|}{P_2(\bmy)| \det(\bmF')|}
    \right] \\
  & = \int \ud \bmx  P_1(\bmx) \ln\left[\dfrac{P_1(\bmx)}{P_2(\bmx)}\right].
\end{aligned}
\end{equation}
As a result, for any measure, independently of the contraction of
volume and flow lines in phase-space, information distance is always
conserved during the classical evolution of a self-gravitating scalar
field. Let us notice, however, that quantum diffusion may violate this
result and lead to further initial conditions erasure.

\subsection{Discussion}

In this section, various measures have been studied, that can be
classified in two types: the field-space $t$-measure, the Hamiltonian
measure, and the field-space $N$-measure, belong to category I, while
the Hamiltonian induced measure~\cite{Remmen:2013eja} or the
field-space $\eta$-measure (that was not analyzed in details here but
simply mentioned in \Sec{sec:tm}) belongs to category II. Only the
measures belonging to the first category are properly defined in the
strict mathematical sense. It is thus quite remarkable that, for all
these measures, regardless of whether an attractor is defined in terms
of volume shrinking or flow compression, it was found that the
slow-roll regime is \emph{always} an attractor (while the status of
ultra-slow roll varies, and with the Hamiltonian measure, phase-space
volumes are conserved by definition, but the flow-line compression is
as large as with the other measures), and that the convergence time
toward slow roll is given by one-third of an e-fold.

In fact, the result easily generalizes to all measures of
category I, since they are related through $x'=F(x,y)$ and $y'=G(x,y)$,
where $F$ and $G$ do not depend explicitly on time (otherwise one
would be considering a measure of category II). The presence of an
attractor is characterized by the fact that, at late time, $y$
asymptotes a given function of $x$, which translates into $y'$
approaching some function of $x'$ as well, at least if $F$ and $G$ are
continuous, with the same relaxation timescale.

Without specifying a physical mechanism that sets initial conditions
for the field-metric system, the choice of a measure is a subjective
prejudice. From a Bayesian perspective, a phase-space measure plays no
more role than a priori. However, in the present case, we have found
that the attracting behavior of slow-roll inflation is robust under
theoretical prejudices as encoded by the phase-space measure, if one
restricts to measures that are unambiguously defined (category
I). This remarkable property is what protects inflation from
phase-space fine-tuning issues.

\section{The Model Building Problem}
\label{sec:building}

In most models that have been proposed so far, inflation is driven by
one (or several) scalar field(s). The physical nature of this scalar
field is still unknown but implementing inflation in high-energy
physics is an important question. In doing so, we face challenges,
that are very briefly described below, and that are sometimes used
against inflation. Before discussing them, one cannot help mentioning
that a criticism that was very often made, namely that inflation needs
a scalar field (which is, by the way, not completely true since there
exist other mechanisms; see, for instance, \Ref{Golovnev:2008cf}) and
that no fundamental scalar field has ever been observed in Nature, has
been proven wrong thanks to the Higgs boson discovery at the Centre
Europ\'een de Recherche Nucl\'eaire (CERN)~\cite{Aad:2012tfa}. It is,
however, true that, from a certain point of view, no uncontroversial
UV-complete model of inflation has been proposed so far, which makes
the model building problem an important one.

Two (related) questions are usually discussed. The first one concerns
the values of the parameters that one needs to assume in order for a
given model to fit the data. The prototypical example is $\lfi{}_4$
with $V(\phi)=\lambda \phi^4$, $\lambda$ being a dimensionless
constant. When the CMB normalization is taken into account, one
obtains $\lambda \sim 10^{-12}$ which violates the standard lore that
a model is ``natural'' if all dimensionless quantities are of order
one. For $\lfi{}_2$, one finds $m \sim 10^{-6}\Mp$ which might be
viewed as ``better''. Since these two models are disfavored, it is
interesting to see what happens for $\hi$, where the non-minimal
coupling constant must be given by $\xi \sim 46000 \sqrt{\lambda}$.
The fact that $\xi/\sqrt{\lambda} \gg 1$ can be problematic for model
building. Defining the naturalness of the value of a parameter is a
difficult problem, and in any case, it should be done on a
model-by-model basis. In fact, as argued in \Sec{sec:discuss}, the
only ``universal'' and objective quantity that quantifies the
fine-tuning of a model is the Bayesian evidence. From the CMB point of
view, by definition, the best model of inflation is therefore the
least fine-tuned one.

The second approach to the model-building issue concerns the
flatness of the potential. Inflation requires flatness in the
logarithm of the potential in order for the effective pressure of the
system to be negative. But protecting the flatness from corrections is
usually challenging. Indeed, on very general grounds, the mass $m$ of
the inflaton field receives corrections given by
\begin{align}
m^2\rightarrow m^2+gM^2\ln\left(\frac{\Lambda}{\mu}\right), 
\end{align}
where $\mu$ is the normalization scale, $M>\Lambda$ the energy scale
of heavy fields, $\Lambda$ the cutoff of the effective theory in
which the model in embedded and $g$ the coupling constant. We see that
this can lead to $m/H\sim 1$ (unless $g$ is extremely small). This
problem is known as the $\eta$-problem of
inflation~\cite{Baumann:2014nda}. It is fair to recognize that this
issue is quite generic and can be problematic. However, there are also
known methods to fix it, typically by requiring symmetries to be
preserved, the prototypical example being shift symmetry, see
\Ref{Kawasaki:2000yn}.

The characteristic scale of the inflationary theory can be as high as
$10^{16}\, \mbox{GeV}$, that is to say $13$ orders of magnitude above
what has been tested at CERN. Whether one prefers to conclude that the
questions mentioned above challenge the ``naturalness'' of the
inflationary paradigm, or that inflation is an opportunity to learn
about physics at those scales, is, of course, subjective. However, it seems
fair to say that these challenges are most probably due to our
lack of knowledge of particle physics at high-energy scales, rather
than due to inconsistencies in the inflation theory itself.

Another argument against inflation considers that it is so flexible
that it can account for any observations. It is true that the standard
predictions (spatial flatness, adiabatic, Gaussian and almost scale
invariant perturbations) are valid for the simplest class of models
(single field with minimal kinetic terms slow-roll models). If, say, a
small contribution originating from non-adiabatic modes is, one day,
discovered in the CMB, this will rule out the simplest class of
scenarios but not inflation itself. Indeed, models of inflation with
several scalar fields could easily explain the presence of
non-adiabatic modes. The worry is then that any new observation could
be explained in this way, thus rendering inflation not
falsifiable. Even the robust prediction that the inflationary tensor
power spectrum should have a red tilt has been challenged in the
context of more complicated models, see, e.g., \Refs{Cai:2014uka,
  Jimenez:2015jqa}. This situation is, however, not uncommon. In
particle physics for instance, it seems difficult to rule out gauge
theories in general. It is only if one specifies the gauge group that
one comes with a version that can be falsified. Inflation is similar
in the sense that it is only by specifying a model that one obtains a
series of accurate predictions. In particle physics, no one would
discard gauge theories because a choice of gauge group is needed and
the same is true for inflation.

\section{The Flatness problem}
\label{sec:flatness}

Cosmic inflation predicts that the universe should be spatially flat
today and this has been one of its first successful predictions, well
before any accurate measurements of the actual
curvature~\cite{Guth:1980zm, Linde:1981mu}. The current bound on
$\OmegaKo$, the density parameter of curvature today, is set by Planck
2018 complemented with other cosmological observables. From
\Ref{Aghanim:2018eyx}, it is
\begin{equation}
  \OmegaKo = 0.0007 \pm 0.0037\,.
  \label{eq:omegako}
\end{equation}
In the past decade, various works have claimed that measuring a very
small curvature today should not be considered as an argument in
favor of cosmic inflation or its alternatives, as it could also be
``natural'' in a purely decelerating Friedmann-Lema\^itre
universe~\cite{Lake:2004xg, Helbig:2011aa, 2018arXiv180102176H}. In
other words, there is no need for cosmological inflation to solve the
flatness problem because there is no flatness problem at all.

Different reasons have been offered to support this opinion. Let us,
for instance, mention the claim that there is no problem if the spatial
curvature is \emph{exactly} vanishing.  Although correct in principle,
this idea is, however, hardly reconcilable with the existence of
curvature fluctuations on Hubble scales.

Another claim consists in denying that there is a ``fine-tuning''
issue based on the idea that one cannot compare models of the universe
because there is one universe only. These are the issues of
frequentist statistics. Model comparison is however standard routine
in cosmology, by means of Bayesian statistics~\cite{Trotta:2008qt,
  March:2010ex, Martin:2014lra, Martin:2014rqa}. It is possible to
compare models of the universe and quantify by how much one is
preferred by the data, using their Bayesian evidence.

Other works state that, in the absence of a well-motivated measure for
$\OmegaKo$, one cannot claim that a tiny curvature at the onset of the
Friedmann-Lema\^itre epoch is unlikely. Again, this objection is
better addressed in the framework of Bayesian statistics where it
boils down to the usual issue of choosing the priors.

To quantitatively address what is meant by the flatness
problem, let us consider the following simple model, where the recent
determination of the cosmological parameters is ignored and only the
measurement of $\OmegaKo$ is considered.  To illustrate the method, we
model \Eq{eq:omegako} by a steplike likelihood
\begin{equation}
\calL(D|\OmegaKo,I) = \calLmax \left[\heaviside{\OmegaKo + \sigmam} -
  \heaviside{\OmegaKo - \sigmap} \right], 
\label{eq:like}
\end{equation}
with $\sigmam = \sigmap = 0.0037$. The likelihood is the probability
of measuring the data $D$ given the theoretical value of $\OmegaKo$
within some prior hypothesis $I$. This is precisely in $I$ that the
choice of a model of the universe affects the inference problem. Let
us then consider only the homogeneous cosmological model of Friedmann
and Lema\^itre (FL) containing a gravitating fluid $P=w \rho$, plus
spatial curvature $\calK$. The Friedmann-Lema\^itre equations [already
given in \Eqs{eq:hubble} and (\ref{eq:hubbledot}) for the particular
case of a scalar field sourcing the geometry] read
\begin{equation}
  \begin{aligned}
    H^2 & = \dfrac{\rho}{3 \Mp^2} - \dfrac{\calK}{a^2}\,,\\
    H^2 + \dot{H} & = -\dfrac{1}{6\Mp^2} \left(\rho + 3 P\right).
  \end{aligned}
\label{eq:FLfluid}
\end{equation}
Provided $H^2 \ne 0$, one can define
\begin{equation}
\OmegaK \equiv -\dfrac{\calK}{a^2 H^2}\,,\qquad
\Omega \equiv \dfrac{\rho}{3 \Mp^2 H^2}\,,
\label{eq:densparams}
\end{equation}
which allows one to recast \Eqs{eq:FLfluid} in terms of the density
parameters
\begin{equation}
\begin{aligned}
  \Omega + \OmegaK & = 1, \qquad \dfrac{1}{H} 
\dfrac{\ud H}{\ud N} + 1 = -\dfrac{1 + 3w}{2} \Omega.
\end{aligned}
\label{eq:FLefolds}
\end{equation}
Here, $N\equiv \ln a$ is, as before, the e-fold time
variable. Because the spatial curvature $\calK$ is constant,
\Eq{eq:densparams} implies that
\begin{equation}
\dfrac{\ud \ln(a H)}{\ud N} = -\dfrac{1}{2} \dfrac{\ud \ln
  |\OmegaK|}{\ud N}\,.
\end{equation}
Plugging this expression into \Eq{eq:FLefolds} gives a closed equation
for the curvature density parameter
\begin{equation}
\dfrac{\ud \ln| \OmegaK|}{\ud N} = \left(1 + 3 w\right)\left(1 -
\OmegaK \right).
\label{eq:evolOmegaK}
\end{equation}
Let us define the absolute relative spatial curvature
\begin{equation}
\varpi \equiv \dfrac{\left|\OmegaK\right|}{1 - \OmegaK}\,,
\label{eq:defvarpi}
\end{equation}
which, from \Eq{eq:evolOmegaK}, satisfies the trivial equation
\begin{equation}
\dfrac{\ud \ln \varpi}{\ud N} = 1 + 3w.
\label{eq:evolvarpi}
\end{equation}
This equation is valid as long as $H^2\ne 0$ and for any
$w(N)$. Because $\Omega > 0$, one has $\OmegaK < 1$ and
$\varpi>0$. However, the distinction has to be made between positively
curved universe, $\calK>0$, for which $\varpi \in ]0,1[$, and,
$\calK<0$, for which $\varpi \in ]0,+\infty[$. The case
$\calK\equiv0$ is singular and will not be considered. Let us
notice that for all values of $\varpi$ close to the upper boundary of
these intervals, the curvature is \emph{initially} dominating the
energy budget. In this situation, the universe remains curvature
dominated, or recollapses, and these situations are ruled out. As
discussed in \Sec{sec:beyondhomo}, for all cosmologically viable
models, one must assume that curvature is not initially dominating.

\subsection{Maximally uninformative prior distribution}

Our theoretical model of the universe therefore involves a parameter
that is the initial value for the absolute relative curvature $\varpi$
at some remote time within the radiation-dominated era. Its prior
probability distribution, $p(\varpi|I)$, could be chosen according to
any theoretical prior knowledge of the model, such as taking a flat
prior distribution if one expects this parameter to be of order
unity. However, in the absence of any theoretical prior knowledge,
there is a well-defined Bayesian way to choose a prior distribution,
which consists in maximizing ignorance. To do so, one should first
identify a transformation group that let the physical equations
invariant and that represents our state of
ignorance~\cite{JaynesBook}. The way we have written \Eq{eq:evolvarpi}
encodes all these desiderata. This equation is invariant under both
scale factor and curvature rescaling, $N\rightarrow N'+\lambda$ and
$\ln \varpi \rightarrow \ln \varpi' + \mu$, $\lambda$ and $\mu$ being
any constant. Therefore, an uninformative prior on the curvature is,
for any value of the scale factor (at constant $w$), a Jeffreys' prior
on $\varpi$, namely a flat prior on $\ln\varpi$:
\begin{equation}
p(\ln \varpi|I) = \dfrac{\heaviside{\ln \varpi-\ln\Cmin} -
  \heaviside{\ln\varpi - \ln\Cmax}}{\ln \Cmax - \ln \Cmin}\,.
\label{eq:uninfprior}
\end{equation}
The boundaries $\Cmin$ and $\Cmax$ must be provided as a minimal state
of knowledge and the choice of the uninformative prior finally boils
down to deciding what the theoretically acceptable extreme values for
$\varpi$ are. Let us stress that \Eq{eq:uninfprior} matches the
uninformative prior derived by \'Evrard and Coles in
\Ref{Evrard:1995ha}.

Let us now consider two models, $\Minf$ and $\MFL$. Both scenarios
assume that the expansion of the universe is standard at energy scales
below $10\,\MeV$ so as not to spoil big-bang nucleosynthesis
(BBN). However, $\Minf$ assumes that, in addition, there is a period
of quasi-de Sitter acceleration at larger energies lasting
$\Delta \Ninf$ e-folds. In the inflationary model, the initial
curvature has to be set before inflation and will be referred to as
$\varpiini$. In the model $\MFL$, the initial curvature is set at
$\rhobbn^{1/4} \equiv 10\,\MeV$, and it will be referred to as
$\varpibbn$.

The value of $\varpio$ today can immediately be solved from
\Eq{eq:evolvarpi} and reads
\begin{equation}
\begin{aligned}
  \ln\left(\dfrac{\varpio}{\varpibbn} \right) & = (1+ 3
  \wbarrad)\Delta \Nrad + (1+3 \wbarmat )\Delta \Nmat \\ & \simeq \Delta
  \Ntot + \Delta \Nrad \equiv \No,
\end{aligned}
\label{eq:varpio}
\end{equation}
where $\wbar$ stands for the mean equation of state parameter during
the epoch of interest; i.e.,
\begin{equation}
  \wbar = \dfrac{1}{\Delta N} \int \dfrac{P(N)}{\rho(N)} \ud N.
\end{equation}
Neglecting the recent domination of the cosmological constant, one has
$\wbarrad \simeq 1/3$ and $\wbarmat \simeq 0$, which leads to the
second line of \Eq{eq:varpio}. The quantity
$\Delta \Ntot \equiv \Delta\Nrad + \Delta \Nmat$ is the total number
of e-folds of decelerated expansion, starting in the radiation era
from $\rhobbn^{1/4}$ till today. Putting numbers together, one gets
\begin{equation}
\begin{aligned}
  \Delta \Ntot & \simeq \ln\left(1+\zbbn \right) \simeq 25,\\
  \Delta \Nrad & \simeq \ln \left(\dfrac{1+\zbbn}{1+\zeq} \right) \simeq 16,
\end{aligned}
\end{equation}
and $\No \simeq 41$. The quantities $\zbbn$ and $\zeq$ are the
redshifts of BBN and equality between matter and radiation
respectively.

During the inflationary era in the model $\Minf$, the evolution of
$\varpi$ is still described by \Eq{eq:evolvarpi} with
$\wbar \simeq -1$. We therefore have
\begin{equation}
\ln\left(\dfrac{\varpibbn}{\varpiini} \right) \simeq  -2\Delta\Ninf +
(1+3\wrehbar)\Delta \Nreh + 2 \Delta \Nthird,
\label{eq:varpiini}
\end{equation}
where we have included the terms coming from the reheating era
($\Delta \Nreh$) after inflation~\cite{Martin:2014nya, Martin:2016oyk}
as well as a possible radiation-dominated era after reheating and
before BBN ($\Delta \Nthird$). Cosmic inflation is of cosmological
interest only when the total number of e-folds, $\Delta \Ninf$,
dominates over the other terms in \Eq{eq:varpiini}, and these act
as model-dependent corrections. We can therefore define an effective
number of accelerated e-foldings by
\begin{equation}
\Delta \Ninfbar \equiv \Delta\Ninf - \dfrac{1+3\wrehbar}{2} \Delta
\Nreh - \Delta \Nthird.
\end{equation}
Prior distributions for both $\varpiini$ in model $\Minf$ and
$\varpibbn$ in model $\MFL$ are given by \Eq{eq:uninfprior}. For the
inflationary model, one can therefore derive what is the uninformative
induced prior on $\varpibbn$
\begin{equation}
\begin{aligned}
  &  p(\ln\varpibbn | \Minf)   = \int p(\ln\varpibbn,\ln\varpiini|\Minf) 
\, \ud \ln \varpiini\\
  & =\int p(\ln\varpiini|\Minf) \,
  p(\ln\varpibbn|\ln\varpiini,\Minf)\, \ud \ln\varpiini.
\end{aligned}
\label{eq:priorinf}
\end{equation}
The evolution from $\varpiini$ to $\varpibbn$ being deterministic, one has
\begin{equation}
  p(\ln \varpibbn | \ln\varpiini,\Minf) = \delta \left[\ln\varpibbn 
- f(\ln\varpiini)\right],
\end{equation}
where, from \Eq{eq:varpiini}, $f(x)=x-2\Delta\Ninfbar$. Performing the
integral in \Eq{eq:priorinf} gives
\begin{equation}
\begin{aligned}
  p(\ln\varpibbn|\Minf) & = \dfrac{1}{\ln\Cmax - \ln\Cmin} 
\times \\ & \bigg[\heaviside{\ln\varpibbn -\ln\Cmin +
    2\Delta\Ninfbar}  \\ &  - \heaviside{\ln\varpibbn -\ln\Cmax 
+ 2\Delta\Ninfbar} \bigg].
\end{aligned}
\label{eq:priorinfwrad}
\end{equation}
As expected, one recovers the prior $p(\varpibbn|\MFL)$ by setting
$\Delta\Ninfbar = 0$ in the previous equation.

\subsection{Bayesian evidence}

The posterior probability distributions of the models $I=\Minf$ and
$I=\MFL$, given the data $D$ (here the measurement of present-day
curvature) are given by
\begin{equation}
p(I | D) = \dfrac{p(D|I) \pi(I)}{p(D)}\,,
\end{equation}
where $\pi(I)$ is the prior belief in model $I$ and $p(D)$ a
normalization constant. The global likelihood, or evidence, $p(D|I)$,
is obtained by marginalizing the likelihood over the model parameters,
here $\ln\varpibbn$:
\begin{equation}
p(D|I) = \int \calL(D|\OmegaKo,I)\, p(\ln\varpibbn|I)\, \ud \ln\varpibbn,
\label{eq:defevid}
\end{equation}
where $\OmegaKo(\ln\varpibbn)$ is a deterministic function of $\ln
\varpibbn$ given by \Eq{eq:varpio}.

Both models $\Minf$ and $\MFL$ do not allow the curvature $\OmegaK$ to
change sign during its evolution and they can be partitioned into two
submodels, $\Minf^\pm$ and $\MFL^\pm$, for each sign of $\OmegaK = \pm
|\OmegaK|$\footnote{Therefore, the $+$ represents open universes while the $-$
corresponds to closed universes.}. For the inflationary model, plugging
  \Eqs{eq:like}, \eqref{eq:varpio} and \eqref{eq:priorinfwrad} into
  \Eq{eq:defevid} yields
\begin{equation}
  \begin{aligned}
    & p\left(D|\Minf^{\pm}\right) = \dfrac{\calLmax}{\ln\Cmax -
      \ln\Cmin} \\ & \times \left[\ln\left(\dfrac{\sigma_\pm}{1 \mp
          \sigma_\pm}\right) -\No + 2\Delta\Ninfbar -\ln\Cmin \right]
    \\ & \times \left\{ \heavisideb{\ln\left(\dfrac{\sigma_\pm}{1 \mp
            \sigma_\pm}\right) - \No + 2\Delta\Ninfbar -
        \ln\Cmin}\right. \\ & -
    \left. \heavisideb{\ln\left(\dfrac{\sigma_\pm}{1 \mp
            \sigma_\pm}\right) - \No + 2\Delta\Ninfbar - \ln
        \Cmax}\right\} \\ & + \calLmax
    \heavisideb{\ln\left(\dfrac{\sigma_\pm}{1 \mp \sigma_\pm}\right) -
      \No + 2\Delta\Ninfbar - \ln\Cmax}.
  \end{aligned}
\label{eq:evidinf}
\end{equation}
Taking $\Delta\Ninfbar=0$ in this equation gives
$p(D|\MFL^{\pm})$. Within a given model, \Eq{eq:evidinf} shows that,
even under a maximally uninformative prior, the evidence depends on
$\Cmin$, $\Cmax$ and, for inflation, on $\Delta\Ninfbar$. For the
extreme values of curvature, one can remark that, if $\calK > 0$,
there is an absolute maximum for $\varpi$ which corresponds to
$\Cmax = 1$ (and $\OmegaK \rightarrow -\infty$). For $\calK < 0$,
$\varpi$ is not bounded when $\OmegaK \rightarrow 1$.

In order to simplify the discussion, let us add a minimal amount of
information and choose $\Cmax=1$ for both cases. For all models,
$\Minf^\pm$ and $\MFL^{\pm}$, this bound corresponds to quite an
extreme case for which the universe has initially an unreasonable
amount of curvature and will never provide a viable cosmological
model. The two remaining parameters are $\Cmin$ and $\Delta\Ninfbar$.

\begin{figure}
  \begin{center}
    \includegraphics[width=\figmaxw]{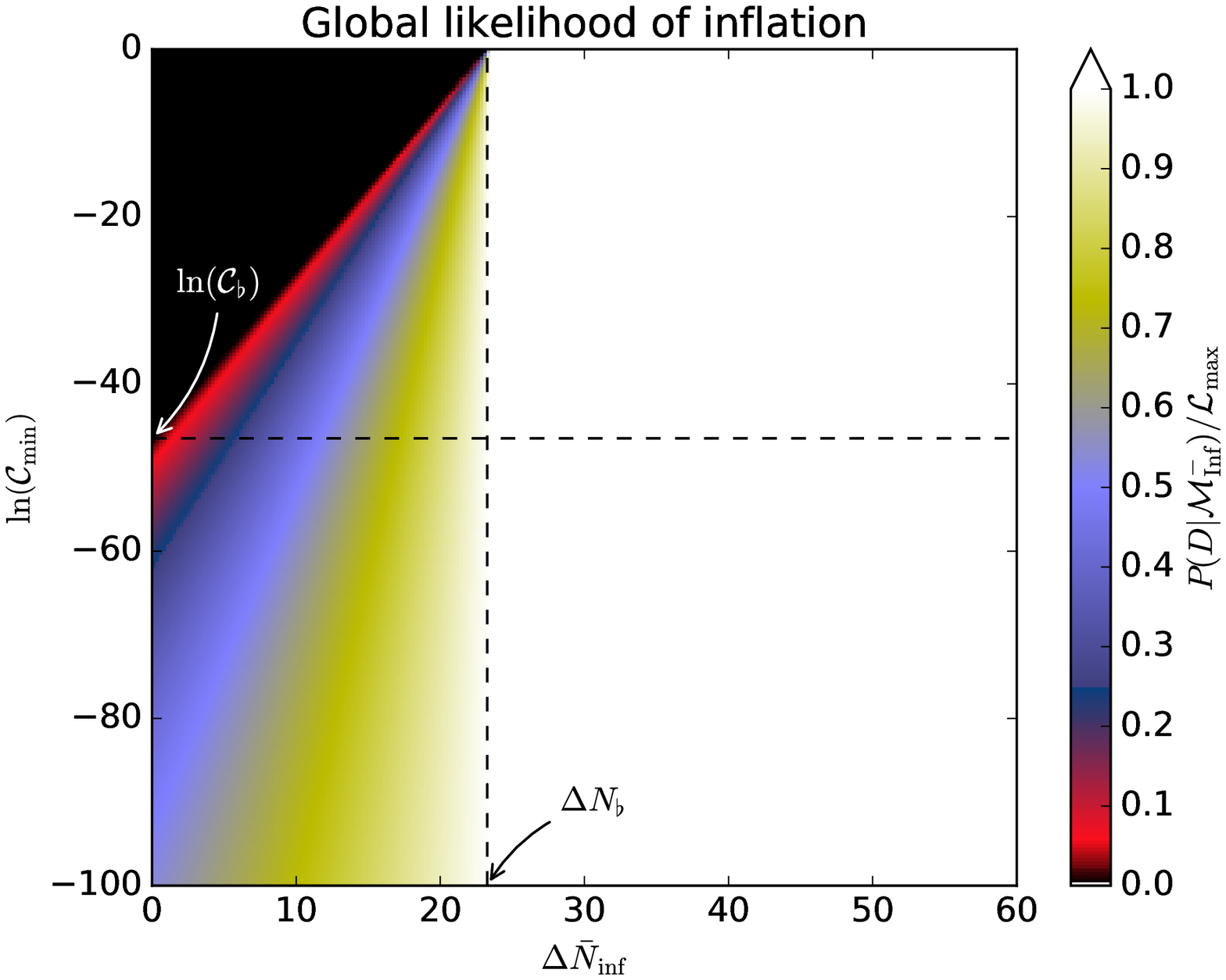}
    \caption{Evidences $p(D|\MFL^{-})$ for the purely decelerating
      Friedmann-Lema\^itre model and $p(D|\Minf^{-})$ for the
      inflationary model, given the spatial curvature $\OmegaKo$
      measured today (see text). For the FL model, $p(D|\MFL^{-})$ is
      obtained for $ \Delta\Ninfbar =0$ and, hence, corresponds to the
      vertical axis. The quantity $\Cmin$ is the lowest possible value
      associated with a maximally uninformative prior on the initial
      absolute relative curvature $\varpi$. The black regions
      correspond to vanishing evidence for which the model cannot
      explain the data. In the case of inflation, as soon as $
      \Delta\Ninfbar > \Delta \Nflat$, the evidence saturates to
      $p(D|\Minf^{-})=\calLmax$.}
\label{fig:evidences}
  \end{center} 
\end{figure}

The dependence of $p(D|\Minf^-)$ and $p(D|\MFL^-)$ with respect to
$\Cmin$ and $\Delta\Ninfbar$ has been plotted in
\Fig{fig:evidences}. Black regions in these plots correspond to
vanishing evidence and the model cannot explain the current data,
i.e., the actual value of $\OmegaKo$. For $\MFL$, all values of
$\Cmin$ larger than $\Cflat$, with
\begin{equation}
  \ln \Cflat \equiv \ln\left(\dfrac{\sigma_\pm}
{1 \mp \sigma_{\pm}}\right) - \No \simeq -47,
\end{equation}
are ruled out. For $\Minf$, as soon as the number of e-folds
$\Delta\Ninfbar$ is greater than
\begin{equation}
\Delta \Nflat \equiv \dfrac{1}{2} \left[\No -
  \ln\left(\dfrac{\sigma_\pm}{1 \mp \sigma_{\pm}}\right) \right]
\simeq 24,
\end{equation}
the evidence saturates to its maximal value, $\calLmax$, independently
of $\Cmin$. The evidence for the submodel $\Minf^{+}$ and $\MFL^{+}$
has not be represented because $\sigma_\pm \ll 1$ and they are
indistinguishable from the ones plotted in \Fig{fig:evidences}.

\begin{figure}
  \begin{center}
    \includegraphics[width=\figmaxw]{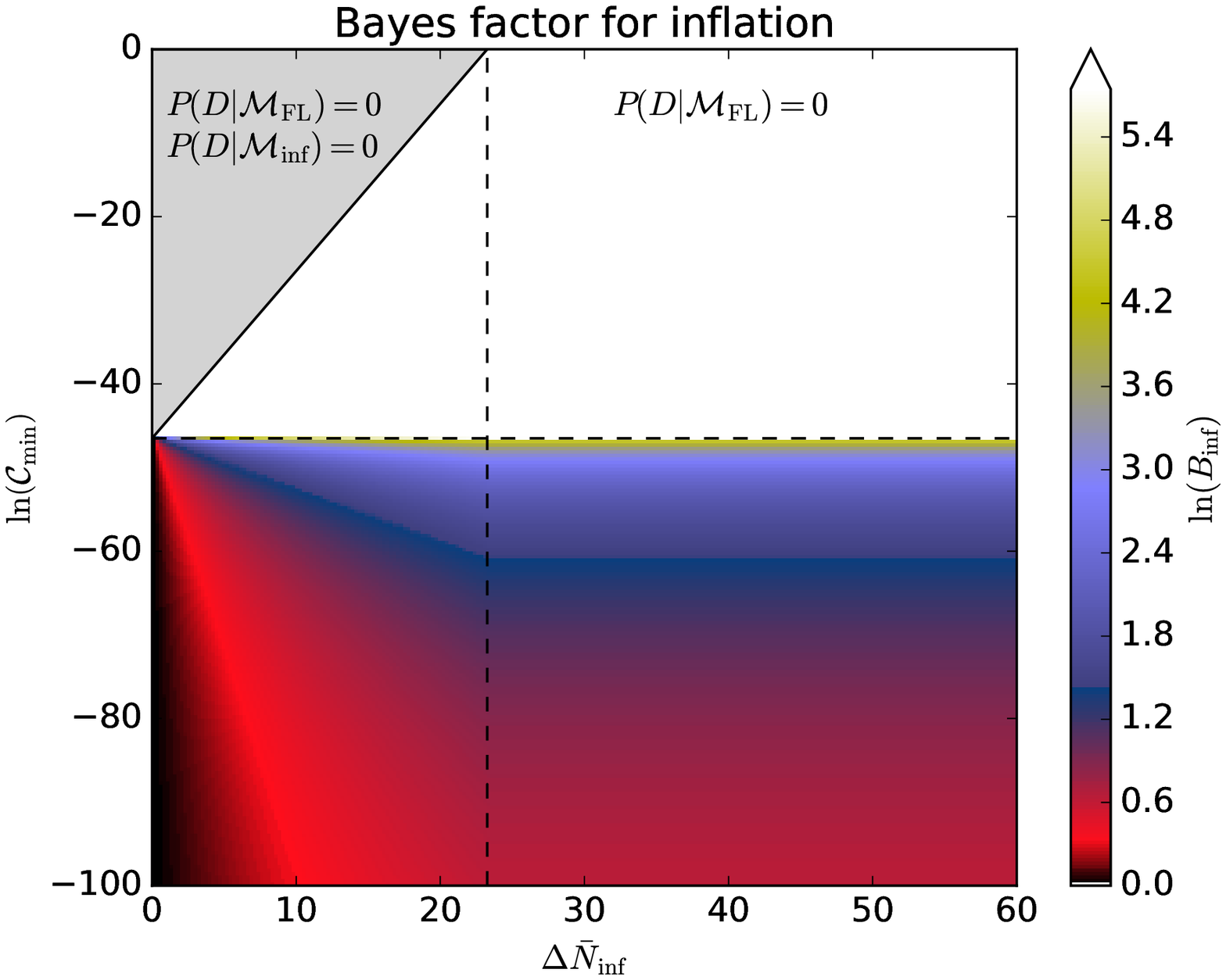}
    \caption{Bayes factor $\Binf \equiv p(D|\Minf)/p(D|\MFL)$ in
      favor of inflation in the prior parameters plane
      $(\Cmin,\Delta\Ninfbar)$. Up to the small upper-left corner
      where neither $\MFL$ nor $\Minf$ are compatible with the current
      measurement of curvature, inflation is $\emph{always}$
      favored. In the white region, inflation is compatible with the
      data whereas $\MFL$ is not and the Bayes factor for inflation is
      infinite. For very small values of $\Cmin$, $\ln\Binf < 1$ and
      the evidence in favor of inflation becomes ``inconclusive''
      according to the Jeffreys' scale (see Table~\ref{tab:js}).}
  \label{fig:bayes}
  \end{center} 
\end{figure}

\begin{table}
  \begin{center}
    \begin{tabular}{l l  l} \hline 
      $\ln \Binf$ & Odds  & Strength of evidence \\\hline 
      $<1.0$ & $\lesssim 3:1$ &  Inconclusive \\
      $1.0$ & $\sim 3:1$ &  Weak  \\
      $2.5$ & $\sim 12:1$ & Moderate  \\
      $5.0$ & $\sim 150:1$ &  Strong \\
      \hline
    \end{tabular}
    \caption{Jeffreys' scale in the strength of evidence in favor of
      inflation (from \Refs{Gordon:2007xm, Trotta:2008qt}).}
    \label{tab:js} 
  \end{center}
\end{table}

In \Fig{fig:bayes}, we have represented the logarithm of the Bayes factor
\begin{equation}
\Binf^{-} \equiv \dfrac{p(D|\Minf^{-})}{p(D|\MFL^{-})}\,,
\end{equation}
in favor of inflation. Starting with non-committal model priors,
i.e., $\pi(\Minf) = \pi(\MFL)$, $\Binf$ is the factor by how much
inflation is more probable than the always decelerating FL model. The
correspondence between the values of $\Binf$ and the so-called
Jeffreys' scale is reported in Table~\ref{tab:js}. As can be checked
in \Fig{fig:bayes}, up to the upper-left corner in which both models
are ruled out, this factor is \emph{always} greater than one. In the
white region for which $\Cmin > \Cflat$, $\Binf^{-}$ is infinite and
$\MFL^{-}$ is ruled out. For $\Cmin < \Cflat$, the evidence in favor
of inflation rapidly decreases from ``strong evidence'' around
$\Cflat$ to weak and inconclusive evidence around $\Cmin = e^{-100}$.

One would obtain almost the same results for the other sign of the
curvature and this is why we have not represented the Bayes factor in
favor of $\Minf^{+}$ with respect to $\MFL^{+}$. Therefore,
\Fig{fig:bayes} is also what one would obtain by adding up the
contribution of the two submodels and represents the overall Bayes
factor in favor of inflation.

\subsection{Discussion}

Let us stress again that the previous results are valid if one only
considers the measurement of $\OmegaKo$ today. Some values of $\varpi$
in the previous plots are actually incompatible with many other
measurements, such as the age of the universe, the density of matter,
\dots. Moreover, the values for $\No$, $\Delta\Nflat$ and $\Cflat$
reported before have been derived under the very conservative
hypothesis that $\MFL$ describes energies lower than $10\,\MeV$
only. If one assumes instead that the FL decelerated evolution starts
at higher energies, then $\Cflat$ would be pushed to much lower values
while $\Delta\Nflat$ would be approaching its fiducial value, around
$60$, the value usually reported for inflation to solve the flatness
and horizon problems.

\Fig{fig:bayes} shows that inflation is \emph{always a better model}
hypothesis than a purely decelerating FL model to solve the flatness
problem. Both models become equally probable only when one has a
theoretical prejudice in which exponentially low values of $\varpi$
within $\MFL$ are allowed. In particular, taking a flat prior over
$\varpi$ effectively consists in taking $\Cmin
\sim\order{1}$ where $\Binf$ is infinite.

Being agnostic, the prior maximizing ignorance is the one derived in
\Eq{eq:uninfprior} and requires to specify, at least, one input
parameter $\Cmin$ (in addition to $\Delta\Ninfbar$). The value of
$\Cmin$ mostly determines the strength of evidence by which inflation
is solving the flatness problem compared to standard FL decelerated
eras.  Let us remark that improving the accuracy by which $\OmegaKo$ is
measured, namely reducing the value of $\sigma_\pm$, pushes down the
value of $\Cflat$. However, this will not change the evidence for
inflation if one already has a prior with $\Cmin \ll
\Cflat$.

In short, if one believes that (or has a theoretical reason why)
having an initial curvature smaller than $\varpibbn = e^{-100}$
is natural within FL cosmology, then the evidence for inflation is
indeed ``inconclusive''. Otherwise, it seems inevitable.

\subsection{Other instabilities}

In the previous section, we have seen that using the prior (or
measure) that maximizes ignorance, the Bayesian evidence for inflation
to solve the flatness problem still depends on the minimal value
allowed for the initial curvature in the decelerating FL model. This
one has to be very small, $\varpibbn<e^{-100}$, to put both models on
an equal Bayesian footing.

Various attempts have been made to justify other priors that would
render very low values of the initial curvature more probable. As an
illustration of this line of reasoning, let us consider again the
measure used in \Ref{Gibbons:2006pa} and given by
\Eq{eq:gt}. Following \Ref{Carroll:2010aj} and using the fact that
$\OmegaK=-\calK/(a^2H^2)$, this measure can be rewritten as
\begin{align}
\label{eq:gtomegak}
\omega_{_\uGHS,\Ham=0,H=H_*}&=
- 3\sqrt{\frac{3}{2}}\frac{\Mp}{H_*^2\vert \OmegaK\vert^{5/2}}
\nonumber \\ & \times 
\frac{1-\Omega_V-2\OmegaK/3}{\sqrt{1-\Omega_V-\OmegaK}}
\dd \vert \OmegaK\vert \wedge \dd \phi, 
\end{align}
in agreement with Eq.~(39) of \Ref{Carroll:2010aj}. In the above
expression, one has defined $\Omega_V\equiv
V(\phi)/(3H_*^2\Mp^2)$. This measure has several problems, in
particular the fact that it blows up when $\OmegaK\to 0$. In
\Ref{Carroll:2010aj}, a scheme of regularization has been proposed
which leads to [see their Eq.~(56)]
\begin{align}
\label{eq:gtomegakreg}
\omega_{_{\mathrm{GHS}},\Ham=0,H=H_*}&\propto
\lim\limits_{\epsilon \rightarrow 0}\epsilon^{-3/2}
F_{\epsilon}(\OmegaK)
\nonumber \\ & \times 
\frac{1-\Omega_V-2\OmegaK/3}{\sqrt{1-\Omega_V-\OmegaK}}
\dd \vert \OmegaK\vert \wedge \dd \phi, 
\end{align}
where the function $F_\epsilon$ tends to a Dirac delta function when
$\epsilon$ goes to zero. As a consequence, \Ref{Carroll:2010aj} obtains
\begin{align}
\label{eq:gtomegakreg2}
\omega_{\uGHS,\Ham=0,H=H_*} \propto \sqrt{1-\Omega_V}\delta(\OmegaK) \, \dd \phi .
\end{align}
Then, \Ref{Carroll:2010aj} concludes
that there is no flatness problem at all. Of course, this argument is
based on the use of a regularized GHS measure and,
in \Sec{sec:probainf}, we have explained why this is difficult to
justify. But the point here is different and the previous
considerations simply illustrate that, if one is given a
measure strongly peaked at $\OmegaK=0$, then the flatness problem can
be alleviated.

Let us, however, stress that, in principle, the curvature parameter
cannot be taken arbitrarily small due to the presence of unavoidable
quantum or backreaction corrections~\cite{Bolejko:2017lai,
  Bolejko:2017gnj}. To understand this point, let us use an analogy
with another unstable system, namely a pencil balancing on its tip. If
one comes up with a measure in phase-space that is peaked at vanishing
tilt angle and initial velocity, it might be tempting to say that the
pencil never falls. One might even argue that, in the absence of a
well-justified measure, one cannot say whether it is surprising to
find a pencil balancing on its tip. In the real world, however, there
are always small fluctuations that cause the pencil to fall. Even if
one manages to completely suppress thermal fluctuations, because of
the Heisenberg uncertainty principle, both the initial angle and velocity
cannot be set to zero simultaneously, which causes the pencil to fall
in a few seconds~\cite{2014arXiv1406.1125L}. The same should happen
for the curvature parameter, although what value of $\Cmin$ quantum
fluctuations or cosmological backreaction imply remains to be
determined.

\section{The Multiverse}
\label{sec:multiverse}

There exists a last type of criticism that, so far, we have not
discussed, although it can play an important role in the
literature~\cite{Ijjas:2013vea,Ijjas:2014nta,Ijjas:2015hcc} (see also
\Ref{Guth:2013sya}). This criticism is related to the claim that
inflation necessarily implies the presence of a multiverse that, in
turn, would imply a total loss of predictive power.

Inflation might indeed create a multiverse structure when it enters
into the regime known as eternal inflation~\cite{Vilenkin:1983xq,
  1986PhLB..175..395L, Goncharov:1987ir}. Eternal inflation relies on
the presence of non-perturbative quantum fluctuations that could
affect the background geometry on scales much larger than the Hubble
radius. In this picture, our universe could be a small part of an
extremely large and inhomogeneous structure, filled with
self-reproducing inflationary bubbles. This picture, combined with the
effective-field-theory landscape, leads to a multiverse in which the landscape
vacua are populated by the eternal inflation
mechanism~\cite{Kachru:2003aw}, and the fundamental constants could
vary on the largest length scales.

Let us first notice that, while eternal inflation occurs in some
inflationary models, it is not necessarily present in all scenarios,
as concretely demonstrated \eg~in \Refs{Mukhanov:2014uwa,
  Dimopoulos:2016yep}. Reference~\cite{Ijjas:2015hcc} argues that this
type of models is ultra-sensitive to initial conditions. But,
according to the considerations in \Sec{sec:beyondI}, this should not
be the case since, even in the presence of corrections, there is no reason
not to join the slow-roll attractor solution on the plateau.

Second, the presence of eternal inflation is based on an extrapolation
of the inflationary potential very far from the observational window:
even if one has established that the potential is, say, a plateau in a
regime relevant for cosmology (for which eternal inflation could
occur), assuming it remains a plateau everywhere appears to be a
strong assumption. If one expects the potential to receive corrections
at high energies, then, the question becomes whether these corrections
allow the mechanism of eternal inflation to occur, a difficult
question since these corrections are usually hard to calculate in
detail.

Third, the fact that what we assume at the eternal inflation scale
could affect lower energy physics, shows (as it was already discussed
for the model building problem) that, to a certain extent, inflation
could have UV sensitivities. Although this can be seen as an undesired
feature of inflation, let us notice that the standard model of
particle physics also has some UV sensitivity (the hierarchy problem
for the Higgs mass) but is not discarded for this reason.

Fourth, as explicitly shown in \Refs{Ringeval:2010hf,
  Hardwick:2017qcw}, there are instances where super-Hubble quantum
fluctuations actually reinforce inflation's predictiveness, by washing
away dependence on initial conditions that would
otherwise be present. 

Fifth, in order to properly assess whether the predictive power of
inflation is increased or reduced once depicted in the multiverse
approach, one needs to justify a measure across the spacetime
manifold, and there is no clear procedure for
doing that~\cite{Winitzki:2006rn, Aguirre:2006ak, Linde:2007nm}. 

Our view is that inflation should be thought of as an effective theory
that needs to be tested within its domain of validity only. While the
search for connections with higher-energy physics is an interesting
and necessary endeavor, the fact that simple extrapolations might
cause problems (if they really do) cannot be used as an argument to
dismiss the entire model; for the same reason that discarding particle
physics on the basis that it is difficult to embed it in an unified
framework would appear quite radical.

\section{Conclusions}
\label{sec:conclusion}

With the advent of high-precision cosmological data, various
observational predictions of inflation have been confirmed, which
makes it one of the leading paradigms for describing the physics of
the very early Universe. However, inflation has also received
criticisms, and alternatives (matter bounce~\cite{Finelli:2001sr,
  Brandenberger:2009yt, Battefeld:2014uga, Brandenberger:2016vhg},
ekpyrotic~\cite{Khoury:2001wf, Khoury:2001bz} and cyclic
models~\cite{Steinhardt:2001st}, string gas
cosmology~\cite{Brandenberger:1988aj} -- as part of emergent universe
theories~\cite{Ellis:2003qz} --, etc.) have been proposed.

In this article we have reviewed these criticisms and discussed how
their status evolved (if it did) with the latest Planck measurements
of the CMB temperature and polarization anisotropies. The punchline
is that the class of inflationary potentials now favored by the data,
namely plateau potentials, is precisely the one for which some of the
criticisms raised against inflation are alleviated. This is notably  the case
for the initial condition problem, at least in an homogeneous
situation. We have also shown that this conclusion is robust under UV
completions of the models and alternative phase-space measures, and
we have underlined the crucial role played by the slow-roll attractor in
inflationary dynamics.

Although Planck favors the simplest implementations of inflation, such
as Higgs inflation where one does not need to go beyond the Standard
Model, or the Starobinsky model where a $R^2$ correction is simply
added to General Relativity, model building issues remain (such as,
for instance, the running of the coupling constants within these
scenarios). It would, however, be naive to expect something different
given that the energy scale of inflation can be as high as the Grand
Unified Theory (GUT) scale. The importance of these issues can be
assessed only by gaining more information about Physics at those
scales, and they do not constitute inconsistencies of the inflationary
mechanism per se.

Concerning the trans-Planckian problem of inflation, the fact that no
super-imposed oscillations have been found in the Planck data shows
that this effect is small. This confirms that the calculation of the
correlation function of inflationary perturbations is robust.

Regarding the multiverse and the criticisms that go with
it~\cite{Ijjas:2014nta, Ijjas:2015hcc}, it is true that the models
singled out by the data are usually associated with a multiverse. But
this is the case only if one assumes that the potential remains
unchanged way beyond the observational window, which is an additional
assumption. In any case, the presence of a multiverse, even in
potentials that would allow for it, remains speculative, and does not
necessarily imply a loss of predictive power, as we have argued. It
seems more reasonable to view inflation as an effective theory that
should be used within its domain of validity only. This is similar to
the way we use the standard model of particle physics, regardless of
what new physics lies beyond it, and despite the fact that this new
physics is likely to exist.

There are also criticisms the status of which has not really changed
after the Planck data. But we have found that most of these criticisms
(likelihood of inflation in the GHS measure, absence of a flatness
problem) are not really relevant for inflation. However, questions
related to difficulties in the interpretation of Quantum Mechanics
remain and affect the discussion about the inflationary mechanism for
structure formation.

In our opinion, the major challenge left for inflation is to better
determine the conditions under which it can naturally start from
highly inhomogeneous initial conditions. This is a difficult and
model-dependent question but, clearly, an important issue for having a
viable inflationary model.

It is also interesting to notice that the situation is not frozen and
that new observations could (and will) change how we view
inflation. For instance, the detection of primordial gravitational
waves at a level consistent with the simplest models favored by Planck
(namely $r \simeq 10^{-3}$) would clearly reinforce our confidence in
inflation at high energy. A smoking gun would be the verification of
the consistency relation $r=-8\nT$ but, unfortunately, given the
present bound on $r$ ($r\lesssim 0.07$), this already appears to be
technologically difficult. A measurement of non-Gaussianity would also
deeply affect the status of inflation and its preferred
implementations.

Inflation appears to be in a better shape after Planck than before,
even if, of course, the questions raised by
the inflationary theory have not all found fully satisfactory answers.

\begin{acknowledgments}
  D.~C. thanks the Centre Franco-Indien pour la Promotion de la
  Recherche Avanc\'{e}e (CEFIPRA), for financial support through the
  Raman-Charpak Fellowship 2016, which facilitated her visit to
  Institut d'Astrophysique de Paris, during which this work was
  commenced. D.~C. thanks the Centre for Cosmological Studies
  (University of Oxford), for financial support to allow a subsequent
  visit to Institut d'Astrophysique de Paris, during which this work
  was continued. D.~C. also thanks Institut d'Astrophysique de Paris
  for hospitality during her stay, and Indian Institute of Technology
  Madras and Tata Institute of Fundamental Research for financial
  support.  The work is C.~R. is partially supported by the ``Fonds de
  la Recherche Scientifique - FNRS'' under Grant
  $\mathrm{N^{\circ}T}.0198.19$. V.~V. acknowledges funding from the
  European Union's Horizon 2020 research and innovation program
  under the Marie Sk\l odowska-Curie Grant Agreement
  ${\textrm{N}}^\circ750491$.
\end{acknowledgments}

\begin{widetext}

\appendix

\section{Phase-space volume with the $t$-measure}
\label{app:PhaseSpaceVolume_t-Measure}

Let us recall that we consider two vectors $\bmu_1$ and $\bmu_2$ both
attached at time $t$ to the point $(\phi,\dot{\phi})$ in phase-space
and with components $(\delta\phi_i,\delta\Pi_i)$ for $i=1,2$
respectively. They describe an (algebraic) volume given by
\Eq{eq:V(t)}, namely $\vol\left(t\right)=\bmu_1\left(t\right)\wedge
\bmu_2\left(t\right) = \delta\phi_1 \delta\Pi_2 -
\delta\Pi_1\delta\phi_2$.
At time $t+\dd t$, the point that is located at $(\phi,\Pi)$ at time
$t$ moves to
\begin{equation}
\begin{aligned}
  \phi & \to \phi+\Pi\dd t,\\  
  \Pi & \to \Pi+\ddot{\phi}\left(\phi,\Pi\right)\dd t .
\end{aligned}
\end{equation}
In this expression, $\ddot{\phi}\left(\phi,\Pi\right)$ refers to the
equation of motion. In the present case, it is given by
\Eq{eq:kg}. Under this transformation, one obtains
\begin{align}
\delta\phi_i & \to \delta\phi_i + \delta\Pi_i \dd t\\
\delta\Pi_i & \to \delta\Pi_i 
+ \frac{\partial\ddot{\phi}\left(\phi,\Pi\right)}
{\partial\phi}\delta{\phi}_i \dd t 
+  \frac{\partial\ddot{\phi}\left(\phi,\Pi\right)}{\partial\Pi}
\delta\Pi_i \dd t.
\end{align}
This gives rise to the coordinates of the vectors $\bmu_i(t+\ud t)$
from which one can compute the volume they encompass at time
$t+\dd t$,
\begin{equation}
\begin{aligned}
\vol\left(t+\dd t\right) &  = \left( \delta\phi_1 \delta\Pi_2 -
\delta\Pi_1\delta\phi_2 \right) 
%\\ & \times
\left[1+\frac{\partial\ddot{\phi}\left(\phi,\Pi\right)}{\partial\Pi}
  \dd t +\order{\dd t^2} \right].
\end{aligned}
\end{equation}
Making use of \Eq{eq:V(t)}, one obtains
\begin{equation}
\label{eq:Vdot:tmeasure:1}
\dfrac{\dd \vol}{\dd t} =  \frac{\partial\ddot{\phi}
\left(\phi,\Pi\right)}{\partial\Pi} \vol\, ,
\end{equation}
from which one concludes that attractor behaviors correspond to
$\partial\ddot{\phi}\left(\phi,\Pi\right)/\partial\Pi<0$. For the
field space $t$-measure, \Eq{eq:kg} gives rise to
\begin{equation}
\label{eq:phiddot(phidot)}
\frac{\partial\ddot{\phi}\left(\phi,\Pi\right)}{\partial\Pi} =
-3H\left(1+\frac{ \Pi^2}{6\Mp^2H^2}\right) .
\end{equation}
Combining \Eqs{eq:Vdot:tmeasure:1} and~(\ref{eq:phiddot(phidot)})
gives rise to \Eq{eq:Vdot:tmeasure} in the main text.
 
\section{Volume conservation in Hamiltonian measure}
\label{app:liouville}

Let us now consider a vector $\bmu_i$ in phase-space with components
$\delta\phi_i,\delta p_{\phi,i},\delta a_i, \delta p_{a,i}$. Because of
the motion of the system, this vector changes and, after time $\dd t$,
the new components are given by
\begin{align}
\delta\phi_i & \to \delta\phi_i + \dfrac{\partial
  \dot{\phi}\left(\phi,p_\phi,a,p_a\right)}{\partial\phi}\delta\phi_i
\dd t + \dfrac{\partial
  \dot{\phi}\left(\phi,p_\phi,a,p_a\right)}{\partial p_\phi} \delta
p_{\phi,i} \dd t + \dfrac{\partial
  \dot{\phi}\left(\phi,p_\phi,a,p_a\right)}{\partial a }\delta a_i \dd
t + \dfrac{\partial \dot{\phi}\left(\phi,p_\phi,a,p_a\right)}{\partial
  p_a} \delta p_{a,i} \dd t, \\ \delta p_{\phi,i} & \to \delta
p_{\phi,i} + \dfrac{\partial
  \dot{p_\phi}\left(\phi,p_\phi,a,p_a\right)}{\partial\phi}\delta\phi_i
\dd t + \dfrac{\partial
  \dot{p_\phi}\left(\phi,p_\phi,a,p_a\right)}{\partial p_\phi} \delta
p_{\phi,i} \dd t + \dfrac{\partial
  \dot{p_\phi}\left(\phi,p_\phi,a,p_a\right)}{\partial a }\delta a_i
\dd t + \dfrac{\partial
  \dot{p_\phi}\left(\phi,p_\phi,a,p_a\right)}{\partial p_a} \delta
p_{a,i} \dd t,\\ \delta a_i & \to \delta a_i + \dfrac{\partial
  \dot{a}\left(\phi,p_\phi,a,p_a\right)}{\partial\phi}\delta\phi_i \dd
t + \dfrac{\partial \dot{a}\left(\phi,p_\phi,a,p_a\right)}{\partial
  p_\phi} \delta p_{\phi,i} \dd t + \dfrac{\partial
  \dot{a}\left(\phi,p_\phi,a,p_a\right)}{\partial a }\delta a_i \dd t
+ \dfrac{\partial \dot{a}\left(\phi,p_\phi,a,p_a\right)}{\partial p_a}
\delta p_{a,i} \dd t,\\ \delta p_{a,i} & \to \delta p_{a,i} +
\dfrac{\partial
  \dot{p_a}\left(\phi,p_\phi,a,p_a\right)}{\partial\phi}\delta\phi_i
\dd t + \dfrac{\partial
  \dot{p_a}\left(\phi,p_\phi,a,p_a\right)}{\partial p_\phi} \delta
p_{\phi,i} \dd t + \dfrac{\partial
  \dot{p_a}\left(\phi,p_\phi,a,p_a\right)}{\partial a }\delta a_i \dd
t + \dfrac{\partial \dot{p_a}\left(\phi,p_\phi,a,p_a\right)}{\partial
  p_a} \delta p_{a,i} \dd t.
\end{align}
As a consequence, the change in the volume generated by four vectors
$\bmu_1$, $\bmu_2$, $\bmu_3$ and $\bmu_4$ can be expressed as
\begin{align}
\vol(t) \to \vol(t)\left[1+\dfrac{\partial
    \dot{\phi}\left(\phi,p_\phi,a,p_a\right)}{\partial\phi}+\dfrac{\partial
    \dot{p_\phi}\left(\phi,p_\phi,a,p_a\right)}{\partial p_\phi} +
  \dfrac{\partial \dot{a}\left(\phi,p_\phi,a,p_a\right)}{\partial a
  }+\dfrac{\partial \dot{p_a}\left(\phi,p_\phi,a,p_a\right)}{\partial
    p_a }\right].
\end{align}
We now make use of Hamilton's equations, namely
$\dot{\phi}=\partial\Ham/\partial p_\phi$,
$\dot{p}_\phi=-\partial\Ham/\partial\phi$,
$\dot{a}=\partial\Ham/\partial p_a$ and
$\dot{p}_a=-\partial\Ham/\partial a$ and get
\begin{equation}
\vol(t)\to \vol(t)\left(1+\dfrac{\partial^2\Ham}{\partial\phi \partial
    p_\phi}-\dfrac{\partial^2\Ham}{\partial p_\phi \partial \phi}
  +\dfrac{\partial^2\Ham}{\partial a \partial
    p_a}-\dfrac{\partial^2\Ham}{\partial p_a \partial a} \right) =
\vol(t).
\end{equation}
Therefore, volumes in phase-space are conserved under the Hamiltonian
evolution.

\section{Phase-space volume with the $N$-measure}
\label{app:PhaseSpaceVolume_N-Measure}

For the $N$-measure, the calculation follows in all points the one
presented for the $t$-measure in
\App{app:PhaseSpaceVolume_t-Measure}. Along the phase-space
trajectories, after $\ud N$ e-folds of evolution, one has
\begin{equation}
\begin{aligned}
  \phip &\to \phip + \Gamma \ud N, \\
  \Gamma &\to \Gamma + \DGamma(\phip,\Gamma)\, \ud N,
\end{aligned}
\end{equation}
where $\DGamma(\phip,\Gamma)$ can be read off from \Eq{eq:kgefold} and
is given by
\begin{equation}
  \DGamma(\phip,\Gamma) \equiv -\left(3 -\dfrac{\Gamma^2}{2}
  \right) \left[\Gamma - \Gammasr(\phip) \right].
\label{eq:DGammafunc}
\end{equation}
Similar to \Eq{eq:Vdot:tmeasure:1}, an infinitesimal volume in phase
space $( \phip,\Gamma)$ evolves according to
\begin{equation}
\dfrac{\ud \vol}{\ud N} = \dfrac{\partial
  \DGamma(\phip,\Gamma)}{\partial \Gamma} \vol = \left[-3\left(1 -
\dfrac{1}{2} \Gamma^2 \right) - \Gamma \Gammasr(\phip)\right] \vol.
\end{equation}
Introducing
\begin{equation}
\Gamma_{\pm}(\phip) = \pm \sqrt{2 + \dfrac{\Gammasr^2(\phip)}{9}} 
+ \dfrac{\Gammasr(\phip)}{3}\,,
\end{equation}
this gives rise to \Eq{eq:Vdot:Nmeasure} in the main text.

\section{Flow compression with the $t$-measure}
\label{app:flowCompression:naive}

Let us consider a point $M_0$ in phase-space of coordinates
$M_0(\phi_0,\dot{\phi}_0)$ at time $t$. In the neighborhood of $M_0$,
its orbit is parametrized by
\begin{align}
 \orb\left(M_0\right)\simeq \left\lbrace\left[ \phi_0+\dot{\phi}_0 u,
   \dot{\phi}_0+\ddot{\phi}
\left(\phi_0,\dot{\phi}_0\right)u\right],u\in\mathbb{R}\right\rbrace,
\end{align}
where $u$ is a dummy parameter and, as explained in
\App{app:PhaseSpaceVolume_t-Measure},
$\ddot{\phi}\left(\phi,\dot{\phi}\right)$ is given by \Eq{eq:kg}. At
time $t+\dd t$, this point becomes $M_1$ with coordinates
$M_1[\phi_0+\dot{\phi}_0 \dd t,
\dot{\phi}_0+\ddot{\phi}\left(\phi_0,\dot{\phi}_0\right)\dd t]$
which obviously belongs to $\orb(M_0)$. We now consider another point
$N_0$ infinitesimally displaced away from $M_0$,
$N_0(\phi_0+\delta\phi,\dot{\phi}_0+\delta\dot{\phi})$, at time
$t$. In the neighborhood of $N_0$ (and of $M_0$), its orbit is given
by
\begin{align}
\orb\left(N_0\right)\simeq & \left\lbrace \left[
  \phi_0+\delta\phi+\left(\dot{\phi}_0 +\delta\dot{\phi}\right)u,
%  \right. \right. \nonumber \\ & \left. \left.
  \dot{\phi}_0+\delta\dot{\phi} + \ddot{\phi}\left(\phi_0 +
  \delta\phi,\dot{\phi}_0+\delta\dot{\phi}\right)u\right],
u\in\mathbb{R}\right\rbrace\,
.
\end{align}
At time $t+\dd t$, this point becomes $N_1$ with coordinates
\begin{equation}
\begin{aligned}
  N_1[&\phi_0+\delta\phi+\left(\dot{\phi}_0
    +\delta\dot{\phi}\right)\dd t,
 %   \\ &
    \dot{\phi}_0+\delta\dot{\phi}+\ddot{\phi} \left(\phi_0 +
    \delta\phi,\dot{\phi}_0+\delta\dot{\phi}\right)\dd t],
\end{aligned}
\end{equation}
which obviously belongs to $\orb(N_0)$. Let us now calculate how
the distance $\dist$ between these two points evolves between times
$t$ and $t+\dd t$.

Let us consider a point $N$ belonging to $\orb(N_0)$ with parameter
$u$. The distance between $M_0$ and $N$ is given by
\begin{align}
\label{eq:D4}
d^2\left(M_0,N\right)=\mu^2\left[\delta\phi +
  \left(\dot{\phi}_0+\delta\dot{\phi}\right)u\right]^2
+\left[\delta\dot{\phi}+\ddot{\phi}\left(\phi_0+\delta\phi,\dot{\phi}_0
  +\delta\dot{\phi}\right)u\right]^2.
\end{align}
By minimizing the above quantity over $u$, one finds 
\begin{align}
\label{eq:D5}
d^2\left[M_0,\orb\left(N_0\right)\right] =
\dfrac{\mu^2\left[\delta\dot{\phi}\left(\dot{\phi}_0+\delta\dot{\phi}\right)
    - \delta\phi \ddot{\phi}\left(\phi_0+\delta\phi,\dot{\phi}_0 +
    \delta\dot{\phi}\right)\right]^2}{\mu^2\left(\dot{\phi}_0 +
  \delta\dot{\phi}\right)^2+\ddot{\phi}^2\left(\phi_0+\delta\phi,\dot{\phi}_0
  +\delta\dot{\phi}\right)}\, .
\end{align}
In the same manner, one can consider a point $M$ belonging to
$\orb(M_0)$ with parameter $u$. The distance between $N_0$ and $M$ is
given by
\begin{align}
\label{eq:D6}
d^2\left(N_0,M\right)=\mu^2
\left(
\delta\phi-\dot{\phi}_0 u
\right)^2+
\left[
\delta\dot{\phi}-\ddot{\phi}\left(\phi_0,\dot{\phi}_0\right)u
\right]^2\, .
\end{align}
By minimizing this quantity over $u$, one finds
\begin{align}
\label{eq:D7}
d^2\left[N_0,\orb\left(M_0\right)\right]=\dfrac{\mu^2\left[
    \dot{\phi}_0\delta\dot{\phi}
    -\delta\phi\ddot{\phi}\left(\phi_0,\dot{\phi}_0\right)
    \right]^2}{\mu^2\dot{\phi}_0^2+
  \ddot{\phi}^2\left(\phi_0,\dot{\phi}_0\right)}\, .
\end{align}
Let us now consider again a point $N$ belonging to
$\orb(N_1)=\orb(N_0)$ with parameter $u$. The distance between $M_1$
and $N$ is given by
\begin{align}
\label{eq:D8}
d^2\left(M_1,N\right)=\mu^2\left[\delta\phi
  +\left(\dot{\phi}_0+\delta\dot{\phi}\right)u-\dot{\phi}_0\dd
  t\right]^2
+\left[\delta\dot{\phi}+\ddot{\phi}\left(\phi_0+\delta\phi,\dot{\phi}_0
  +\delta\dot{\phi}\right)u-\ddot{\phi}\left(\phi_0,\dot{\phi}_0\right)\dd
  t\right]^2.
\end{align}
By minimizing this quantity over $u$, one finds
\begin{align}
d^2\left[M_1,\orb\left(N_1\right)\right]=\dfrac{\mu^2\left\lbrace
  \left(\dot{\phi}_0+\delta\dot{\phi}\right)
  \left[\delta\dot{\phi}-\ddot{\phi}\left(\phi_0,\dot{\phi}_0\right)\dd
    t\right] - \left(\delta\phi-\dot{\phi}_0\dd t\right)
  \ddot{\phi}\left(\phi_0+\delta\phi,\dot{\phi}_0
+\delta\dot{\phi}\right)
  \right\rbrace^2}{\mu^2\left(\dot{\phi}_0
+\delta\dot{\phi}\right)^2+\ddot{\phi}^2
\left(\phi_0+\delta\phi,\dot{\phi}_0+\delta\dot{\phi}\right)}\,
.
\end{align}
In the same manner, considering a point $M$ belonging to
$\orb(M_1)=\orb(M_0)$ with parameter $u$, the distance between $N_1$
and $M$ is given by
\begin{align}
d^2\left(N_1,M\right)=\mu^2\left[\delta\phi+\left(\dot{\phi}_0
+\delta\dot{\phi}\right)\dd t - \dot{\phi}_0 u \right]^2
+\left[\delta\dot{\phi}+\ddot{\phi}\left(\phi_0+\delta\phi,\dot{\phi}_0
+\delta\dot{\phi}\right)\dd t-\ddot{\phi}\left(\phi_0,\dot{\phi}_0\right) 
u\right]^2\, .
\end{align}
By minimizing this quantity over $u$, one finds
\begin{align}
\label{eq:D11}
d^2\left[N_1,\orb\left(M_1\right)\right]= \dfrac{\mu^2\left\lbrace
  \dot{\phi}_0\left[\delta\dot{\phi}+\ddot{\phi}
\left(\phi_0+\delta\phi,\dot{\phi}_0+\delta\dot{\phi}\right)\dd
    t\right]
  -\left[\delta\phi+\left(\dot{\phi}_0+\delta\dot{\phi}\right)\dd
    t\right] \ddot{\phi}\left(\phi_0,\dot{\phi}_0\right)
  \right\rbrace^2}{\mu^2\dot{\phi}_0^2 +
  \ddot{\phi}^2\left(\phi_0,\dot{\phi}_0\right)}\, .
\end{align}
Expanding the previous expressions at quadratic order in $\delta\phi$ and
$\delta\dot{\phi}$, and
at leading order in $\dd t$, one obtains
\begin{align}
\label{eq:D12}
d^2\left[M_0,\orb\left(N_0\right)\right]& =
d^2\left[N_0,\orb\left(M_0\right)\right]
=\dfrac{\mu^2\left(\dot{\phi}_0\delta\dot{\phi}
-\delta\phi\ddot{\phi}_0\right)^2}{\mu^2\dot{\phi}_0^2+\ddot{\phi}_0^2}\,,
\\
d^2\left[M_1,\orb\left(N_1\right)\right] &
= d^2\left[N_1,\orb\left(M_1\right)\right] =
\dfrac{\mu^2\left(\dot{\phi}_0\delta\dot{\phi}
  -\delta\phi\ddot{\phi}_0\right)^2}{\mu^2\dot{\phi}_0^2+\ddot{\phi}_0^2}
  %\\ &
 +2\dd t\dfrac{\mu^2 \left(\dot{\phi}_0\delta\dot{\phi}
  -\delta\phi\ddot{\phi}_0\right)}{\mu^2\dot{\phi}_0^2+\ddot{\phi}_0^2}
\left(\dot{\phi}_0\dfrac{\partial\ddot{\phi}}{\partial\phi}\delta\phi
+\dfrac{\partial\ddot{\phi}}{\partial\dot{\phi}}\dot{\phi}_0\delta\dot{\phi}
-\ddot{\phi}_0\delta\dot{\phi}\right)\, ,
\label{eq:D13}
\end{align}
where we have defined
$\ddot{\phi}_0 \equiv \ddot{\phi}\left(\phi_0,\dot{\phi}_0\right)$.
It is interesting to notice that at leading order,
$d^2[M_0,\orb(N_0)] =d^2[N_0,\orb (M_0 )]$ and
$d^2 [M_1,\orb(N_1)] =d^2[N_1,\orb(M_1)]$. This implies that in the
definition \eqref{eq:OrbitDistance:def}, any way to symmetrize the
expression would give the same result below. The above expression
gives rise to
\begin{align}
\dfrac{\dd \dist^2}{\dd t} = 2 \dfrac{\mu^2
  \left(\dot{\phi}_0\delta\dot{\phi}
  -\delta\phi\ddot{\phi}_0\right)}{\mu^2\dot{\phi}_0^2+\ddot{\phi}_0^2}
\left(\dot{\phi}_0\dfrac{\partial\ddot{\phi}}{\partial\phi}\delta\phi
+\dfrac{\partial\ddot{\phi}}{\partial\dot{\phi}}\dot{\phi}_0\delta\dot{\phi}
-\ddot{\phi}_0\delta\dot{\phi}\right),
\end{align}
and one obtains
\begin{align}
\label{eq:d_dist_d_t_t_measure:gen}
\dfrac{\dd \ln \dist}{\dd t} =
\dfrac{\dot{\phi}_0\dfrac{\partial\ddot{\phi}}{\partial\phi}\delta\phi +
  \dfrac{\partial\ddot{\phi}}{\partial\dot{\phi}}\dot{\phi}_0\delta\dot{\phi}
  - \ddot{\phi}_0\delta\dot{\phi}}{\dot{\phi}_0\delta\dot{\phi} -
  \delta\phi\ddot{\phi}_0}\, .
\end{align}
Making use of \Eq{eq:kg} to evaluate the function
$\ddot{\phi}(\phi,\dot{\phi})$, this gives rise to
\Eq{eq:d_dist_d_t_t_measure} in the main text.

\section{Flow compression with the Hamiltonian 
induced measure}
\label{app:flowCompression:Hamiltonian}

As explained in \Sec{sec:him}, in order to be explicitly defined, the
Hamiltonian induced measure needs to come with a choice of
hypersurfaces of constant time in phase-space. Here, this choice is
given by the scale $\mu$, in such a way that distances computed at a
given time $t$ are measured along phase-space directions where the
scale factor is constant and equal to $a(t)$, such that
$d^2 = \mu^2 \delta\phi^2 + a^6(t)\delta\dot{\phi}^2$.

The calculation then proceeds in a way that is very similar to
\App{app:flowCompression:naive}. For instance, \Eq{eq:D4} needs to be
replaced with
\begin{align}
d^2\left(M_0,N\right)=\mu^2\left[\delta\phi +
  \left(\dot{\phi}_0+\delta\dot{\phi}\right)u\right]^2
+a^6(t)\left[\delta\dot{\phi}+\ddot{\phi}\left(\phi_0+\delta\phi,\dot{\phi}_0
  +\delta\dot{\phi}\right)u\right]^2,
\end{align}
which gives rise to
\begin{align}
d^2\left[M_0,\orb\left(N_0\right)\right] =
\dfrac{\mu^2 a^6(t)\left[\delta\dot{\phi}\left(\dot{\phi}_0+\delta\dot{\phi}\right)
    - \delta\phi \ddot{\phi}\left(\phi_0+\delta\phi,\dot{\phi}_0 +
    \delta\dot{\phi}\right)\right]^2}{\mu^2\left(\dot{\phi}_0 +
  \delta\dot{\phi}\right)^2+a^6(t)\ddot{\phi}^2\left(\phi_0+\delta\phi,\dot{\phi}_0
  +\delta\dot{\phi}\right)}
\end{align}
instead of \Eq{eq:D5}. The same modifications apply to \Eqs{eq:D6}
and~(\ref{eq:D7}), as well as to \Eqs{eq:D8}-(\ref{eq:D11}) except
that the scale factor needs to be evaluated at time $t+\dd t$ in the
latter set of equations. Instead of \Eqs{eq:D12} and~(\ref{eq:D13}),
one then has
\begin{align}
d^2\left[M_0,\orb\left(N_0\right)\right] & =
d^2\left[N_0,\orb\left(M_0\right)\right] =
\dfrac{\mu^2a^6(t)\left(\dot{\phi}_0\delta\dot{\phi}
-\delta\phi\ddot{\phi}_0\right)^2}{\mu^2\dot{\phi}_0^2+a^6(t)\ddot{\phi}_0^2}\,,\\
d^2\left[M_1,\orb\left(N_1\right)\right] & =
d^2\left[N_1,\orb\left(M_1\right)\right]=\dfrac{\mu^2a^6(t+\dd t)\left[
    \dot{\phi}_0\delta\dot{\phi} - \delta\phi\ddot{\phi}_0
    +\dot{\phi}_0 \dfrac{\partial\ddot{\phi}}{\partial\phi}
    \delta\phi\dd t
    +\left(\dfrac{\partial\ddot{\phi}}{\partial\dot{\phi}}\dot{\phi}_0
    -\ddot{\phi}_0\right)\delta\dot{\phi}\dd
    t\right]^2}{\mu^2\dot{\phi}_0^2 +
  a^6(t+\dd t)\ddot{\phi}^2_0}
+\orderb{\left(\delta\phi,\delta\dot{\phi}\right)^3}.
\end{align}
One then obtains an additional term compared to the $t$-measure, that does not depend on $\delta\phi$ and $\delta\dot{\phi}$, namely
\begin{align}
\dfrac{\dd \ln \dist}{\dd t} = \dfrac{\dot{\phi}_0
\dfrac{\partial\ddot{\phi}}{\partial\phi}\delta\phi
+\dfrac{\partial\ddot{\phi}}{\partial\dot{\phi}}
\dot{\phi}_0\delta\dot{\phi}-\ddot{\phi}_0\delta
\dot{\phi}}{\dot{\phi}_0\delta\dot{\phi}-\delta\phi\ddot{\phi}_0}
+3\dfrac{\mu^2 H \dot{\phi}_0^2}{\mu^2\dot{\phi}_0^2+\ddot{\phi}_0^2}\, .
\end{align}
In terms of the slow-roll functions, this gives rise to
\Eq{eq:d_dist_dt:Hamiltonian_measure:gen} in the main text.

\end{widetext}

\bibliographystyle{apsrev}
\bibliography{biblio}

\end{document}